\documentclass[a4paper,11pt]{article}
\pdfoutput=1 

\usepackage{jheppub} 

\usepackage[T1]{fontenc} 
\usepackage{hyperref}
\usepackage{amsmath}

\DeclareMathOperator{\arctanh}{arctanh}

\usepackage{float}
\usepackage{graphicx}
\usepackage{epsfig}
\usepackage{psfrag}
\usepackage{color}
\usepackage{slashed}
\usepackage{epstopdf}
\usepackage{amsfonts}
\usepackage{amssymb}
\usepackage{mathtools}

\newcommand*{\be}{\begin{equation}}
\newcommand*{\ee}{\end{equation}}
\newcommand*{\bea}{\begin{eqnarray}}
\newcommand*{\eea}{\end{eqnarray}}

\newcommand{\comment}[1]{}

\newcommand{\cref}[1]{Chapter~\ref{c.#1}}

\newcommand{\barr}{\begin{eqnarray}}
\newcommand{\earr}{\end{eqnarray}}

\def\nn{\nonumber \\}

\def\beq{\begin{equation}}
\def\eeq{\end{equation}}
\def\bea{\begin{eqnarray}}
\def\eea{\end{eqnarray}}
\def\ba{\begin{array}}
\def\ea{\end{array}}
\def\bi{\begin{itemize}}
\def\ei{\end{itemize}}
\def\be{\begin{enumerate}}
\def\ee{\end{enumerate}}
\def\bc{\begin{center}}
\def\ec{\end{center}}
\def\bt{\begin{table}}
\def\et{\end{table}}
\def\btb{\begin{tabular}}
\def\etb{\end{tabular}}

\def\ki{\chi_i^0}
\def\kj{\chi_j^0}
\def\kl{\chi_l^\pm}
\def\kk{\chi_k^\pm}
\def\knk{\chi_k^0}
\def\knl{\chi_l^0}

\def\taua{\tau_{\phi_1\phi_2}}
\def\taub{\tau_{\phi_2\phi_1}}
\def\omegaa{\omega_{\phi_1\phi_2}}
\def\omegab{\omega_{\phi_2\phi_1}}
\def\ia{(m_i^2-m_j^2)}
\def\ka{(m_1^2-m_2^2)}
\def\rua{R_{\phi_1\chi}}

\def\ruaa{R_{\phi_1\chi_1}}
\def\rubb{R_{\phi_2\chi_2}}
\def\ba{b_{\phi_1\phi_2}}
\def\da{d_{\phi_1\phi_2}}

\def\lsim{\raise0.3ex\hbox{$\;<$\kern-0.75em\raise-1.1ex\hbox{$\sim\;$}}}
\def\gsim{\raise0.3ex\hbox{$\;>$\kern-0.75em\raise-1.1ex\hbox{$\sim\;$}}}

\title{Neutralino-Neutralino pair annihilation crosssections with flavour violation}
	\author{V. Suryanarayana Mummidi}
	\emailAdd{soori9@cts.iisc.ernet.in}
	\affiliation{Centre for High Energy Physics, Indian Institute of Science,
		Bangalore 560012}
\abstract{We present the first exact and most general neutralino-neutralino annihilation cross sections in the presence of flavor violation. The existing set of calculations in literature do not include flavour violation. These calculations are important for the calculation of relic density and indirect detection of dark matter in various experiments. }
\begin{document}
\maketitle	
\section{Introduction}
One of the unresolved issues of the twentieth century physics is the nature of dark matter. As per the present knowledge on the content of the Universe, major part of the Universe is not made of protons and neutrons but by a new form of matter called dark matter. It is six times more abundant than the matter in the visible Universe. 

There is sufficient evidence suggesting the existence of dark matter. And it does not fall into the category of any known matter. One of the well studied dark matter candidates are Weakly Interacting Massive Particles(WIMPS). WIMPs are electrically neutral, stable and they interact through weak and gravitational interactions. There is no particle, in Standard Model (SM), which could act as dark matter candidate except for neutrino. As neutrinos are stable, weakly interacting and electrically neutral they are suitable candidate for dark matter. However, neutrinos being relativistic particles they could only contribute to the hot dark matter. But to explain the structure formation in the Universe dark matter has to be cold. One of the popular extensions of SM which has darkmatter candidate is Minimally Supersymmetric Standard Model (MSSM)\cite{Martin:1997ns}. An attractive feature of supersymmetric models is the existence of quantum number called R-symmetry. Under this symmetry non supersymmetric particle has R quantum number +1 while their super partners are assigned -1.  As all the supersymmetric interactions should obey R-symmetry, supersymmetric particles are produced and annihilated in pairs. This makes lightest supersymmetric particles stable. If lightest supersymmetric particle were  neutral it could be a dark matter candidate. In R-parity conserving supersymmetric extensions of the Standard Model there must exist a LSP which could be a dark matter candidate. They exist in the early Universe along with other particles and eventually annihilate and coannihilate into Standard Model particles. As the Universe expands some of the LSPs will cease to interact and will be left over in the present Universe as thermal relics. More detailed calculation of thermal relic density of dark matter is presented in the next section.

In MSSM, owing to the fact that nuetral gauginos bino, wino and Higgsino have identical weak isospin qunatum number, they mix to form four mass eigen states called Neutralinos. They are denoted by ${\chi}_1^0,{\chi}_2^0,\chi_3^0$ and $\chi_4^0$. If R-parity is conserved, lightest neutralino is the LSP in most of the models. It is compelling to consider neutralinos in MSSM as the dark matter candidate. In the mixed basis, $\psi^T=(\tilde B,\tilde W,\tilde H_1,\tilde H_2)$, the neutralino part of the Lagrangian is
\begin{align}
\mathcal{L}_{neutralino}&=-\frac{1}{2}\,\psi^T M_\chi\psi
\end{align}
Nuetralino mass matrix $M_\chi$ is given by
\begin{align}
\left(
\begin{array}{cccc}
M_1 & 0 & -{e\,v_1 \over 2\,c_W} & {e\,v_{2} \over
	2\,c_W} \\ 
0 & M_2 & {ev_{1} \over 2\,s_W} & -{e\,v_{2} \over 
	2\,s_W} \\ 
-{e\,v_{1} \over 2\,c_W} & {e\,v_{1} \over 2\,s_W} & 0
& -\mu \\ 
{e\,v_{2} \over 2\,c_W} & -{e\,v_{2} \over 2\,s_W} &-\mu
& 0
\end{array}\right)
\end{align}
where $M_1$ and $M_2$ are masses of neutral gauginos Bino and Wino respectively. $\mu$ is the higgsino mass parameter, `e' is the gauge coupling constant, $v_1$ and $v_2$ are vacuum expectation values and $\theta_W$ ($c_W=\cos \theta_W$ and $s_W=\sin \theta_W$) is the Weinberg angle. Mass eigen states can be obtained by diagonalizing the neutralino
mass matrix $M_\chi$ by a unitary matrix $Z_N$:
\begin{align}
\chi^0_i&=(Z_N)_{ij}\,\psi_j
\end{align} 
so that
\begin{align}
Z_N^{-1}\,M_\chi Z_N&=\left(
\begin{array}{cccc}
m_{\chi_1^0}&0&0&0\\
0&m_{\chi_2^0}&0&0\\
0&0&m_{\chi_3^0}&0\\
0&0&0&m_{\chi_4^0}
\end{array}\right)
\end{align} 
 With high precision measurements of dark matter relic density, it may not be of distant possibility to constrain supersymmetric models. If heavier supersymmetric states predicted by supersymmetry are discovered at LHC, one can study the possibility that neutralino is the dark matter particle. In this context knowing the thermal relic density precisely has utmost importance. In standard cosmology thermal relic density of dark matter is given by
\begin{align}
\Omega_{\chi}&=\frac{n_\chi\,m_\chi}{\rho_{c}}
\end{align}
where $\rho_c$ is the critical density given by $\frac{3\,H^2}{8\,\pi\,G}$. G and H are gravitational constant and present day Hubble parameter respectively. The solution of Boltzmann equation gives the number density of any particle in the Early Universe as Universe evolves. We have assumed that dark matter is thermally produced and is in thermal equilibrium with the background in the early phase of the Universe, its number density $n_\chi$ can be calculated by solving Boltzmann rate equation 
\begin{align}
\frac{d n_\chi}{dt}+3 H n_\chi=- \langle \sigma|v|\rangle(n_\chi^2-n_{\chi,eq}^2)
\label{boltz}
\end{align}
Where t is time, and $\langle \sigma_{X\bar X}|v|\rangle$ is the thermally averaged cross section times the relative velocity. One can solve eq.~(\ref{boltz}) to obtain the number density at t=0 (today) provided, $ \langle \sigma|v|\rangle$ is known. As it was pointed out in \cite{Griest:1990kh}, if the masses of the LSP and next to lightest supersymmetric particle (nLSP) are degenerate  coannihilations between LSP and nLSP prominently alters the number density. So a general dark matter desity calculation should include coannihilation of neutralino(LSP) with nLSP. In MSSM, nLSP could be neutralino, chargino or a slepton depending on the boundary conditions at the supersymmetry breaking scale\footnote{Stops or gluinos could also be nLSP. One has to do case by case study of cross sections}. Therefore, the general calculation of neutralino number density involves consideration of following cross sections:\\
(i)  Neutralino (LSP)-Neutralino(LSP)\hspace{2cm}
(ii) Neutralino (LSP)-Neutralino (nLSP)\\
(iii)Neutralino (LSP)-Chargino  (nLSP)\hspace{1.7cm}
(iv) Neutralino (LSP)- Slepton    (nLSP)\\

So the problem of calculation of today's neutralino dark matter relic density is usually the calculation of the cross sections with the initial states given in above cases (i), (ii), (iii) and (iv)  into all standard model particles. The relic density of neutralinos in the MSSM has been
calculated by several authors over the years \cite{Griest:1990kh,Srednicki:1988ce,McDonald:1992ee,Mizuta:1992qp,Drees:1992am,Drees:1996pk,Edsjo:1997bg,Nihei:2002ij,Nihei:2002sc,Choudhury:2011um,Herrmann:2011xe}. Nihei etal. \cite{Nihei:2002ij, Nihei:2002sc} have considered Neutralino cross sections extensively except for the case (ii). They also have assumed neutralino mixing matrix $Z_N$ to be real for simplicity and not included flavour violation in the supersymmetric soft sector. Authors in\cite{Choudhury:2011um,Herrmann:2011xe} have considered effect of flavor violation on coannihilation and annihilations. In most calculations of the cross sections flavor violation is not considered seriously because of the existing low energy constraints from rare decays like $\mu~ \rightarrow~ e\gamma$,  $\tau~ \rightarrow~ \mu\gamma$, etc. in the leptonic sector and $b~ \rightarrow~ s\gamma$, $B-\bar B^0$, $B_s$-${\bar B}_s $, in the hadronic sector. However, it is known that there are regions in the parameter space where these constraints are highly weakened. In the leptonic sector cancellations of amplitude of $l_j~\rightarrow l_i \gamma$ are widely noted \cite{Hisano:1995cp}. These parameter regions increase if one considers a more general SUSY breaking framework like the one with the non-universal gauge  masses\cite{Profumo:2003ema,Baer:2005bu}. In the hadronic sector, constraints are weakened typically for a relatively heavy squark spectrum $\sim $1 TeV or so. In fact, so much so that, in mini-split models one expects a little bit of flavor violation to be present in the 2-3 sector\cite{Sundrum:2009gv,ArkaniHamed:2004yi}.

The purpose of our paper is to present most general neutralino(LSP)-neutralino (LSP/nLSP) cross sections without assuming minimal flavor violation or real mixing matrix for neutral gauge eigen state matrix $Z_N$. And we present case (iii) and (iv) in our future work. We hope to extend this analysis Sommerfeld enhancement phenomena in a future work. It should be noted that these cross sections are also useful to study indirect detection of dark matter. With flavor becoming important in the future experiments, the large presence of flavor violation in the supersymmetric soft sector could possibly probed by indirect experiments. 

The neutralino pair annihilation and coannihilation cross section with neutralino in this paper is organized and presented as follows. In table.\ref{cross}, we present the various possible SM final states. For each set of final states, we present the channels and the corresponding mediators. In section.\ref{relic}, we sketch the calculation of thermal relic density. In section.\ref{cal}, we present cross sections corresponding to final stated presented in table.\ref{cross}.
\begin {table}
\setlength{\tabcolsep}{25pt}
\centering
\begin {tabular}{|c |c |c |}
\hline
Process & \multicolumn{2}{|c|} {Exchanged Particle}\\ \cline{2-3} &s-channel&t- and u-channel\\
\hline\hline
$\ki\kj\rightarrow H_m^0H_n^0$&$H_k^0$&$\knk$\\
\hline
$\ki\kj\rightarrow A_1^0A_1^0$&$H_k^0$&$\knk$\\
\hline
$\ki\kj\rightarrow H_1^0A_1^0$&$A_1^0$,Z&$\knk$\\
\hline
$\ki\kj\rightarrow H_2^0A_1^0$&$H_k^0$&$\knk$\\
\hline
$\ki\kj\rightarrow H_1^+H_1^-$&$H_k^0$, Z&$\kk$\\
\hline
$\ki\kj\rightarrow W^\pm H_1^\mp$&$H_k^0$, $ A_1^0$&$\kk$\\
\hline
$\ki\kj\rightarrow ZH_n^0$&Z, $ A_1^0$&$\knk$\\
\hline
$\ki\kj\rightarrow ZA_1^0$&$ H_k^0$&$\knk$\\
\hline
$\ki\kj\rightarrow W^+W^-$&$ H_k^0$, Z&$\kk$\\
\hline
$\ki\kj\rightarrow ZZ$&$ H_k^0$&$\knk$\\
\hline
$\ki\kj\rightarrow f_m\bar f_n$&$ H_k^0, A_1^0$, Z&$\tilde f_k$\\
\hline
\end {tabular}
\caption {A complete set of neutralino pair-annihilation channels into tree-level two-body final
	states in the MSSM. The index k runs as follows: For neutalinos k=1,..,4; for charginos k=1,2; for sfermions k=1,...,6; for Higgs k=1,2 and for fermions m,n=1,..,6}
\label{cross}
\end {table}

\section{Thermal relic density}
\label{relic}
In this section, we outline the calculation of relic density and write it as one dimensional integral \cite{Griest:1990kh,Edsjo:1997bg,Nihei:2002sc}.

Consider annihilation of N supersymmetric particles $\chi_i$ (i=1,2,..,N-1, N). And their mass and number density are given by $m_i$ and $n_i$ respectively. The decay rate of supersymmetric particles $\chi_i$ which are heavier than the LSP is much faster than the age of the Universe. In the case of R-parity conserved models, like the one we have considered here, these particles decay into the LSP. If LSP were the dark matter particle, its final abundance is described by the sum of the number density of all the supersymmetric particles:
\begin{align}
n=\sum_i^N n_i
\end{align}
For n, Boltzmann evolution equation is given by \cite{Edsjo:1997bg}
\begin{align}
\frac{d n}{dt}+3 H n=-\sum_{i,j}^N \langle \sigma_{ij}\,v_{ij}\rangle(n_i\,n_j-n_i^{eq}\,n_{j}^{eq})
\label{boltz1}
\end{align}
where
\begin{align}
\sigma_{ij}&=\sigma(\chi_i\chi_j\rightarrow \text{All})
\end{align}
Taking into the fact that supersymmetric particles are non relativistic and their number density distributions are suppressed by a Boltzmann factor. In this scenario, thermal distribution of density $n_i$ remain in equilibrium. So the ratio $n_i\over n$ is equal to the equilibrium value:
\begin{align}
\frac{n_i}{n}&\sim\frac{n_i^{eq}}{n^{eq}}
\end{align}
then the eq.\ref{boltz1} is simplified to
\begin{align}
\frac{d n}{dt}+3 H n=-\sum_{i,j}^N \langle \sigma_{eff}\,v\rangle(n^2-n_{eq}^2)
\label{boltz2}
\end{align}
where
\begin{align}
\langle\sigma_{eff}v\rangle&=\sum_{i,j}^N\,\langle\sigma_{ij}v_{ij}\rangle\,\frac{n_i^{eq}}{n^{eq}}\,\frac{n_j^{eq}}{n^{eq}}
\label{eff}
\end{align}
By using Maxwel-Boltzmann equilibrium distribution the thermal average $\langle\sigma_{ij}v_{ij}\rangle$ is written as
\begin{align}
\langle \sigma_{ij}v_{ij}\rangle&=\frac{\int d^3 p_i\,d^3p_j\,\sigma\, v_{ij}\, e^{-E_i/T}\,e^{-E_j/T}}{\int d^3 p_i\,d^3p_j\, e^{-E_i/T}\,e^{-E_j/T}}
\end{align}
where $p_i=(E_i,\bold p_i)$ have usual meaning; energy and momenta of the two annihilating particles and T is the equilibrium temperature. Eq.\ref{eff} can be rewritten as
\begin{align}
\langle\sigma_{eff}v\rangle&=\frac{\int_{4\,m_1^2}^{\infty}\,ds\, s^{3/2}\,K_1\left({\sqrt{s}\over T}\right)\,\sum_{ij}^N\,\beta_f^2(s,m_i^2,m_j^2)\,\frac{g_i\,g_j}{g_1^2}\,\sigma_{ij}(s)}{8\,m_1^4\,T\,\left[\sum_i^N\,\frac{g_i}{g_1}\,\frac{m_i^2}{m_1^2}\,K_2\left(m_i\over T\right)\right]^2}
\end{align}
where $g_i$ is number of degrees of freedom, $K_i$ is modified Bessel function of order `i' and $\beta_f(s,m_i,m_j)$ is the kinetic function defined as
\begin{align}
\beta_f(s,m_i,m_j)&=\frac{\sqrt{s-(m_i^2+m_j^2)}\,\sqrt{s-(m_i^2-m_j^2)}}{s}
\end{align}
\section{Cross sections}
\label{cal}
In this section, general cross section for the cases (i) and (ii) mentioned in the introduction is given. To ease the process of comparison of our results with the ones already present in the literature\cite{Nihei:2002sc}, we rewrite $\sigma_{ij}(s)$ as
\begin{align}
\sigma_{ij}(s)&=\frac{s}{16\pi\,\beta_f(s,m_i,m_j)}\sum_{\phi_1\phi_2}\left[c\,\theta\left(s-(m_1^2+m_2^2)^2\right)\,\beta_f(s,m_1,m_2)\,\tilde w_{\phi_1\phi_2}(s)\right]
\end{align}
where$m_1$ and $m_2$ are the masses of the final states $\phi_1$ and $\phi_2$ respectively. And `c' is the colour factor. If $\chi^0_i\chi_j^0~\rightarrow~ \phi_1\phi_2$, then the function $\tilde w(s)_{\phi_1\phi_2}$, in the centre of mass frame, given as
\begin{align}
\tilde w(s)_{\phi_1\phi_2}\equiv \frac{1}{2}\int_{-1}^{1} d \cos\theta_{CM}|\mathcal{A}(\chi^0_i\chi_j^0~\rightarrow~ \phi_1\phi_2)|^2
\end{align}
where $\theta_{CM}$ denotes the scattering. We will follow Table.\ref{cross} in presenting explicit expressions for $\tilde w_{ \phi_1 \phi_ 2} (s)$ for all the two-body final states. 
\subsection{$\chi_i^0\chi_j^0\rightarrow H_m^0 H_n^0$}
This process is the sum of the the s-channel diagram through CP even Higgs boson ($H_1^0$ and $H_2^0$) and the t-
and u-channel neutralino ($\knk$ , k = 1, . . . , 4) exchange. The resulting cross section, including  the s, t, u and possible interference processes among themselves, is given by.
\begin{align}
\tilde{w}_{H_m^0,H_n^0}&=\sum_{k,l=1,2}\tilde{w}_{H_m^0,H_n^0}^{(H_k^0,H_l^0)}+\sum_{k=1,2}\sum_{l=1,4}\tilde{w}_{H_m^0,H_n^0}^{(H_k^0,\chi_l^0)}+\sum_{k,l=1,4}\tilde{w}_{H_m^0,H_n^0}^{(\knk,\chi_l^0)}
\end{align}
\begin{itemize}
\item \underline{CP-even Higgs boson ($H_k^0,H_l^0$) exchange}:
\begin{equation}
\tilde{w}_{H_m^0,H_n^0}^{(H_k^0,H_l^0)}=\frac{1}{2}\frac{C_{H_k^0H_m^0H_n^0}\,C_{H_l^0H_m^0H_n^0}^*}{(s-m_{H_k^0}^2)\,(s-m_{H_l^0}^2)}\left(f_{\ki \kj}^{(H_k^0-H_l^0)}(s-m_i^2-m_j^2)-2\,m_i\,m_j\,g_{\ki \kj}^{(H_k^0-H_l^0)}\right)
\label{ver1}
\end{equation}
\item  \underline{Neutralino ($\knk-\knl$) exchange}:
\begin{eqnarray}
\tilde{w}_{H_m^0,H_n^0}^{t(\chi_k^0,\chi_l^0)}&=& 4\, d \,m_{\chi_k^0}\, m_{\knl}\,\Gamma_{H_m^0H_n^0}^{1 (\knk-\knl)}+\beta_1(H_m^0,H_n^0,\chi_k^0,\chi_l^0)\, \Gamma_{H_m^0H_n^0}^{2 (\knk-\knl)}\nn
&-&4\, m_i\, m_j\, m_{\knk}\, m_{\knl}\,\Gamma_{H_m^0H_n^0}^{3 (\knk-\knl)}-m_i\,m_j\,\beta_4(H_m^0,H_n^0,\chi_k^0,\chi_l^0)\,\Gamma_{H_m^0H_n^0}^{4 (\knk-\knl)}\nn
&-&m_j\,(m_{\knk}\,\Gamma_{H_m^0H_n^0}^{5 (\knk-\knl)}+m_{\knl}\,\Gamma_{H_m^0H_n^0}^{6 (\knk-\knl)})\,\beta_3(H_m^0,H_n^0,\chi_k^0,\chi_l^0)\nn
&+&(4\, d\, m_i\,-\frac{m_i\,\omega_{H_m^0H_n^0}}{s})\,(m_{\knk}\,\Gamma_{H_m^0H_n^0}^{7 (\knk-\knl)}+m_{\knl}\,\Gamma_{H_m^0H_n^0}^{8 (\knk-\knl)})\\
\tilde{w}_{H_n^0,H_m^0}^{u(\chi_k^0,\chi_l^0)}&=& 4\, d \,m_{\chi_k^0}\, m_{\knl}\,\Gamma_{H_n^0H_m^0}^{1 (\knk-\knl)}+\beta_1(H_n^0,H_m^0,\chi_k^0,\chi_l^0)\, \Gamma_{H_n^0H_m^0}^{2 (\knk-\knl)}\nn
&-&4\, m_i\, m_j\, m_{\knk}\, m_{\knl}\,\Gamma_{H_n^0H_m^0}^{3 (\knk-\knl)}-m_i\,m_j\,\beta_2(H_n^0,H_m^0,\chi_k^0,\chi_l^0)\,\Gamma_{H_n^0H_m^0}^{4 (\knk-\knl)}\nn
&-&m_j\,(m_{\knk}\,\Gamma_{H_n^0H_m^0}^{5 (\knk-\knl)}+m_{\knl}\,\Gamma_{H_n^0H_m^0}^{6 (\knk-\knl)})\,\beta_3(H_n^0,H_m^0,\chi_k^0,\chi_l^0)\nn
&+&(4\, d\, m_i\,-\frac{m_i\,\omega_{H_n^0H_m^0}}{s})\,(m_{\knk}\,\Gamma_{H_n^0H_m^0}^{7 (\knk-\knl)}+m_{\knl}\,\Gamma_{H_n^0H_m^0}^{8 (\knk-\knl)})
\end{eqnarray}
\begin{eqnarray}
\tilde{w}_{H_m^0,H_n^0}^{t-u(\chi_k^0,\chi_l^0)}&=&d\,m_{\knk}\,m_{\knl}\,s^2 J_1(H_m^0,H_n^0,\knk,\knl)\,I_{H_m^0 H_n^0}^{1 (\knk-\knl)}+J_2(H_m^0,H_n^0,\knk,\knl)\nn
&&I_{H_m^0 H_n^0}^{2 (\knk-\knl)}-m_i\,m_j\,m_{\knk}\, m_{\knl} s^2\,J_1(H_m^0,H_n^0,\knk,\knl)\,I_{H_m^0 H_n^0}^{3 (\knk-\knl)}\nn
&+&m_i\, m_j\, s^2\,(l_i/2-d_{H_m^0H_n^0}-m_i^2)\,J_1(H_m^0,H_n^0,\knk,\knl)I_{H_m^0 H_n^0}^{4 (\knk-\knl)}\nn
&+&\frac{1}{4}\,m_j\,m_{\knk}\,s\,J_3({H_m^0,H_n^0,\knk,\knl})I_{H_m^0 H_n^0}^{5 (\knk-\knl)}+\frac{1}{4}\,m_j\,m_{\knl}\,s\nn&&J_3({H_m^0,H_n^0,\knl,\knk})\,I_{H_m^0 H_n^0}^{6 (\knk-\knl)}-\frac{1}{4}m_i\,m_{\knk}\,s\,J_4(H_m^0,H_n^0,\knk,\knl)\nn&&I_{H_m^0 H_n^0}^{7 (\knk-\knl)}
+\frac{1}{4}m_i\,m_{\knl}\,s\,J_5(H_m^0,H_n^0,\knk,\knl)\,I_{H_m^0 H_n^0}^{8 (\knk-\knl)}\\
\tilde{w}_{H_m^0,H_n^0}^{(\chi_k^0,\chi_l^0)}&=&\tilde{w}_{H_m^0,H_n^0}^{t(\chi_k^0,\chi_l^0)}+\tilde{w}_{H_m^0,H_n^0}^{u(\chi_k^0,\chi_l^0)}+\left(\tilde{w}_{H_m^0,H_n^0}^{t-u(\chi_k^0,\chi_l^0)}+c.c.\right)  
\end{eqnarray}
\item  \underline{CP-even Higgs boson ($H_k^0$) and Neutralino ($\knl$) Interference}:
\begin{eqnarray}
\tilde{w}_{H_m^0,H_n^0}^{st(H_k^0,\chi_l^0)}&=&-\frac{1}{2} m_i\, m_j\, m_{\knl}\, s\,L(\knl,H_m^0,H_n^0)J_{H_m^0 H_n^0}^{(1)(H_k^0 - \knl)}\nn
&-&\frac{m_j}{4}\left(b_{H_m^0H_n^0}-(R_{H_m^0\knl}-4\,m_i^2\,s)\right)\,\mathcal{A}(\knl,H_m^0,H_n^0)J_{H_m^0 H_n^0}^{(2)(H_k^0 - \knl)}\nn
&+&\frac{1}{2}\,d\, m_{\knl}\,s\,L(\knl,H_m^0,H_n^0)\,J_{H_m^0 H_n^0}^{(3)(H_k^0 - \knl)}-\frac{m_i}{4}\left(b_{H_m^0H_n^0}-(R_{H_m^0\knl}\right.\nn&+&\left.4\,d\,s-\tau_{H_m^0H_n^0}-\omega_{H_m^0H_n^0})\mathcal{A}(\knl,H_m^0,H_n^0)\right)J_{H_m^0 H_n^0}^{(4)(H_k^0 - \knl)}\\
\tilde{w}_{H_m^0,H_n^0}^{su(H_k^0,\chi_l^0)}&=&-\frac{1}{2} m_i\, m_j\, m_{\knl}\, s\,L(\knl,H_n^0,H_m^0)J_{H_n^0H_m^0}^{(1)(H_k^0 - \knl)}\nn
&-&\frac{m_j}{4}\left(b_{H_m^0H_n^0}-(R_{H_n^0\knl}-4\,m_i^2\,s)\right)\,\mathcal{A}(\knl,H_n^0,H_m^0)\,J_{H_n^0H_m^0}^{(2)(H_k^0 - \knl)}\nn
&+&\frac{1}{2}\,d\, m_{\knl}\,s\,L(\knl,H_n^0,H_m^0)\,J_{H_m^0H_n^0}^{(3)(H_k^0 - \knl)}-\frac{m_i}{4}\left(b_{H_m^0H_n^0}-(R_{H_n^0\knl}\right.\nn&+&\left.4\,d\,s-\tau_{H_m^0H_n^0}-\omega_{H_m^0H_n^0})\mathcal{A}(\knl,H_n^0,H_m^0)\right)J_{A_1^0 H_1^0}^{(4)(H_k^0 - \knl)}\\
\tilde{w}_{H_m^0,H_n^0}^{(H_k^0,\chi_l^0)}&=&\frac{1}{b_{H_m^0H_n^0}(s-m_{H_k^0}^2)}\left(\tilde{w}_{H_m^0,H_n^0}^{st(H_k^0,\chi_l^0)}+\tilde{w}_{H_m^0,H_n^0}^{su(H_k^0,\chi_l^0)}\right)+c.c.
\end{eqnarray}
The functions $\Gamma$'s, J's, I's, f's and g's, used in the above expressions depend on  mixing matrices and the couplings in the MSSM and are presented in the Appendix.\ref{aux}. From the expression given above one can obtain specific final states by taking m,n=1,2. Please note that in case of identical final states ($H_1^0H_1^0$ and $H^0_2H_2^0$) $\tilde w$ has to be multiplied by a symmetry factor $1\over 2$.
\end{itemize}
\subsection{$\ki\kj \rightarrow A_1^0 A_1^0$}
This process is the sum of s-channel diagrams though CP even Higgs boson ($H_1^0$ and $H_2^0$), the t
and u-channel neutralino ($\knk$ , k = 1, . . . , 4) exchange. The resulting cross section, including  the s, t, u and possible interference processes among themselves, is given by (Similar to the $\chi_i^0\chi_j^0\rightarrow H_m^0 H_n^0$ )
\begin{align}
\tilde{w}_{A_1^0,A_1^0}&=\sum_{k,l=1,2}\tilde{w}_{A_1^0,A_1^0}^{(H_k^0,H_l^0)}+\sum_{k=1,2}\sum_{l=1,4}\tilde{w}_{A_1^0,A_1^0}^{(H_k^0,\chi_l^0)}+\sum_{k,l=1,4}\tilde{w}_{A_1^0,A_1^0}^{(\knk,\chi_l^0)}
\end{align}
\begin{itemize}
	\item \underline{CP-odd Higgs boson ($A_1^0-A_1^0$) exchange}:
	\begin{equation}
	\tilde{w}_{A_1^0,A_1^0}^{(H_k^0,H_l^0)}=\frac{1}{2}\frac{C_{H_k^0A_1^0A_1^0}\,C_{H_l^0A_1^0A_1^0}^*}{(s-m_{H_k^0}^2)\,(s-m_{H_l^0}^2)}\left(f_{\ki \kj}^{(H_k^0-H_l^0)}(s-m_i^2-m_j^2)-2\,m_i\,m_j\,g_{\ki \kj}^{(H_k^0-H_l^0)}\right)
	\end{equation}
	\item  \underline{Neutralino  ($\knk-\knl$)  exchange}:
	\begin{eqnarray}
	\tilde{w}_{A_1^0,A_1^0}^{t(\chi_k^0,\chi_l^0)}&=& 4\, d \,m_{\chi_k^0}\, m_{\knl}\,\Gamma_{A_1^0A_1^0}^{1 (\knk-\knl)}+\beta_1(A_1^0,A_1^0,\chi_k^0,\chi_l^0)\, \Gamma_{A_1^0A_1^0}^{2 (\knk-\knl)}\nn
	&-&4\, m_i\, m_j\, m_{\knk}\, m_{\knl}\,\Gamma_{A_1^0A_1^0}^{3 (\knk-\knl)}-m_i\,m_j\,\beta_2(A_1^0,A_1^0,\chi_k^0,\chi_l^0)\,\Gamma_{A_1^0A_1^0}^{4 (\knk-\knl)}\nn
	&-&m_j\,(m_{\knk}\,\Gamma_{A_1^0A_1^0}^{5 (\knk-\knl)}+m_{\knl}\,\Gamma_{A_1^0A_1^0}^{6 (\knk-\knl)})\,\beta_3(A_1^0,A_1^0,\chi_k^0,\chi_l^0)\nn
	&+&(4\, d\, m_i\,-\frac{m_i\,\omega_{A_1^0A_1^0}}{s})\,(m_{\knk}\,\Gamma_{A_1^0A_1^0}^{7 (\knk-\knl)}+m_{\knl}\,\Gamma_{A_1^0A_1^0}^{8 (\knk-\knl)})\\
	\tilde{w}_{A_1^0,A_1^0}^{u(\chi_k^0,\chi_l^0)}&=& 4\, d \,m_{\chi_k^0}\, m_{\knl}\,\Gamma_{A_1^0A_1^0}^{1 (\knk-\knl)}+\beta_1(A_1^0,A_1^0,\chi_k^0,\chi_l^0)\, \Gamma_{A_1^0A_1^0}^{2 (\knk-\knl)}\nn
	&-&4\, m_i\, m_j\, m_{\knk}\, m_{\knl}\,\Gamma_{A_1^0A_1^0}^{3 (\knk-\knl)}-m_i\,m_j\,\beta_2(A_1^0,A_1^0,\chi_k^0,\chi_l^0)\,\Gamma_{A_1^0A_1^0}^{4 (\knk-\knl)}\nn
	&-&m_j\,(m_{\knk}\,\Gamma_{A_1^0A_1^0}^{5 (\knk-\knl)}+m_{\knl}\,\Gamma_{A_1^0A_1^0}^{6 (\knk-\knl)})\,\beta_3(A_1^0,A_1^0,\chi_k^0,\chi_l^0)\nn
	&+&(4\, d\, m_i\,-\frac{m_i\,\omega_{A_1^0A_1^0}}{s})\,(m_{\knk}\,\Gamma_{A_1^0A_1^0}^{7 (\knk-\knl)}+m_{\knl}\,\Gamma_{A_1^0A_1^0}^{8 (\knk-\knl)})\\
	\tilde{w}_{A_1^0,A_1^0}^{t-u(\chi_k^0,\chi_l^0)}&=&d\,m_{\knk}\,m_{\knl}\,s^2 J_1(A_1^0,A_1^0,\knk,\knl)\,I_{A_1^0A_1^0}^{1 (\knk-\knl)}+J_2(A_1^0,A_1^0,\knk,\knl)\nn
	&&I_{A_1^0A_1^0}^{2 (\knk-\knl)}-m_i\,m_j\,m_{\knk}\, m_{\knl} s^2\,J_1(A_1^0,A_1^0,\knk,\knl)\,I_{A_1^0A_1^0}^{3 (\knk-\knl)}\nn
	&+&m_i\, m_j\, s^2\,(l_i/2-d_{A_1^0A_1^0}-m_i^2)\,J_1(A_1^0,A_1^0,\knk,\knl)I_{A_1^0A_1^0}^{4 (\knk-\knl)}\nn
	&+&\frac{1}{4}\,m_j\,m_{\knk}\,s\,J_3({A_1^0,A_1^0,\knk,\knl})I_{A_1^0A_1^0}^{5 (\knk-\knl)}+\frac{1}{4}\,m_j\,m_{\knl}\,s\nn&&J_3({A_1^0,A_1^0,\knl,\knk})\,
	I_{A_1^0A_1^0}^{6 (\knk-\knl)}-\frac{1}{4}m_i\,m_{\knk}\,s\,J_4(A_1^0,A_1^0,\knk,\knl)\nn&&I_{A_1^0A_1^0}^{7 (\knk-\knl)}
	+\frac{1}{4}m_i\,m_{\knl}\,s\,J_5(A_1^0,A_1^0,\knk,\knl)\,I_{A_1^0A_1^0}^{8 (\knk-\knl)}\\
	\tilde{w}_{A_1^0,A_1^0}^{(\chi_k^0,\chi_l^0)}&=&\tilde{w}_{A_1^0,A_1^0}^{t(\chi_k^0,\chi_l^0)}+\tilde{w}_{A_1^0,A_1^0}^{u(\chi_k^0,\chi_l^0)}+\left(\tilde{w}_{A_1^0,A_1^0}^{t-u(\chi_k^0,\chi_l^0)}+c.c.\right)  
	\end{eqnarray}
	\item  \underline{CP-odd Higgs boson ($A_1^0$) and Neutralino ($\knl$) Interference}:
	\begin{eqnarray}
	\tilde{w}_{A_1^0,A_1^0}^{st(H_k^0,\chi_l^0)}&=&-\frac{1}{2} m_i\, m_j\, m_{\knl}\, s\,L(\knl,A_1^0,A_1^0)\,J_{A_1^0A_1^0}^{(1)(H_k^0 - \knl)}\nn
	&-&\frac{m_j}{4}\left(b_{A_1^0A_1^0}-(R_{H_m^0\knl}-4\,m_i^2\,s)\right)\,\mathcal{A}(\knl,A_1^0,A_1^0)\,J_{A_1^0A_1^0}^{(2)(H_k^0 - \knl)}\nn
	&+&\frac{1}{2}\,d\, m_{\knl}\,s\,L(\knl,A_1^0,A_1^0)\,J_{A_1^0A_1^0}^{(3)(H_k^0 - \knl)}-\frac{m_i}{4}\left(b_{A_1^0A_1^0}-(R_{A_1^0\knl}\right.\nn&+&\left.4\,d\,s-\tau_{A_1^0A_1^0}-\omega_{A_1^0A_1^0})\,\mathcal{A}(\knl,A_1^0,A_1^0)\right)\,J_{A_1^0A_1^0}^{(4)(H_k^0 - \knl)}
	\end{eqnarray}
	\begin{eqnarray}
	\tilde{w}_{A_1^0,A_1^0}^{su(H_k^0,\chi_l^0)}&=&-\frac{1}{2}\, m_i\, m_j\, m_{\knl}\, s\,L(\knl,A_1^0,A_1^0)\,J_{A_1^0 A_1^0}^{(1)(H_k^0 - \knl)}\nn
	&-&\frac{m_j}{4}\left(b_{A_1^0A_1^0}-(R_{A_1^0\knl}-4\,m_i^2\,s)\right)\,\mathcal{A}(\knl,A_1^0,A_1^0)\,J_{A_1^0 A_1^0}^{(2)(H_k^0 - \knl)}\nn
	&+&\frac{1}{2}\,d\, m_{\knl}\,s\,L(\knl,A_1^0,A_1^0)\,J_{A_1^0 A_1^0}^{(3)(H_k^0 - \knl)}-\frac{m_i}{4}\left(b_{A_1^0A_1^0}-(R_{A_1^0\knl}\right.\nn&+&\left.4\,d\,s-\tau_{A_1^0A_1^0}-\omega_{A_1^0A_1^0})\,\mathcal{A}(\knl,A_1^0,A_1^0)\right)\,J_{A_1^0 A_1^0}^{(4)(H_k^0 - \knl)}\\
	\tilde{w}_{A_1^0,A_1^0}^{(H_k^0,\chi_l^0)}&=&\frac{1}{b_{A_1^0A_1^0}(s-m_{H_k^0}^2)}\left(\tilde{w}_{A_1^0,A_1^0}^{st(H_k^0,\chi_l^0)}+\tilde{w}_{A_1^0,A_1^0}^{su(H_k^0,\chi_l^0)}\right)+c.c.
	\end{eqnarray}
\end{itemize}
Note: As there are identical final states total cross section has to be multiplied by a factor $1\over 2$
\subsection{$\chi_i^0\chi_j^0\rightarrow H_1^0 A_1^0$}
This process is sum of s-channel diagrams through CP-odd Higgs boson $A_1^0$ and Z boson; t and u channel diagrams with neutralino exchange. The resulting cross section, including  the s, t, u and possible interference processes among themselves, is given by
\begin{align}
\tilde{w}_{H_1^0,A_1^0}&=\tilde{w}_{H_1^0,A_1^0}^{(A_1^0,A_1^0)}+\tilde{w}_{H_1^0,A_1^0}^{(A_1^0,A_1^0)}+\tilde{w}_{H_1^0,A_1^0}^{(A_1^0,Z)}+\sum_{k,l=1,4}\tilde{w}_{H_1^0,A_1^0}^{(\knk,\chi_l^0)}+\sum_{l=1,4}\tilde{w}_{H_1^0,A_1^0}^{(A_1^0,\chi_l^0)}+\sum_{l=1,4}\tilde{w}_{H_1^0,A_1^0}^{(Z,\chi_l^0)}
\end{align}
\begin{itemize}
\item \underline{CP-odd Higgs boson $A_1^0$ and Z exchange}:
\begin{align}
\tilde{w}_{H_1^0,A_1^0}^{(A_1^0,A_1^0)}&=\frac{1}{2}\frac{|C_{A_1^0H_1^0A_1^0}|^2}{(s-m_{A_1^0}^2)^2}\left(f_{\ki \kj}^{(A_1^0-A_1^0)}(s-m_i^2-m_j^2)-2\,m_i\,m_j\,g_{\ki \kj}^{(A_1^0-A_1^0)}\right)\\
\tilde{w}_{H_1^0,A_1^0}^{(Z,Z)}&=\frac{1}{2}\frac{|C_{ZH_1^0A_1^0}|^2}{(s-M_Z^2)^2\,M_Z^4}\left(f_{\ki \kj}^{(Z-Z)}\,\mathcal{B}_1(Z,H_1^0,A_1^0)-g_{\ki \kj}^{(Z-Z)}\,\mathcal{B}_2(Z,H_1^0,A_1^0)\right)\nn
\tilde{w}_{H_1^0,A_1^0}^{(A_1^0,Z)}&=\frac{1}{2}\frac{C_{A_1^0H_1^0A_1^0}\,C_{ZH_1^0A_1^0}^*}{(s-m_{A_1^0}^2)^2\,M_Z^2\,s}\left(f_{\ki \kj}^{(A_1^0-Z)}\,m_j\,\tau_{H_1^0A_1^0}-g_{\ki \kj}^{(Z-Z)}\,m_i\,\tau_{A_1^0H_1^0}\right)\nn&(m_{H_1^0}^2-m_{A_1^0}^2)+c.c.
\end{align}
\item  \underline{Neutralino ($\knk-\knl$) exchange}:
\begin{eqnarray}
\tilde{w}_{H_1^0, A_1^0}^{t(\chi_k^0,\chi_l^0)}&=& 4\, d \,m_{\chi_k^0}\, m_{\knl}\,\Gamma_{H_1^0 A_1^0}^{1 (\knk-\knl)}+\beta_1(H_1^0, A_1^0,\chi_k^0,\chi_l^0)\, \Gamma_{H_1^0 A_1^0}^{2 (\knk-\knl)}\nn
&-&4\, m_i\, m_j\, m_{\knk}\, m_{\knl}\,\Gamma_{H_1^0 A_1^0}^{3 (\knk-\knl)}-m_i\,m_j\,\beta_2(H_1^0, A_1^0,\chi_k^0,\chi_l^0)\,\Gamma_{H_1^0 A_1^0}^{4 (\knk-\knl)}\nn
&-&m_j\,(m_{\knk}\,\Gamma_{H_1^0 A_1^0}^{5 (\knk-\knl)}+m_{\knl}\,\Gamma_{H_1^0 A_1^0}^{6 (\knk-\knl)})\,\beta_3(H_1^0, A_1^0,\chi_k^0,\chi_l^0)\nn
&+&\left(4\, d\, m_i\,-\frac{m_i\,\omega_{H_1^0 A_1^0}}{s}\right)\,\left(m_{\knk}\,\Gamma_{H_1^0 A_1^0}^{7 (\knk-\knl)}+m_{\knl}\,\Gamma_{H_1^0 A_1^0}^{8 (\knk-\knl)}\right)
\end{eqnarray}
\begin{eqnarray}
\tilde{w}_{A_1^0,H_1^0 }^{u(\chi_k^0,\chi_l^0)}&=& 4\, d \,m_{\chi_k^0}\, m_{\knl}\,\Gamma_{A_1^0A_1^0}^{1 (\knk-\knl)}+\beta_1(A_1^0,H_1^0 ,\chi_k^0,\chi_l^0)\, \Gamma_{A_1^0A_1^0}^{2 (\knk-\knl)}\nn
&-&4\, m_i\, m_j\, m_{\knk}\, m_{\knl}\,\Gamma_{A_1^0A_1^0}^{3 (\knk-\knl)}-m_i\,m_j\,\beta_2(A_1^0,H_1^0 ,\chi_k^0,\chi_l^0)\,\Gamma_{A_1^0A_1^0}^{4 (\knk-\knl)}\nn
&-&m_j\,(m_{\knk}\,\Gamma_{A_1^0A_1^0}^{5 (\knk-\knl)}+m_{\knl}\,\Gamma_{A_1^0A_1^0}^{6 (\knk-\knl)})\,\beta_3(A_1^0,H_1^0 ,\chi_k^0,\chi_l^0)\nn
&+&(4\, d\, m_i\,-\frac{m_i\,\omega_{A_1^0A_1^0}}{s})\,(m_{\knk}\,\Gamma_{A_1^0A_1^0}^{7 (\knk-\knl)}+m_{\knl}\,\Gamma_{A_1^0A_1^0}^{8 (\knk-\knl)})
\end{eqnarray}
\begin{eqnarray}
\tilde{w}_{H_1^0, A_1^0}^{t-u(\chi_k^0,\chi_l^0)}&=&d\,m_{\knk}\,m_{\knl}\,s^2 J_1(H_1^0, A_1^0,\knk,\knl)\,I_{H_1^0 A_1^0}^{1 (\knk-\knl)}+J_2(H_1^0, A_1^0,\knk,\knl)\nn
&&I_{H_1^0 A_1^0}^{2 (\knk-\knl)}-m_i\,m_j\,m_{\knk}\, m_{\knl}\, s^2\,J_1(H_1^0, A_1^0,\knk,\knl)\,I_{H_1^0 A_1^0}^{3 (\knk-\knl)}\nn
&+&m_i\, m_j\, s^2\,(l_i/2-d_{H_1^0 A_1^0}-m_i^2)\,J_1(H_1^0, A_1^0,\knk,\knl)\,I_{H_1^0 A_1^0}^{4 (\knk-\knl)}\nn
&+&\frac{1}{4}\,m_j\,m_{\knk}\,s\,J_3({H_1^0, A_1^0,\knk,\knl})\,I_{H_1^0 A_1^0}^{5 (\knk-\knl)}+\frac{1}{4}\,m_j\,m_{\knl}\,s\nn&&J_3({H_1^0, A_1^0,\knl,\knk})\,
I_{H_1^0 A_1^0}^{6 (\knk-\knl)}-\frac{1}{4}m_i\,m_{\knk}\,s\,J_4(H_1^0, A_1^0,\knk,\knl)\nn&&I_{H_1^0 A_1^0}^{7 (\knk-\knl)}
+\frac{1}{4}m_i\,m_{\knl}\,s\,J_5(H_1^0, A_1^0,\knk,\knl)\,I_{ H_1^0  A_1^0}^{8 (\knk-\knl)}\\
\tilde{w}_{H_1^0, A_1^0}^{(\chi_k^0,\chi_l^0)}&=&\tilde{w}_{H_1^0, A_1^0}^{t(\chi_k^0,\chi_l^0)}+\tilde{w}_{H_1^0, A_1^0}^{u(\chi_k^0,\chi_l^0)}+\left(\tilde{w}_{H_1^0, A_1^0}^{t-u(\chi_k^0,\chi_l^0)}+c.c.\right)  
\end{eqnarray}
\item  \underline{CP-odd Higgs boson $A_1^0$, Z boson and Neutralino ($\knl$) Interference}:
\begin{align}
\tilde{w}_{H_1^0,A_1^0}^{st(Z,\knk)}&=\frac{C_{ZH_1^0A_1^0}}{s-M_Z^2}\left(\frac{m_j\,m_{\knk}}{2\,b_{H_1^0 A_1^0}\,M_Z^2}\,\mathcal{B}_3(Z,\knk,H_1^0,A_1^0)\,J_{H_1^0 A_1^0}^{(5)(Z - \knk)}\right.\nn
&+\left.\frac{m_j\,m_{\knk}}{4\,b_{H_1^0 A_1^0}\,M_Z^2}\,\mathcal{B}_4(Z,\knk,H_1^0,A_1^0)\,J_{H_1^0 A_1^0}^{(6)(Z - \knk)}+\mathcal{B}_5(Z,\knk,H_1^0,A_1^0)\right.\nn&\left.J_{H_1^0 A_1^0}^{(7)(Z - \knk)}
+\frac{m_i\,m_j}{2\,b_{H_1^0 A_1^0}\,M_Z^2}\mathcal{B}_6(Z,\knk,H_1^0,A_1^0)\,J_{H_1^0 A_1^0}^{(8)(Z - \knk)}\right)\\
\tilde{w}_{H_1^0,A_1^0}^{su(Z,\knk)}&=\frac{C_{ZH_1^0A_1^0}}{s-M_Z^2}\left(-\frac{m_j\,m_{\knk}}{2\,b_{H_1^0 A_1^0}\,M_Z^2}\,\mathcal{B}_3(Z,\knk,A_1^0,H_1^0)\,J_{ A_1^0 H_1^0}^{(5)(Z - \knk)}\right.\nn
&-\left.\frac{m_j\,m_{\knk}}{4\,b_{H_1^0 A_1^0}\,M_Z^2}\,\mathcal{B}_4(Z,\knk,A_1^0,H_1^0)\,J_{ A_1^0H_1^0}^{(6)(Z - \knk)}-\mathcal{B}_5(Z,\knk,A_1^0,H_1^0)\right.\nn&\left.J_{ A_1^0H_1^0}^{(7)(Z - \knk)}
+\frac{m_i\,m_j}{2\,b_{H_1^0 A_1^0}\,M_Z^2}\mathcal{B}_6(Z,\knk,A_1^0,H_1^0)\,J_{ A_1^0H_1^0}^{(8)(Z - \knk)}\right)\\
\tilde{w}_{H_1^0,A_1^0}^{st(A_1^0,\chi_l^0)}&=-\frac{1}{2} m_i\, m_j\, m_{\knl}\, s\,L(\knl,H_1^0,A_1^0)J_{H_1^0 A_1^0}^{(1)(H_k^0 - \knl)}\nn
&-\frac{m_j}{4}\left(b_{H_1^0A_1^0}-(R_{H_1^0\knl}-4\,m_i^2\,s)\right)\,\mathcal{A}(\knl,H_1^0,A_1^0)J_{H_1^0 A_1^0}^{(2)(H_k^0 - \knl)}\nn
&+\frac{1}{2}\,d\, m_{\knl}\,s\,L(\knl,H_1^0,A_1^0)\,J_{H_1^0 A_1^0}^{(3)(H_k^0 - \knl)}-\frac{m_i}{4}\left(b_{H_1^0A_1^0}-(R_{H_1^0\knl}\right.\nn&+\left.4\,d\,s-\tau_{H_1^0A_1^0}-\omega_{H_1^0A_1^0})\mathcal{A}(\knl,H_1^0,A_1^0)\right)J_{H_1^0 A_1^0}^{(4)(H_k^0 - \knl)}
\end{align}
\begin{align}
\tilde{w}_{H_1^0,A_1^0}^{su(H_k^0,\chi_l^0)}&=-\frac{1}{2} m_i\, m_j\, m_{\knl}\, s\,L(\knl,A_1^0,H_1^0)\,J_{A_1^0H_1^0}^{(1)(H_k^0 - \knl)}\nn
&-\frac{m_j}{4}\left(b_{H_1^0A_1^0}-(R_{A_1^0\knl}-4\,m_i^2\,s)\right)\,\mathcal{A}(\knl,A_1^0,H_1^0)\,J_{A_1^0H_1^0}^{(2)(H_k^0 - \knl)}\nn
&+\frac{1}{2}\,d\, m_{\knl}\,s\,L(\knl,A_1^0,H_1^0)\,J_{H_1^0A_1^0}^{(3)(H_k^0 - \knl)}-\frac{m_i}{4}\left(b_{H_1^0A_1^0}-(R_{A_1^0\knl}\right.\nn&+\left.4\,d\,s-\tau_{H_1^0A_1^0}-\omega_{H_1^0A_1^0})\,\mathcal{A}(\knl,A_1^0,H_1^0)\right)J_{A_1^0 H_1^0}^{(4)(H_k^0 - \knl)}\\
\tilde{w}_{H_1^0,A_1^0}^{(A_1^0,\chi_l^0)}&=\frac{1}{b_{H_1^0A_1^0}\,(s-m_{A_1^0}^2)}\left(\tilde{w}_{H_1^0,A_1^0}^{st(A_1^0,\chi_l^0)}+\tilde{w}_{H_1^0,A_1^0}^{su(A_1^0,\chi_l^0)}\right)+c.c.\\
\tilde{w}_{H_1^0,A_1^0}^{(Z,\chi_l^0)}&=\frac{1}{b_{H_1^0A_1^0}\,(s-M_Z^2)}\left(\tilde{w}_{H_1^0,A_1^0}^{st(Z,\chi_l^0)}+\tilde{w}_{H_1^0,A_1^0}^{su(Z,\chi_l^0)}\right)+c.c.
\end{align}
\end{itemize}
\subsection{$\chi_i^0\chi_j^0\rightarrow H_2^0 A_1^0$}
This process is the sum of CP even Higgs boson $H_k^0$ in s-channel and neutralino exchange in t and u channel. The resulting cross section, including  the s, t, u and possible interference processes among themselves, is given by:
\begin{align}
\tilde{w}_{H_2^0,A_1^0}&=\sum_{k,l=1,2}\tilde{w}_{H_2^0,A_1^0}^{(H_k^0,H_l^0)}+\sum_{k,l=1,4}\tilde{w}_{H_2^0,A_1^0}^{(\knk,\chi_l^0)}+\sum_{k=1,2}\sum_{l=1,4}\tilde{w}_{H_2^0,A_1^0}^{(H_k^0,\chi_l^0)}
\end{align}
\begin{itemize}
	\item \underline{CP-even Higgs boson ($H_k^0,H_l^0$) exchange}:
	\begin{equation}
	\tilde{w}_{ H_2^0  A_1^0}^{(H_k^0,H_l^0)}=\frac{1}{2}\frac{C_{H_k^0 H_2^0 A_1^0}\,C_{H_l^0 H_2^0 A_1^0}^*}{(s-m_{H_k^0}^2)\,(s-m_{H_l^0}^2)}\left(f_{\ki \kj}^{(H_k^0-H_l^0)}(s-m_i^2-m_j^2)-2\,m_i\,m_j\,g_{\ki \kj}^{(H_k^0-H_l^0)}\right)
	\end{equation}
	\item  \underline{Neutralino ($\chi_k^0-\chi_l^0$) exchange}:
	\begin{eqnarray}
	\tilde{w}_{H_2^0,A_1^0}^{t(\chi_k^0,\chi_l^0)}&=& 4\, d \,m_{\chi_k^0}\, m_{\knl}\,\Gamma_{H_2^0A_1^0}^{1 (\knk-\knl)}+\beta_1( H_2^0, A_1^0,\chi_k^0,\chi_l^0)\, \Gamma_{H_2^0A_1^0}^{2 (\knk-\knl)}\nn
	&-&4\, m_i\, m_j\, m_{\knk}\, m_{\knl}\,\Gamma_{H_2^0A_1^0}^{3 (\knk-\knl)}-m_i\,m_j\,\beta_2( H_2^0, A_1^0,\chi_k^0,\chi_l^0)\,\Gamma_{H_2^0A_1^0}^{4 (\knk-\knl)}\nn
	&-&m_j\,(m_{\knk}\,\Gamma_{H_2^0A_1^0}^{5 (\knk-\knl)}+m_{\knl}\,\Gamma_{H_2^0A_1^0}^{6 (\knk-\knl)})\,\beta_3( H_2^0 A_1^0,\chi_k^0,\chi_l^0)\nn
	&+&(4\, d\, m_i\,-\frac{m_i\,\omega_{H_2^0A_1^0}}{s})\,(m_{\knk}\,\Gamma_{H_2^0A_1^0}^{7 (\knk-\knl)}+m_{\knl}\,\Gamma_{H_2^0A_1^0}^{8 (\knk-\knl)})\\
	\tilde{w}_{A_1^0,H_2^0}^{u(\chi_k^0,\chi_l^0)}&=& 4\, d \,m_{\chi_k^0}\, m_{\knl}\,\Gamma_{A_1^0H_2^0}^{1 (\knk-\knl)}+\beta_1(A_1^0,H_2^0,\chi_k^0,\chi_l^0)\, \Gamma_{A_1^0H_2^0}^{2 (\knk-\knl)}\nn
	&-&4\, m_i\, m_j\, m_{\knk}\, m_{\knl}\,\Gamma_{A_1^0H_2^0}^{3 (\knk-\knl)}-m_i\,m_j\,\beta_2(A_1^0,H_2^0,\chi_k^0,\chi_l^0)\,\Gamma_{A_1^0H_2^0}^{4 (\knk-\knl)}\nn
	&-&m_j\,(m_{\knk}\,\Gamma_{A_1^0H_2^0}^{5 (\knk-\knl)}+m_{\knl}\,\Gamma_{A_1^0H_2^0}^{6 (\knk-\knl)})\,\beta_3(A_1^0,H_2^0,\chi_k^0,\chi_l^0)\nn
	&+&(4\, d\, m_i\,-\frac{m_i\,\omega_{A_1^0H_2^0}}{s})\,(m_{\knk}\,\Gamma_{A_1^0H_2^0}^{7 (\knk-\knl)}+m_{\knl}\,\Gamma_{A_1^0H_2^0}^{8 (\knk-\knl)})
	\end{eqnarray}
	\begin{eqnarray}
	\tilde{w}_{ H_2^0, A_1^0}^{t-u(\chi_k^0,\chi_l^0)}&=&d\,m_{\knk}\,m_{\knl}\,s^2 J_1( H_2^0, A_1^0,\knk,\knl)\,I_{H_2^0 A_1^0}^{1 (\knk-\knl)}+J_2( H_2^0, A_1^0,\knk,\knl)\nn
	&&I_{H_2^0 A_1^0}^{2 (\knk-\knl)}-m_i\,m_j\,m_{\knk}\, m_{\knl}\, s^2\,J_1( H_2^0, A_1^0,\knk,\knl)\,I_{H_2^0 A_1^0}^{3 (\knk-\knl)}\nn
	&+&m_i\, m_j\, s^2\,(l_i/2-d_{ H_2^0A_1^0}-m_i^2)\,J_1( H_2^0, A_1^0,\knk,\knl)\,I_{H_2^0 A_1^0}^{4 (\knk-\knl)}\nn
	&+&\frac{1}{4}\,m_j\,m_{\knk}\,s\,J_3({ H_2^0, A_1^0,\knk,\knl})\,I_{H_2^0 A_1^0}^{5 (\knk-\knl)}+\frac{1}{4}\,m_j\,m_{\knl}\,s\nn&&J_3({ H_2^0, A_1^0,\knl,\knk})\,I_{H_2^0 A_1^0}^{6 (\knk-\knl)}-\frac{1}{4}\,m_i\,m_{\knk}\,s\,J_4( H_2^0, A_1^0,\knk,\knl)\nn&&I_{H_2^0 A_1^0}^{7 (\knk-\knl)}
	+\frac{1}{4}\,m_i\,m_{\knl}\,s\,J_5( H_2^0, A_1^0,\knk,\knl)\,I_{H_2^0 A_1^0}^{8 (\knk-\knl)}\\
	\tilde{w}_{ H_2^0, A_1^0}^{(\chi_k^0,\chi_l^0)}&=&\tilde{w}_{ H_2^0, A_1^0}^{t(\chi_k^0,\chi_l^0)}+\tilde{w}_{ H_2^0, A_1^0}^{u(\chi_k^0,\chi_l^0)}+\left(\tilde{w}_{ H_2^0, A_1^0}^{t-u(\chi_k^0,\chi_l^0)}+c.c.\right)  
	\end{eqnarray}
	\item  \underline{CP-even Higgs boson ($H_k^0$) and Neutralino ($\knl$) Interference}:
	\begin{eqnarray}
	\tilde{w}_{ H_2^0, A_1^0}^{st(H_k^0,\chi_l^0)}&=&-\frac{1}{2} m_i\, m_j\, m_{\knl}\, s\,L(\knl, H_2^0, A_1^0)\,J_{H_2^0 A_1^0}^{(1)(H_k^0 - \knl)}\nn
	&-&\frac{m_j}{4}\left(b_{ H_2^0A_1^0}-(R_{H_2^0\knl}-4\,m_i^2\,s)\right)\,\mathcal{A}(\knl, H_2^0, A_1^0)\,J_{H_2^0 A_1^0}^{(2)(H_k^0 - \knl)}\nn
	&+&\frac{1}{2}\,d\, m_{\knl}\,s\,L(\knl, H_2^0, A_1^0)\,J_{H_2^0 A_1^0}^{(3)(H_k^0 - \knl)}-\frac{m_i}{4}\left(b_{ H_2^0A_1^0}-(R_{H_2^0\knl}\right.\nn&+&\left.4\,d\,s-\tau_{ H_2^0A_1^0}-\omega_{ H_2^0A_1^0})\,\mathcal{A}(\knl, H_2^0, A_1^0)\right)\,J_{H_2^0 A_1^0}^{(4)(H_k^0 - \knl)}\\
	\tilde{w}_{ H_2^0, A_1^0}^{su(H_k^0,\chi_l^0)}&=&-\frac{1}{2}\, m_i\, m_j\, m_{\knl}\, s\,L(\knl,A_1^0,H_2^)\,J_{A_1^0A_1^0}^{(1)(H_k^0 - \knl)}\nn
	&-&\frac{m_j}{4}\,\left(b_{ H_2^0A_1^0}-(R_{H_2^0\knl}-4\,m_i^2\,s)\right)\,\mathcal{A}(\knl,A_1^0,H_2^0)\,J_{A_1^0A_1^0}^{(2)(H_k^0 - \knl)}\nn
	&+&\frac{1}{2}\,d\, m_{\knl}\,s\,L(\knl,A_1^0,H_2^0)\,J_{A_1^0A_1^0}^{(3)(H_k^0 - \knl)}-\frac{m_i}{4}\,\left(b_{ H_2^0A_1^0}-(R_{H_2^0\knl}\right.\nn&+&\left.4\,d\,s-\tau_{ H_2^0A_1^0}-\omega_{ H_2^0A_1^0})\,\mathcal{A}(\knl,A_1^0,H_2^0)\right)\,J_{A_1^0 H_2^0}^{(4)(H_k^0 - \knl)}\\
	\tilde{w}_{ H_2^0, A_1^0}^{(H_k^0,\chi_l^0)}&=&\frac{1}{b_{ H_2^0A_1^0}\,(s-m_{H_k^0}^2)}\left(\tilde{w}_{ H_2^0, A_1^0}^{st(H_k^0,\chi_l^0)}+\tilde{w}_{ H_2^0, A_1^0}^{su(H_k^0,\chi_l^0)}\right)+c.c.
	\end{eqnarray}
\end{itemize}
\subsection{$\chi_i^0\chi_j^0\rightarrow H^+ H^-$}
This process is the sum of the s-channel Z and CP even Higgs boson ($H_1^0$, $H_2^0$) exchange and  t- and u-channel chargino ($\kk$ , k = 1, 2) exchange. Resulting cross section including considering interference via s, t and u channel diagrams is given by:
\begin{align}
\tilde{w}_{H_1^+,H_1^-}&=\sum_{k,l=1,2}\tilde{w}_{H_1^+,H_1^-}^{(H_k^0,H_l^0)}+\tilde{w}_{H_1^+,H_1^-}^{(Z,Z)}+\sum_{k=1,2}\sum_{l=1,2}\tilde{w}_{H_1^+,H_1^-}^{(H_k^0,\kl)}+\sum_{k,l=1,2}\tilde{w}_{H_1^+,H_1^-}^{(\kk,\kl)}
\end{align}
\begin{itemize}
\item \underline{CP even Higgs boson ($H_k^0$,$H_l^0$) and Z exchange}:
\begin{align}
\tilde{w}_{H_1^+,H_1^-}^{s(H_k^0,H_l^0)}&=\frac{1}{2}\,\frac{C_{H_k^0H_1^+H_1^-}\,C_{H_l^0H_1^+H_1^-}^*}{(s-m_{H_k^0}^2)\,(s-m_{H_l^0}^2)}\left(f_{\ki \kj}^{(H_k^0-H_l^0)}(s-m_i^2-m_j^2)-2\,m_i\,m_j\,g_{\ki \kj}^{(H_k^0-H_l^0)}\right)
\end{align}
\begin{align}
\tilde{w}_{H_1^+,H_1^-}^{s(Z,Z)}&=\frac{1}{2}\frac{|C_{ZH_1^+H_1^-}|^2}{(s-M_Z^2)^2\,M_Z^4}\left(f_{\ki \kj}^{(Z-Z)}\,\mathcal{B}_1(Z,H_1^+H_1^-)-g_{\ki \kj}^{(Z-Z)}\,\mathcal{B}_2(Z,H_1^+H_1^-)\right)\\
\tilde{w}_{H_1^+,H_1^-}^{s(H_k^0,Z)}&=0
\end{align}
\item \underline{Chargino ($\kk$,$\kl$) exchange}:
\begin{eqnarray}
\tilde{w}_{H_1^+,H_1^-}^{t(\kk,\kl)}&=& 4\, d \,m_{\kk}\, m_{\kl}\,\Gamma_{H_1^+H_1^-}^{1 (\kk-\kl)}+\beta_1(H_1^+,H_1^-,\kk,\kl)\, \Gamma_{H_1^+H_1^-}^{2 (\kk-\kl)}\nn
&-&4\, m_i\, m_j\, m_{\kk}\, m_{\kl}\,\Gamma_{H_1^+H_1^-}^{3 (\kk-\kl)}-m_i\,m_j\,\beta_2(H_1^+,H_1^-,\kk,\kl)\,\Gamma_{H_1^+H_1^-}^{4 (\kk-\kl)}\nn
&-&m_j\,\left(m_{\kk}\,\Gamma_{H_1^+H_1^-}^{5 (\kk-\kl)}+m_{\kl}\,\Gamma_{H_1^+H_1^-}^{6 (\kk-\kl)}\right)\,\beta_3(H_1^+,H_1^-,\kk,\kl)\nn
&+&\left(4\, d\, m_i\,-\frac{m_i\,\omega_{H_1^+H_1^-}}{s}\right)\,\left(m_{\kk}\,\Gamma_{H_1^+H_1^-}^{7 (\kk-\kl)}+m_{\kl}\,\Gamma_{H_1^+H_1^-}^{8 (\kk-\kl)}\right)\\
\tilde{w}_{H_1^-,H_1^+}^{u(\kk,\kl)}&=& 4\, d \,m_{\kk}\, m_{\kl}\,\Gamma_{H_1^-H_1^+}^{1 (\kk-\kl)*}+\beta_1(H_1^-,H_1^+,\kk,\kl)\, \Gamma_{H_1^-,H_1^+}^{2 (\kk-\kl)*}\nn
&-&4\, m_i\, m_j\, m_{\kk}\, m_{\kl}\,\Gamma_{H_1^-,H_1^+}^{3 (\kk-\kl)*}-m_i\,m_j\,\beta_2(H_1^-,H_1^+,\kk,\kl)\,\Gamma_{H_1^-H_1^+}^{4 (\kk-\kl)*}\nn
&-&m_j\,\left(m_{\kk}\,\Gamma_{H_1^-H_1^+}^{5 (\kk-\kl)*}+m_{\kl}\,\Gamma_{H_1^-H_1^+}^{6 (\kk-\kl)*}\right)\,\beta_3(H_1^-,H_1^+,\kk,\kl)\nn
&+&\left(4\, d\, m_i\,-\frac{m_i\,\omega_{H_1^-H_1^+}}{s}\right)\,\left(m_{\kk}\,\Gamma_{H_1^-H_1^+}^{7 (\kk-\kl)*}+m_{\kl}\,\Gamma_{H_1^-H_1^+}^{8 (\kk-\kl)*}\right)\\
\tilde{w}_{H_1^+,H_1^-}^{t-u(\kk,\kl)}&=&d\,m_{\kk}\,m_{\kl}\,s^2\, J_1(H_1^+,H_1^-,\kk,\kl)\,I_{H_1^+,H_1^-}^{1 (\kk-\kl)}+J_2(H_1^+,H_1^-,\kk,\kl)\nn
&&I_{H_m^0 H_n^0}^{2 (\kk-\kl)}-m_i\,m_j\,m_{\kk}\, m_{\kl} s^2\,J_1(H_1^+,H_1^-,\kk,\kl)\,I_{H_1^+,H_1^-}^{3 (\kk-\kl)}\nn
&+&m_i\, m_j\, s^2\,(l_i/2-d_{H_1^+,H_1^-}-m_i^2)\,J_1(H_1^+,H_1^-,\kk,\kl)I_{H_1^+H_1^-}^{4 (\kk-\kl)}\nn
&+&\frac{1}{4}\,m_j\,m_{\kk}\,s\,J_3({H_1^+,H_1^-,\kk,\kl})I_{H_1^+H_1^-}^{5 (\kk-\kl)}+\frac{1}{4}\,m_j\,m_{\kl}\,s\nn&&J_3({H_1^+,H_1^-,\kl,\kk})\,
I_{H_1^+H_1^-}^{6 (\kk-\kl)}-\frac{1}{4}m_i\,m_{\kk}\,s\,J_4(H_1^+,H_1^-,\kk,\kl)\nn&&I_{H_1^+H_1^-}^{7 (\kk-\kl)}
+\frac{1}{4}m_i\,m_{\kl}\,s\,J_5(H_1^+,H_1^-,\kk,\kl)\,I_{H_1^+H_1^-}^{8 (\kk-\kl)}\\
\tilde{w}_{H_1^+,H_1^-}^{(\kk,\kl)}&=&\tilde{w}_{H_1^+,H_1^-}^{t(\kk,\kl)}+\tilde{w}_{H_1^+,H_1^-}^{u(\kk,\kl)}+\left(\tilde{w}_{H_1^+,H_1^-}^{t-u(\kk,\kl)}+c.c.\right)  
\end{eqnarray}
\item  \underline{CP-even Higgs boson ($H_k^0$) and Neutralino ($\kl$) Interference}:
\begin{eqnarray}
\tilde{w}_{H_1^+,H_1^-}^{st(H_k^0,\kl)}&=&-\frac{1}{2} m_i\, m_j\, m_{\kl}\, s\,L(\kl,H_1^+,H_1^-)\,J_{H_1^+H_1^-}^{(1)(H_k^0 - \kl)}\nn
&-&\frac{m_j}{4}\,\left(b_{H_1^+H_1^-}-(R_{H_1^+\kl}-4\,m_i^2\,s)\right)\,\mathcal{A}(\kl,H_1^+,H_1^-)\,J_{H_1^+H_1^-}^{(2)(H_k^0 - \knl)}\nn
&+&\frac{1}{2}\,d\, m_{\kl}\,s\,L(\kl,H_1^+,H_1^-)\,J_{H_1^+H_1^-}^{(3)(H_k^0 - \kl)}-\frac{m_i}{4}\,\left(b_{H_1^+H_1^-}-(R_{H_1^+\kl}\right.\nn&+&\left.4\,d\,s-\tau_{H_1^+H_1^-}-\omega_{H_1^+H_1^-})\,\mathcal{A}(\kl,H_1^+,H_1^-)\right)\,J_{H_1^+H_1^-}^{(4)(H_k^0 - \kl)}
\end{eqnarray}
\begin{eqnarray}
\tilde{w}_{H_1^+,H_1^-}^{su(H_k^0,\kl)}&=&-\frac{1}{2}\, m_i\, m_j\, m_{\kl}\, s\,L(\kl,H_1^-,H_1^+)\,J_{H_1^-H_1^+}^{(1)(H_k^0 - \kl)}\nn
&-&\frac{m_j}{4}\left(b_{H_1^+H_1^-}-(R_{H_1^-\kl}-4\,m_i^2\,s)\right)\,\mathcal{A}(\kl,H_1^-,H_1^+)\,J_{H_1^-H_1^+}^{(2)(H_k^0 - \kl)}\nn
&+&\frac{1}{2}\,d\, m_{\kl}\,s\,L(\kl,H_1^-,H_1^+)\,J_{H_1^+H_1^-}^{(3)(H_k^0 - \kl)}-\frac{m_i}{4}\left(b_{H_1^+H_1^-}-(R_{H_1^-\kl}\right.\nn&+&\left.4\,d\,s-\tau_{H_1^+H_1^-}-\omega_{H_1^+H_1^-})\,\mathcal{A}(\kl,H_1^-,H_1^+)\right)\,J_{A_1^0 H_1^0}^{(4)(H_k^0 - \kl)}\\
\tilde{w}_{H_1^+,H_1^-}^{(H_k^0,\kl)}&=&\frac{1}{b_{H_1^+H_1^-}(s-m_{H_k^0}^2)}\left(\tilde{w}_{H_1^+,H_1^-}^{st(H_k^0,\kl)}+\tilde{w}_{H_1^+,H_1^-}^{su(H_k^0,\kl)}\right)+c.c.
\end{eqnarray}
\end{itemize}
\subsection{$\chi_i^0\chi_j^0\rightarrow W^{\pm} H^{\mp}$}
This process is the sum of the s-channel CP even ($H_1^0$, $H_2^0$), CP odd ($A_1^0$) Higgs boson exchange and  the t- and u-channel chargino ( $\kk$, k = 1, 2) exchange
\begin{align}
\tilde{w}_{W^+H_1^-}&=\tilde{w}_{W^+H_1^-}^{(A_1^0,A_1^0)}+\sum_{k,l=1,2}\tilde{w}_{W^+H_1^-}^{(H_k^0,H_l^0)}+\sum_{k=1,2}\tilde{w}_{W^+H_1^-}^{(H_k^0,A_1^0)}+\sum_{k,l=1,2}\left(\tilde{w}_{W^+H_1^-}^{(\kk,\kl)}+\tilde{w}_{W^+H_1^-}^{(H_k^0,\kl)}\right)
\end{align}
\begin{itemize}
	\item  \underline{CP-even Higgs boson ($H_k^0-H_l^0$) and CP-odd Higgs boson ($A_1^0$)}:
	\begin{align}
	\tilde{w}_{W^+H_1^-}^{(A_1^0,A_1^0)}&=\frac{1}{2}\frac{|C_{A_1^0W^+H_1^-}|^2}{(s-m_{A_1^0}^2)^2}\left(f_{\ki \kj}^{(A_1^0-A_1^0)}\,d-2\,m_i\,m_j\,g_{\ki \kj}^{(A_1^0-A_1^0)}\right)(s-l_{W^+H^-})\\
	\tilde{w}_{W^+H_1^-}^{(H_k^0,H_l^0)}&=\frac{1}{2}\frac{C_{H_k^0W^+H_1^-}\,C_{H_l^0W^+H_1^-}^*}{(s-m_{H_k^0}^2)\,(s-m_{H_l^0}^2)}\left(f_{\ki \kj}^{(H_k^0-H_l^0)}\,d-2\,m_i\,m_j\,g_{\ki \kj}^{(H_k^0-H_l^0)}\right)(s-l_{W^+H^-})\\
	\tilde{w}_{W^+H_1^-}^{(H_k^0,A_1^0)}&=\frac{1}{2}\frac{C_{H_k^0W^+H_1^-}\,C_{A_1^0W^+H_1^-}^*}{(s-m_{H_k^0}^2)\,(s-m_{A_1^0}^2)}\left(f_{\ki \kj}^{(H_k^0-A_1^0)}\,d-2\,m_i\,m_j\,g_{\ki \kj}^{(H_k^0-A_1^0)}\right)(s-l_{W^+H^-})
	\end{align}
	\item  \underline{Chargino ($\kk-\kl$) exchange}:
	\begin{align}
	\tilde{w}_{W^+H_1^-}^{t(\kk,\kl)}&=\Gamma_{W^+ H_1^-}^{1 (\kk-\kl)}\,\mathcal{C}_1(\kk,\kl,W^+,H_1^-)+\Gamma_{W^+ H_1^-}^{2 (\kk-\kl)}\,\mathcal{C}_2(\kk,\kl,W^+,H_1^-)\nn&-3\,m_i\,m_j\,s^2\,\left(M_w^2\,\Gamma_{W^+ H_1^-}^{3 (\kk-\kl)}+m_{\kk}\,m_{\kl}\,\Gamma_{W^+ H_1^-}^{4 (\kk-\kl)}\right)\nn&\mathcal{C}_3(\kk,\kl,W^+,H_1^-)-\left(\Gamma_{W^+ H_1^-}^{5 (\kk-\kl)}\,m_{\kl}+\Gamma_{W^+ H_1^-}^{6 (\kk-\kl)}\,m_{\kk}\right)\nn&\mathcal{C}_4(\kk,\kl,W^+,H_1^-)+\left(\Gamma_{W^+ H_1^-}^{7 (\kk-\kl)}\,m_{\kl}+\Gamma_{W^+ H_1^-}^{8 (\kk-\kl)}\,m_{\kk}\right)\nn
	&\mathcal{C}_5(\kk,\kl,W^+,H_1^-)+\left(\Gamma_{W^+ H_1^-}^{5 (\kk-\kl)}\,m_{\kl}+\Gamma_{W^+ H_1^-}^{6 (\kk-\kl)}\,m_{\kk}\right.\nn&\left.-2\,\Gamma_{W^+ H_1^-}^{3 (\kk-\kl)}\,m_i\right)\,\mathcal{C}_6(\kk,\kl,W^+,H_1^-)
	\end{align}
	\begin{align}
	\tilde{w}_{W^+H_1^-}^{u(\kk,\kl)}&=\left(\Gamma_{ H_1^-W^+}^{1 (\kk-\kl)}\,m_{\kk}+\Gamma_{ H_1^-W^+}^{2 (\kk-\kl)}\,m_{\kl}\right)\,\mathcal{C}_{11}(\kk,\kl,H_1^-,W^+)\nn&+\left(\Gamma_{ H_1^-W^+}^{3 (\kk-\kl)}\,m_{\kk}+\Gamma_{ H_1^-W^+}^{4 (\kk-\kl)}\,m_{\kl}\right)\,\mathcal{C}_{10}(\kk,\kl,H_1^-,W^+)\nn&+\left(\Gamma_{ H_1^-W^+}^{3 (\kk-\kl)}\,m_i\,m_{\kk}+\Gamma_{ H_1^-W^+}^{4 (\kk-\kl)}\,m_i\,m_{\kl}+\Gamma_{ H_1^-W^+}^{5 (\kk-\kl)}\,m_{\kl}\,m_{\kk}\right)\nn&\mathcal{C}_9(\kk,\kl,H_1^-,W^+)+\Gamma_{ H_1^-W^+}^{6 (\kk-\kl)}\,\mathcal{C}_8(\kk,\kl,H_1^-,W^+)+\Gamma_{ H_1^-W^+}^{7 (\kk-\kl)}\nn&\mathcal{C}_6(\kk,\kl,H_1^-,W^+)+\Gamma_{H_1^-W^+ }^{8 (\kk-\kl)}\,\mathcal{C}_7(\kk,\kl,H_1^-,W^+)\\
		\tilde{w}_{W^+H_1^-}^{tu(\kk,\kl)}&=I_{W^+H_1^-}^{1 (\kk-\kl)}\,J_6(W^+,H_1^-,\kk,\kl)+I_{W^+H_1^-}^{2 (\kk-\kl)}\,J_7(W^+,H_1^-,\kk,\kl)\nn&+I_{W^+H_1^-}^{3 (\kk-\kl)}\,J_8(W^+,H_1^-,\kk,\kl)+I_{W^+H_1^-}^{4 (\kk-\kl)}\,J_9(W^+,H_1^-,\kk,\kl)\nn&+I_{W^+H_1^-}^{5 (\kk-\kl)}\,J_{10}(W^+,H_1^-,\kk,\kl)+I_{W^+H_1^-}^{6 (\kk-\kl)}\,J_{11}(W^+,H_1^-,\kk,\kl)\nn&+I_{W^+H_1^-}^{7 (\kk-\kl)}\,J_{12}(W^+,H_1^-,\kk,\kl)+I_{W^+H_1^-}^{8 (\kk-\kl)}\,J_{13}(W^+,H_1^-,\kk,\kl)\\
	\tilde{w}_{W^+H_1^-}^{(\kk,\kl)}&=\tilde{w}_{W^+H_1^-}^{t(\kk,\kl)}+\tilde{w}_{W^+H_1^-}^{u(\kk,\kl)}+\left(\tilde{w}_{W^+H_1^-}^{tu(\kk,\kl)}+c.c.\right)
	\end{align}
	\item  \underline{Chargino ($H_k^0-\kl$) exchange}:
	\begin{align}
	\tilde{w}_{W^+H_1^-}^{st(H_k^0,\kl)}&=\left((-2\,d\,l_i+l_{ia}\,m_i^2)\, J_{W^+H_1^-}^{(5)(H_k^0 - \kl)} +J_{W^+H_1^-}^{(6)(H_k^0 - \kl)}\, l_i\, m_i\, m_j + 
	J_{W^+H_1^-}^{(7)(H_k^0 - \kl)}\right.\nn&\left. l_i\, m_j\, m_{\kl}
	- J_{W^+H_1^-}^{(8)(H_k^0 - \kl)}\, l_{ia}\, m_i\, m_{\kl}\right)\,\mathcal{C}_{12}(H_k^0,\kl,H_1^-,W^+)\nn&+\left( \left(2\, l_{H_1^-W^+} (l_i - l_{ia})\, M_W^2 - 
	2\,d_{W^+H_1^-}\, l_i\,s\right)\, J_{W^+H_1^-}^{(5)(H_k^0 - \kl)} -2\,d_{W^+H_1^-}\,s\right.\nn&\left.\left( 2\,J_{W^+H_1^-}^{(6)(H_k^0 - \kl)}\, m_i\, m_j- 2\, J_{W^+H_1^-}^{(8)(H_k^0 - \kl)}\, m_i\, m_{\kl}+2\,J_{W^+H_1^-}^{(7)(H_k^0 - \kl)}\, m_j\, m_{\kl} \right)\, \right)\nn&\frac{-b_{W^+H_1^-}+(\tau_{W^+H_1^-}-R_{W^+\kk})\,\mathcal{A}(\kl,W^+,H^-)}{16\, b_{W^+H_1^-}\, M_W^2\,s\, (m_{H_k^0}^2 - s)}
	\end{align}
	\begin{align}
	\tilde{w}_{W^+H_1^-}^{su(H_k^0,\kl)}&=\left(4\, s\, \left( J_{H_1^-W^+}^{(6)(H_k^0 - \kl)}\, \mathcal{C}_{13}(H_k^0,\kl,H_1^-,W^+)\, l_{ia}\, m_i\, m_{\kl} + 
	J_{H_1^-W^+}^{(5)(H_k^0 - \kl)}\right.\right.\nn&\left.\left.m_j\, s\, l_i\, m_{\kl}\,\left(l_{W^+H_1^-} - 6\, M_W^2\right) + 
	J_{H_1^-W^+}^{(8)(H_k^0 - \kl)}\, m_i\,m_j\, s\, \,\left((M_W^2-m_{H_1^-}^2)\, l_i \right.\right.\right.\nn
	&+\left.\left.\left.  6\, M_W^2\,(M_W^2-m_{H_1^-}^2-m_i^2+m_j^2) + 2\, b_{W^+H_1^-}\, s - l_{ia}\, s\right)\right) + 
	2\,J_{H_1^-W^+}^{(5)(H_k^0 - \kl)}\right.\nn&\left. \left(2\, d\,s\, \mathcal{C}_{14}(H_k^0,\kl,H_1^-,W^+)  + 2\, \mathcal{C}_{13}(H_k^0,\kl,H_1^-,W^+)\, l_{ia}\, m_i^2\, s \right.\right.\nn&+\left.\left.l_{H_1^-W^+}\, l_i\, l_{ia}\,(M_W^2-m_{H_1^-}^2)\,(M_W^2-s)\right)\right)\,\frac{\mathcal{A}(\kl,H^-,W^+)}{16\, b_{W^+H_1^-}\, M_W^2\, (m_{H_k^0}^2 - s)\, s}\nn
	&\left(\left(J_{H_1^-W^+}^{(8)(H_k^0 - \kl)}\, m_i\, m_j -J_{H_1^-W^+}^{(6)(H_k^0 - \kl)}\, m_i\, m_{\kl} + J_{H_1^-W^+}^{(5)(H_k^0 - \kl)}\, m_j m_{\kl}\right)\right.\nn&\left.4\, s\, (2\, M_W^2 - s) - 
	2\, J_{H_1^-W^+}^{(7)(H_k^0 - \kl)}\, \left(l_i\,(M_W^2-m_{H_1^-}^2)\,(M_W^2 - s) + 4\, M_W^2\,s\right.\right.\nn&\left.\left. (2\,m_i^2 -m_j^2) - 2\, m_i^2\, s^2\right)\right)\nn&\frac{-b_{W^+H_1^-}+(\tau_{H_1^-W^+}-R_{H_1^-\kk})\,\mathcal{A}(\kl,H^-,W^+)}{16\, b_{W^+H_1^-}\, M_W^2\, (m_{H_k^0}^2 - s)\, s}
	\end{align}
	\begin{align}
	\tilde{w}_{W^+H_1^-}^{(H_k^0,\kl)}&=\left(\tilde{w}_{W^+H_1^-}^{st(H_k^0,\kl)}+\tilde{w}_{W^+H_1^-}^{su(H_k^0,\kl)}\right)+c.c. 
	\end{align}	
	Note that the contributions to the final state $W^-$ $H^+$ and the final state $W^+$ $H^-$ are same.
\end{itemize}
\subsection{$\chi_i^0\chi_j^0\rightarrow Z H_n^0$}
This process is sum of the s-channel CP odd Higgs boson and Z boson besides t and u channel neutralino exchange. Resulting cross section is obtained by taking the sum of the s, t, and u channel cross sections and possible cross sections obtained considering interference among themselves. And it is given by:
\begin{align}
\tilde{w}_{ Z, H_n^0}&=\tilde{w}_{ Z, H_n^0}^{s(A_1^0,A_1^0)}+\tilde{w}_{ Z ,H_n^0}^{s(Z,Z)}+\tilde{w}_{ Z, H_n^0}^{s(A_1^0,Z)}+\sum_{k,l=1,4}	\tilde{w}_{ZH_n^0}^{(\knk,\knl)}+\sum_{l=1,4}\tilde{w}_{ Z, H_n^0}^{(Z,\knl)}+\sum_{l=1,4}\tilde{w}_{ Z, H_n^0}^{(A_1^0,\knl)}
\end{align}
\begin{itemize}
	\item \underline{CP-odd Higgs boson $A_1^0$ and Z exchange}:
	\begin{align}
	\tilde{w}_{ Z, H_n^0}^{s(A_1^0,A_1^0)}&=\frac{1}{2}\frac{|C_{A_1^0ZH_n^0}|^2}{(s-m_{A_1^0}^2)^2}\left(f_{\ki \kj}^{(A_1^0-A_1^0)}\,d-2\,m_i\,m_j\,g_{\ki \kj}^{(A_1^0-A_1^0)}\right)(s-l_{ZH_n^0})\\
	\tilde{w}_{ Z ,H_n^0}^{s(Z,Z)}&=\frac{|C_{ZZH_n^0}|^2}{(s-M_Z^2)^2} \left(f_{\ki \kj}^{(Z-Z)}\,\mathcal{B}_7(Z,Z,H_n^0)+m_i\,m_j\,g_{\ki \kj}^{(Z-Z)}\,\mathcal{B}_8(Z,Z,H_n^0)\right)	\\
	\tilde{w}_{ Z, H_n^0}^{s(A_1^0,Z)}&=\frac{C_{ZZH_n^0}\,C_{A_1^0ZH_n^0}^*}{16\,M_Z^4\,s\,(s-m_A^2)}\,\left(-l_i\,m_j\,f_{\ki \kj}^{(Z-A_1^0)}+l_{ia}\,m_i\,g_{\ki \kj}^{(Z-A_1^0)}\right)\nn&\left(2\, d_{ZH_n^0}\, l_{ZH_n^0} + l_{ZH_n^0}^2 - 2\, M_Z^2\, (l_{H_n^0Z} + 2\, s)\right)
	\end{align}
	\item \underline{ Neutralino ($\knk-\knl$) exchange}:
	\begin{align}
	\tilde{w}_{ZH_n^0}^{t(\knk,\knl)}&=\Gamma_{ZH_n^0}^{1 (\knk-\knl)}\,\mathcal{C}_1(\knk,\knl,Z,H_n^0)+\Gamma_{ZH_n^0}^{2 (\knk-\knl)}\,\mathcal{C}_2(\knk,\knl,Z,H_n^0)-3\,m_i\,m_j\,s^2\nn
	&\left(M_Z^2\,\Gamma_{ZH_n^0}^{3 (\knk-\knl)}+m_{\knk}\,m_{\knl}\,\Gamma_{ZH_n^0}^{4 (\knk-\knl)}\right)\mathcal{C}_3(\knk,\knl,Z,H_n^0)-\left(\Gamma_{ZH_n^0}^{5 (\knk-\knl)}\,m_{\knl}\right.\nn
	&\left.+\Gamma_{ZH_n^0}^{6 (\kk-\kl)}\,m_{\kk}\right)\,\mathcal{C}_4(\knk,\knl,Z_n,H_n^0)+\left(\Gamma_{ZH_n^0}^{7 (\knk-\knl)}\,m_{\knl}+\Gamma_{ZH_n^0}^{8 (\knk-\knl)}\,m_{\knk}\right)\nn
	&\mathcal{C}_5(\knk,\knl,Z_n,H_n^0)+\left(\Gamma_{ZH_n^0}^{5 (\knk-\knl)}\,m_{\knl}+\Gamma_{ZH_n^0}^{6 (\knk-\knl)}\,m_{\knk}-2\,\Gamma_{ZH_n^0}^{3 (\knk-\knl)}\,m_i\right)\nn
	&\mathcal{C}_6(\knk,\knl,Z_n,H_n^0)\\
	\tilde{w}_{ZH_n^0}^{u(\knk,\knl)}&=\left(\Gamma_{ H_n^0Z}^{1 (\knk-\knl)}\,m_{\knk}+\Gamma_{ H_n^0Z}^{2 (\knk-\knl)}\,m_{\knl}\right)\,\mathcal{C}_{11}(\knk,\knl,H_n^0,Z)\nn
	&+\left(\Gamma_{ H_n^0Z}^{3 (\knk-\knl)}\,m_{\knk}+\Gamma_{ H_n^0Z}^{4 (\knk-\knl)}\,m_{\knl}\right)\,\mathcal{C}_{10}(\knk,\knl,H_n^0,Z)+\left(\Gamma_{ H_n^0Z}^{3 (\knk-\knl)}\right.\nn
	&\left.m_i\,m_{\knk}+\Gamma_{ H_n^0Z}^{4 (\knk-\knl)}\,m_i\,m_{\knl}+\Gamma_{ H_n^0Z}^{5 (\knk-\knl)}\,m_{\knl}\,m_{\knk}\right)\,\mathcal{C}_9(\knk,\knl,H_n^0,Z)\nn
	&+\Gamma_{ H_n^0Z}^{6 (\knk-\knl)}\,\mathcal{C}_8(\knk,\knl,H_n^0,Z)+\Gamma_{ H_n^0Z}^{7 (\knk-\knl)}\,\mathcal{C}_6(\knk,\knl,H_n^0,Z)\nn
	&+\Gamma_{H_n^0Z }^{8 (\knk-\knl)}\,\mathcal{C}_7(\knk,\knl,H_n^0,Z)\\
		\tilde{w}_{ZH_n^0}^{tu(\knk,\knl)}&=I_{ZH_n^0}^{1 (\knk-\knl)}\,J_{6}(Z,H_n^0,\knk,\knl)+I_{ZH_n^0}^{2 (\knk-\knl)}\,J_7(Z,H_n^0,\knk,\knl)\nn&+I_{ZH_n^0}^{3 (\knk-\knl)}\,J_8(Z,H_n^0,\knk,\knl)+I_{ZH_n^0}^{4 (\knk-\knl)}\,J_9(Z,H_n^0,\knk,\knl)\nn&+I_{ZH_n^0}^{5 (\knk-\knl)}\,J_{10}(Z,H_n^0,\knk,\knl)+I_{ZH_n^0}^{6 (\knk-\knl)}\,J_{11}(Z,H_n^0,\knk,\knl)\nn&+I_{ZH_n^0}^{7 (\knk-\knl)}\,J_{12}(Z,H_n^0,\knk,\knl)+I_{ZH_n^0}^{8 (\knk-\knl)}\,J_{13}(Z,H_n^0,\knk,\knl)\\
	\tilde{w}_{ZH_n^0}^{(\knk,\knl)}&=\tilde{w}_{ZH_n^0}^{t(\knk,\knl)}+\tilde{w}_{ZH_n^0}^{u(\knk,\knl)}+\left(\tilde{w}_{ZH_n^0}^{tu(\knk,\knl)}+c.c.\right)
	\end{align}
	\item \underline{Z boson, CP odd Higgs boson ($A_1^0$) and neutalino $(\knl)$ exchange}:
	\begin{align}
	\tilde{w}_{ Z, H_n^0}^{st(Z,\knl)}&=\frac{C_{ZZH_n^0}}{s-M_Z^2}\left(\mathcal{B}_{9}(Z,\knl,Z,H_n^0)\,J_{ZH_n^0}^{(5)(Z - \knl)}+\mathcal{B}_{10}(Z,\knl,Z,H_n^0)\,J_{ZH_n^0}^{(6)(Z - \knl)}\right.\nn&\left.
	+\mathcal{B}_{11}(Z,\knl,Z,H_n^0)\,J_{ZH_n^0}^{(7)(Z - \knl)}+\mathcal{B}_{12}(Z,\knl,Z,H_n^0)\,J_{ZH_n^0}^{(8)(Z - \knl)}\right)\\
	\tilde{w}_{ Z, H_n^0}^{su(Z,\knl)}&=\frac{C_{ZZH_n^0}}{s-M_Z^2}\,\left(\mathcal{B}_{13}(Z,\knl,H_n^0,Z)\,J_{H_n^0Z}^{(5)(Z - \knl)}+\mathcal{B}_{14}(Z,\knl,H_n^0,Z)\,J_{H_n^0Z}^{(6)(Z - \knl)}\right.\nn&\left.
	+\mathcal{B}_{15}(Z,\knl,H_n^0,Z)\,J_{H_n^0Z}^{(7)(Z - \knl)}+\mathcal{B}_{16}(Z,\knl,H_n^0,Z)\,J_{H_n^0Z}^{(8)(Z - \knl)}\right)
	\end{align}
	\begin{align}
	\tilde{w}_{ZH_n^0}^{st(A_1^0,\knl)}&=\left((-2\,d\,l_i+l_{ia}\,m_i^2)\, J_{ZH_n^0}^{(5)(A_1^0 - \knl)} +J_{ZH_n^0}^{(6)(A_1^0 - \knl)}\, l_i\, m_i\, m_j + 
	J_{ZH_n^0}^{(7)(A_1^0 - \knl)}\right.\nn&\left. l_i\, m_j\, m_{\knl}
	- J_{ZH_n^0}^{(8)(A_1^0 - \knl)}\, l_{ia}\, m_i\, m_{\knl}\right)\,\mathcal{C}_{12}(A_1^0,\knl,H_n^0,Z)\nn&+\left( \left(2\, l_{H_n^0Z}\, (l_i - l_{ia})\, M_Z^2 - 
	2\,d_{ZH_n^0}\, l_i\,s\right)\, J_{ZH_n^0}^{(5)(A_1^0 - \knl)} -2\,d_{ZH_n^0}\,s\right.\nn&\left.\left( 2\,J_{ZH_n^0}^{(6)(A_1^0 - \knl)}\, m_i\, m_j- 2\, J_{ZH_n^0}^{(8)(A_1^0 - \knl)}\, m_i\, m_{\knl}+2\,J_{ZH_n^0}^{(7)(A_1^0 - \knl)}\, m_j\, m_{\knl} \right)\, \right)\nn&\frac{-b_{ZH_n^0}+(\tau_{ZH_n^0}-R_{Z\knk})\,\mathcal{A}(\knl,Z,H_n^0)}{16\, b_{ZH_n^0}\, M_Z^2\,s\, (m_{A_1^0}^2 - s)}
	\end{align}
	\begin{align}
	\tilde{w}_{ZH_n^0}^{su(A_1^0,\knl)}&=\left(4\, s\, \left( J_{H_n^0Z}^{(6)(A_1^0 - \knl)}\, \mathcal{C}_{13}(A_1^0,\knl,H_n^0,Z)\, l_{ia}\, m_i\, m_{\knl} + 
	J_{H_n^0Z}^{(5)(A_1^0 - \knl)}\right.\right.\nn&\left.\left.m_j\, s\, l_i\, m_{\knl}\,\left(l_{ZH_n^0} - 6\, M_Z^2\right) + 
	J_{H_n^0Z}^{(8)(A_1^0 - \knl)}\, m_i\,m_j\, s\, \,\left((M_Z^2-m_{H_n^0}^2)\, l_i \right.\right.\right.\nn
	&+\left.\left.\left.  6\, M_Z^2\,(M_Z^2-m_{H_n^0}^2-m_i^2+m_j^2) + 2\, b_{ZH_n^0}\, s - l_{ia}\, s\right)\right) + 
	2\,J_{H_n^0Z}^{(5)(A_1^0 - \knl)}\right.\nn&\left. \left(2\, d\,s\, \mathcal{C}_{14}(A_1^0,\knl,H_n^0,Z)  + 2\, \mathcal{C}_{13}(A_1^0,\knl,H_n^0,Z)\, l_{ia}\, m_i^2\, s \right.\right.\nn&+\left.\left.l_{H_n^0Z}\, l_i\, l_{ia}\,(M_Z^2-m_{H_n^0}^2)\,(M_Z^2-s)\right)\right)\,\frac{\mathcal{A}(\knl,H_n^0,Z)}{16\, b_{ZH_n^0}\, M_Z^2\, (m_{A_1^0}^2 - s)\, s}\nn
	&\left(\left(J_{H_n^0Z}^{(8)(A_1^0 - \knl)}\, m_i\, m_j -J_{H_n^0Z}^{(6)(A_1^0 - \knl)}\, m_i\, m_{\knl} + J_{H_n^0Z}^{(5)(A_1^0 - \knl)}\, m_j m_{\knl}\right)\right.\nn&\left.4\, s\, (2\, M_Z^2 - s) - 
	2\, J_{H_n^0Z}^{(7)(A_1^0 - \knl)}\, \left(l_i\,(M_Z^2-m_{H_n^0}^2)\,(M_Z^2 - s) + 4\, M_Z^2\,s\right.\right.\nn&\left.\left. (2\,m_i^2 -m_j^2) - 2\, m_i^2\, s^2\right)\right)\nn&\frac{-b_{ZH_n^0}+(\tau_{H_n^0Z}-R_{H_n^0\knk})\,\mathcal{A}(\knl,H_n^0,Z)}{16\, b_{ZH_n^0}\, M_Z^2\, (m_{A_1^0}^2 - s)\, s}\\
	\tilde{w}_{ZH_n^0}^{(A_1^0,\knl)}&=\left(\tilde{w}_{ZH_n^0}^{st(A_1^0,\knl)}+\tilde{w}_{ZH_n^0}^{su(A_1^0,\knl)}\right)+c.c. 
	\end{align}	
	\begin{align}
	\tilde{w}_{ Z, H_n^0}^{(Z,\knl)}&=\left(\tilde{w}_{ Z, H_n^0}^{st(Z,\knl)}+\tilde{w}_{ Z, H_n^0}^{su(Z,\knl)}\right)+c.c.\\
	\tilde{w}_{ Z, H_n^0}^{(A_1^0,\knl)}&=\left(\tilde{w}_{ Z, H_n^0}^{st(A_1^0,\knl)}+\tilde{w}_{ Z, H_n^0}^{su(A_1^0,\knl)}\right)+c.c.
	\end{align}		
\end{itemize}
\subsection{$\chi_i^0\chi_j^0\rightarrow Z H_n^0$}
This process is sum of the s-channel CP even Higgs boson besides t and u channel neutralino exchange. Resulting cross section is obtained by taking the sum of the s, t, and u channel cross sections and possible cross sections obtained considering interference among themselves. And it is given by:
\begin{align}
\tilde{w}_{ Z, A_1^0}&=\sum_{k,l=1,2}\tilde{w}_{ Z, A_1^0}^{s(H_k^0,H_l^0)}+\sum_{k,l=1,4}	\tilde{w}_{ZH_n^0}^{(\knk,\knl)}+\sum_{k=1,2}\sum_{l=1,4}\tilde{w}_{H_k^0, A_1^0}^{(H_k^0,\knl)}
\end{align}	
\begin{itemize}
	\item \underline{CP-even Higgs boson $H_k^0-H_l^0$  exchange}:
	\begin{align}
	\tilde{w}_{ZA_1^0}^{(H_k^0,H_l^0)}&=\frac{1}{2}\frac{C_{H_k^0ZA_1^0}\,C_{H_l^0ZA_1^0}^*}{(s-m_{H_k^0}^2)\,(s-m_{H_l^0}^2)}\left(f_{\ki \kj}^{(H_k^0-H_l^0)}\,d-2\,m_i\,m_j\,g_{\ki \kj}^{(H_k^0-H_l^0)}\right)(s-l_{ZA_1^0})	
	\end{align}
	\item \underline{Neutralino ($\knk-\knl$) exchange}:
	\begin{align}
	\tilde{w}_{ZA_1^0}^{t(\knk,\knl)}&=\Gamma_{ZA_1^0}^{1 (\knk-\knl)}\,\mathcal{C}_1(\knk,\knl,Z,A_1^0)+\Gamma_{ZA_1^0}^{2 (\knk-\knl)}\,\mathcal{C}_2(\knk,\knl,Z,A_1^0)-3\,m_i\,m_j\,s^2\nn
	&\left(M_Z^2\,\Gamma_{ZA_1^0}^{3 (\knk-\knl)}+m_{\knk}\,m_{\knl}\,\Gamma_{ZA_1^0}^{4 (\knk-\knl)}\right)\mathcal{C}_3(\knk,\knl,Z,A_1^0)-\left(\Gamma_{ZA_1^0}^{5 (\knk-\knl)}\,m_{\knl}\right.\nn
	&\left.+\Gamma_{ZA_1^0}^{6 (\kk-\kl)}\,m_{\kk}\right)\,\mathcal{C}_4(\knk,\knl,Z_n,A_1^0)+\left(\Gamma_{ZA_1^0}^{7 (\knk-\knl)}\,m_{\knl}+\Gamma_{ZA_1^0}^{8 (\knk-\knl)}\,m_{\knk}\right)\nn
	&\mathcal{C}_5(\knk,\knl,Z_n,A_1^0)+\left(\Gamma_{ZA_1^0}^{5 (\knk-\knl)}\,m_{\knl}+\Gamma_{ZA_1^0}^{6 (\knk-\knl)}\,m_{\knk}-2\,\Gamma_{ZA_1^0}^{3 (\knk-\knl)}\,m_i\right)\nn
	&\mathcal{C}_6(\knk,\knl,Z_n,A_1^0)
	\end{align}
	\begin{align}
	\tilde{w}_{ZA_1^0}^{u(\knk,\knl)}&=\left(\Gamma_{ A_1^0Z}^{1 (\knk-\knl)}\,m_{\knk}+\Gamma_{ A_1^0Z}^{2 (\knk-\knl)}\,m_{\knl}\right)\,\mathcal{C}_{11}(\knk,\knl,A_1^0,Z)\nn
	&+\left(\Gamma_{ A_1^0Z}^{3 (\knk-\knl)}\,m_{\knk}+\Gamma_{ A_1^0Z}^{4 (\knk-\knl)}\,m_{\knl}\right)\,\mathcal{C}_{10}(\knk,\knl,A_1^0,Z)+\left(\Gamma_{ A_1^0Z}^{3 (\knk-\knl)}\right.\nn
	&\left.m_i\,m_{\knk}+\Gamma_{ A_1^0Z}^{4 (\knk-\knl)}\,m_i\,m_{\knl}+\Gamma_{ A_1^0Z}^{5 (\knk-\knl)}\,m_{\knl}\,m_{\knk}\right)\,\mathcal{C}_9(\knk,\knl,A_1^0,Z)\nn
	&+\Gamma_{ A_1^0Z}^{6 (\knk-\knl)}\,\mathcal{C}_8(\knk,\knl,A_1^0,Z)+\Gamma_{ A_1^0Z}^{7 (\knk-\knl)}\,\mathcal{C}_6(\knk,\knl,A_1^0,Z)\nn
	&+\Gamma_{A_1^0Z }^{8 (\knk-\knl)}\,\mathcal{C}_7(\knk,\knl,A_1^0,Z)\\
	\tilde{w}_{ZA_1^0}^{tu(\knk,\knl)}&=I_{ZA_1^0}^{1 (\knk-\knl)}\,J_{6}(Z,A_1^0,\knk,\knl)+I_{ZA_1^0}^{2 (\knk-\knl)}\,J_7(Z,A_1^0,\knk,\knl)\nn&+I_{ZA_1^0}^{3 (\knk-\knl)}\,J_8(Z,A_1^0,\knk,\knl)+I_{ZA_1^0}^{4 (\knk-\knl)}\,J_9(Z,A_1^0,\knk,\knl)\nn&+I_{ZA_1^0}^{5 (\knk-\knl)}\,J_{10}(Z,A_1^0,\knk,\knl)+I_{ZA_1^0}^{6 (\knk-\knl)}\,J_{11}(Z,A_1^0,\knk,\knl)\nn&+I_{ZA_1^0}^{7 (\knk-\knl)}\,J_{12}(Z,A_1^0,\knk,\knl)+I_{ZA_1^0}^{8 (\knk-\knl)}\,J_{13}(Z,A_1^0,\knk,\knl)\\
	\tilde{w}_{ZA_1^0}^{(\knk,\knl)}&=\tilde{w}_{ZA_1^0}^{t(\knk,\knl)}+\tilde{w}_{ZA_1^0}^{u(\knk,\knl)}+\left(\tilde{w}_{ZA_1^0}^{tu(\knk,\knl)}+c.c.\right)
	\end{align}	
	\item \underline{CP even Higgs boson ($H_k^0$) and neutralino ($\knl$)  exchange}:
	\begin{align}
	\tilde{w}_{ZA_1^0}^{st(H_k^0,\knl)}&=\left((-2\,d\,l_i+l_{ia}\,m_i^2)\, J_{ZA_1^0}^{(5)(H_k^0 - \knl)} +J_{ZA_1^0}^{(6)(H_k^0 - \knl)}\, l_i\, m_i\, m_j + 
	J_{ZA_1^0}^{(7)(H_k^0 - \knl)}\right.\nn&\left. l_i\, m_j\, m_{\knl}
	- J_{ZA_1^0}^{(8)(H_k^0 - \knl)}\, l_{ia}\, m_i\, m_{\knl}\right)\,\mathcal{C}_{12}(H_k^0,\knl,A_1^0,Z)\nn&+\left( \left(2\, l_{A_1^0Z}\, (l_i - l_{ia})\, M_Z^2 - 
	2\,d_{ZA_1^0}\, l_i\,s\right)\, J_{ZA_1^0}^{(5)(H_k^0 - \knl)} -2\,d_{ZA_1^0}\,s\right.\nn&\left.\left( 2\,J_{ZA_1^0}^{(6)(H_k^0 - \knl)}\, m_i\, m_j- 2\, J_{ZA_1^0}^{(8)(H_k^0 - \knl)}\, m_i\, m_{\knl}+2\,J_{ZA_1^0}^{(7)(H_k^0 - \knl)}\, m_j\, m_{\knl} \right)\, \right)\nn&\frac{-b_{ZA_1^0}+(\tau_{ZA_1^0}-R_{Z\knk})\,\mathcal{A}(\knl,Z,A_1^0)}{16\, b_{ZA_1^0}\, M_Z^2\,s\, (m_{H_k^0}^2 - s)}
	\end{align}
	\begin{align}
	\tilde{w}_{ZA_1^0}^{su(H_k^0,\knl)}&=\left(4\, s\, \left( J_{A_1^0Z}^{(6)(H_k^0 - \knl)}\, \mathcal{C}_{13}(H_k^0,\knl,A_1^0,Z)\, l_{ia}\, m_i\, m_{\knl} + 
	J_{A_1^0Z}^{(5)(H_k^0 - \knl)}\right.\right.\nn&\left.\left.m_j\, s\, l_i\, m_{\knl}\,\left(l_{ZA_1^0} - 6\, M_Z^2\right) + 
	J_{A_1^0Z}^{(8)(H_k^0 - \knl)}\, m_i\,m_j\, s\, \,\left((M_Z^2-m_{A_1^0}^2)\, l_i \right.\right.\right.\nn
	&+\left.\left.\left.  6\, M_Z^2\,(M_Z^2-m_{A_1^0}^2-m_i^2+m_j^2) + 2\, b_{ZA_1^0}\, s - l_{ia}\, s\right)\right) + 
	2\,J_{A_1^0Z}^{(5)(H_k^0 - \knl)}\right.\nn&\left. \left(2\, d\,s\, \mathcal{C}_{14}(H_k^0,\knl,A_1^0,Z)  + 2\, \mathcal{C}_{13}(H_k^0,\knl,A_1^0,Z)\, l_{ia}\, m_i^2\, s \right.\right.\nn&+\left.\left.l_{A_1^0Z}\, l_i\, l_{ia}\,(M_Z^2-m_{A_1^0}^2)\,(M_Z^2-s)\right)\right)\,\frac{\mathcal{A}(\knl,A_1^0,Z)}{16\, b_{ZA_1^0}\, M_Z^2\, (m_{H_k^0}^2 - s)\, s}\nn
	&\left(\left(J_{A_1^0Z}^{(8)(H_k^0 - \knl)}\, m_i\, m_j -J_{A_1^0Z}^{(6)(H_k^0 - \knl)}\, m_i\, m_{\knl} + J_{A_1^0Z}^{(5)(H_k^0 - \knl)}\, m_j m_{\knl}\right)\right.\nn&\left.4\, s\, (2\, M_Z^2 - s) - 
	2\, J_{A_1^0Z}^{(7)(H_k^0 - \knl)}\, \left(l_i\,(M_Z^2-m_{A_1^0}^2)\,(M_Z^2 - s) + 4\, M_Z^2\,s\right.\right.\nn&\left.\left. (2\,m_i^2 -m_j^2) - 2\, m_i^2\, s^2\right)\right)\nn&\frac{-b_{ZA_1^0}+(\tau_{A_1^0Z}-R_{A_1^0\knk})\,\mathcal{A}(\knl,A_1^0,Z)}{16\, b_{ZA_1^0}\, M_Z^2\, (m_{H_k^0}^2 - s)\, s}\\
	\tilde{w}_{ZA_1^0}^{(H_k^0,\knl)}&=\left(\tilde{w}_{ZA_1^0}^{st(H_k^0,\knl)}+\tilde{w}_{ZA_1^0}^{su(H_k^0,\knl)}\right)+c.c. 
	\end{align}	
\end{itemize}
\subsection{$\chi_i^0\chi_j^0\rightarrow W^{\pm} W^{\mp}$}
This process is the sum of s-channel diagram through CP even ($H_1^0$, $H_2^0$), Z boson and  t- and u-channel chargino ( $\kk$, k = 1, 2) exchange.
\begin{align}
\tilde{w}_{W^\pm W^\mp}&=\sum_{k,l=1,2}\tilde{w}_{W^\pm W^\mp}^{(H_k^0,H_l^0)}+\tilde{w}_{W^\pm W^\mp}^{(Z,Z)}+\sum_{k,l=1,2}\tilde{w}_{W^\pm W^\mp}^{(\kk,\kl)}+\sum_{k=1,2}\sum_{l=1,4}\tilde{w}_{W^\pm W^\mp}^{(H_k^0,\kl)}+\sum_{l=1,4}\tilde{w}_{W^\pm W^\mp}^{(Z,\kl)}
\end{align}
\begin{itemize}
	\item  \underline{CP-even Higgs boson ($H_k^0-H_l^0$) and CP-odd Higgs boson (Z)}:
	\begin{align}
	\tilde w_{W W}^{(H_k^0-H_l^0)}&=\frac{C_{H_k^0 W W} C_{H_l^0 W W}^*}{(S-m_{H_k^0}^2)\,(S-m_{H_l^0}^2)} \frac{1}{8\, M_W^4}(12\, M_W^4-4\, M_W^2 S+S^2)\nn&(f_{\ki \kj}^{(H_k^0-H_l^0)} d-g_{\ki \kj}^{(H_k^0-H_l^0)}\, m_i\, m_j)
	\end{align}
	\begin{align}
	\tilde w_{W W}^{(Z-Z)}&=\frac{C_{Z W W} C_{Z W W}^*}{(S-M_Z^2)^2}\,\left(-\frac{f_{\ki \kj}^{(Z-Z)}}{48\,M_W^2\,S^2}\left(24\, l_i\, l_{ia}\, M_W^2\, (4\, M_W^2 - S)\, S^2 \right.\right.\nn&+\left.\left. b_{WW}^2\, (12\, M_W^4 - 4\, M_W^2\, S + S^2)\right)+\frac{g_{\ki \kj}^{(Z-Z)}\,m_i\,m_j}{M_W^2}\,S\,(S-4\,M_W^2)\right.\nn&+\left.\frac{f_{\ki \kj}^{(Z-Z)}\,d+g_{\ki \kj}^{(Z-Z)}\,m_i\,m_j}{8\,M_W^4}\,\left(80\, M_W^6 - 36\, M_W^4\, S - 8\, M_W^2\, S^2 + 3\, S^3\right) 
	\right)\\
	\tilde w_{W W}^{(Z-H_l^0)}&=0	
	\end{align}
	\item  \underline{Chargino exchange}:
	\begin{align}
	\tilde w_{W W}^{t(\kk-\kl)}&=\Gamma_{W W}^{1 (\kk-\kl)}\,\mathcal{C}_{12}(\kk,\kl,W,W)+\Gamma_{W W}^{2 (\kk-\kl)}\,\mathcal{C}_{13}(\kk,\kl,W,W)\nn&+\Gamma_{W W}^{3 (\kk-\kl)}\,\mathcal{C}_{14}(\kk,\kl,W,W)+\Gamma_{W W}^{4 (\kk-\kl)}\,\mathcal{C}_{15}(\kk,\kl,W,W)\nn
	&+\left(m_{\kl}\,\Gamma_{W W}^{5 (\kk-\kl)}+m_{\kk}\,\Gamma_{W W}^{6 (\kk-\kl)}\right)\,\mathcal{C}_{16}(\kk,\kl,W,W)\nn&+\left(m_{\kl}\,\Gamma_{W W}^{7 (\kk-\kl)}+m_{\kk}\,\Gamma_{W W}^{8 (\kk-\kl)}\right)\,\mathcal{C}_{17}(\kk,\kl,W,W)\\
	\tilde w_{W W}^{u(\kk-\kl)}&=\Gamma_{W W}^{1 (\kk-\kl)}\,\mathcal{C}_{12}(\kk,\kl,W,W)+\Gamma_{W W}^{2 (\kk-\kl)}\,\mathcal{C}_{13}(\kk,\kl,W,W)\nn&+\Gamma_{W W}^{3 (\kk-\kl)}\,\mathcal{C}_{14}(\kk,\kl,W,W)+\Gamma_{W W}^{4 (\kk-\kl)}\,\mathcal{C}_{15}(\kk,\kl,W,W)\nn
	&+\left(m_{\kl}\,\Gamma_{W W}^{5 (\kk-\kl)}+m_{\kk}\,\Gamma_{W W}^{6 (\kk-\kl)}\right)\,\mathcal{C}_{16}(\kk,\kl,W,W)\nn&+\left(m_{\kl}\,\Gamma_{W W}^{7 (\kk-\kl)}+m_{\kk}\,\Gamma_{W W}^{8 (\kk-\kl)}\right)\,\mathcal{C}_{17}(\kk,\kl,W,W)\\
	\tilde w_{W W}^{tu(\kk-\kl)}&=I_{W W}^{1 (\kk-\kl)}\,J_{14}(\kk,\kl,W,W)+I_{W W}^{2 (\kk-\kl)}\,J_{15}(\kk,\kl,W,W)\nn&+I_{W W}^{3 (\kk-\kl)}\,J_{16}(\kk,\kl,W,W)+I_{W W}^{4 (\kk-\kl)}\,J_{17}(\kk,\kl,W,W)\nn
	&+I_{W W}^{5 (\kk-\kl)}\,J_{18}(\kk,\kl,W,W)+I_{W W}^{6 (\kk-\kl)}\,J_{19}(\kk,\kl,W,W)\nn&+I_{W W}^{7 (\kk-\kl)}\,J_{20}(\kk,\kl,W,W)+I_{W W}^{8 (\kk-\kl)}\,J_{21}(\kk,\kl,W,W)
	\end{align}	
		\item  \underline{CP-even Higgs boson ($H_k^0$), Z boson interference with chargino ($\kl$) }:
\begin{align}
\tilde w_{W W}^{st(H_k^0-\kl)}&=\frac{C_{H_k^0 W W}}{s-m_{H_k^0}^2} \,\left( J_{WW}^{(5)(H_k^0 - \kl)}\,\mathcal{B}_{20}(H_k^0,\kl,W,W)+J_{WW}^{(6)(H_k^0 - \kl)}\right.\nn&\left.\mathcal{B}_{21}(H_k^0,\kl,W,W)+J_{WW}^{(7)(H_k^0 - \kl)}\,\mathcal{B}_{22}(H_k^0,\kl,W,W)+J_{WW}^{(8)(H_k^0 - \kl)}\right.\nn&\left.\mathcal{B}_{23}(H_k^0,\kl,W,W)\right)\\	
\tilde w_{W W}^{su(H_k^0-\kl)}&=\frac{C_{H_k^0 W W}}{s-m_{H_k^0}^2} \,\left( J_{WW}^{(5)(H_k^0 - \kl)*}\,\mathcal{B}_{20}(H_k^0,\kl,W,W)+J_{WW}^{(6)(H_k^0 - \kl)*}\right.\nn&\left.\mathcal{B}_{21}(H_k^0,\kl,W,W)+J_{WW}^{(7)(H_k^0 - \kl)*}\,\mathcal{B}_{22}(H_k^0,\kl,W,W)+J_{WW}^{(8)(H_k^0 - \kl)*}\right.\nn&\left.\mathcal{B}_{23}(H_k^0,\kl,W,W)\right)\end{align}
\begin{align}
\tilde w_{W W}^{st(Z-\kl)}&=\frac{C_{Z W W}}{s-M_Z^2} \,\left( J_{WW}^{(5)(Z - \kl)}\,\mathcal{B}_{24}(Z,\kl,W,W)+J_{WW}^{(6)(Z - \kl)}\right.\nn&\left.\mathcal{B}_{25}(Z,\kl,W,W)+J_{WW}^{(7)(Z - \kl)}\,\mathcal{B}_{26}(Z,\kl,W,W)+J_{WW}^{(8)(Z- \kl)}\right.\nn&\left.\mathcal{B}_{27}(Z,\kl,W,W)\right)
\end{align}
\begin{align}
\tilde w_{W W}^{su(Z-\kl)}&=-\frac{C_{Z W W}}{s-M_Z^2} \,\left( J_{WW}^{(5)(Z - \kl)*}\,\mathcal{B}_{24}(Z,\kl,W,W)+J_{WW}^{(6)(Z - \kl)*}\right.\nn&\left.\mathcal{B}_{25}(Z,\kl,W,W)+J_{WW}^{(7)(Z - \kl)*}\,\mathcal{B}_{26}(Z,\kl,W,W)+J_{WW}^{(8)(Z- \kl)*}\right.\nn&\left.\mathcal{B}_{27}(Z,\kl,W,W)\right)\\
\tilde{w}_{WW}^{(H_k^0,\kl)}&=\left(\tilde{w}_{WW}^{st(H_k^0,\kl)}+\tilde{w}_{WW}^{su(H_k^0,\kl)}\right)+c.c. \\
\tilde{w}_{WW}^{(Z,\kl)}&=\left(\tilde{w}_{WW}^{st(Z,\kl)}+\tilde{w}_{WW}^{su(Z,\kl)}\right)+c.c.
\end{align}
\end{itemize}
\subsection{$\chi_i^0\chi_j^0\rightarrow Z Z$}
This process is the sum of s-channel CP even ($H_1^0$, $H_2^0$) and the t- and u-channel neutralino ( $\knk$, k = 1, 4) exchange.  Resulting cross section is obtained by taking the sum of the s, t, and u channel cross sections and possible cross sections obtained considering interference among themselves. And it is given by:
\begin{align}
\tilde{w}_{ZZ}&=\sum_{k,l=1,2}\tilde{w}_{ZZ}^{(H_k^0,H_l^0)}+\sum_{k,l=1,2}\tilde{w}_{ZZ}^{(\knk,\knl)}+\sum_{l=1,4}\tilde{w}_{ZZ}^{(H_k^0,\knl)}
\end{align}
\begin{itemize}
	\item  \underline{CP-even Higgs boson ($H_k^0-H_l^0$) }:
	\begin{align}
	\tilde w_{ZZ}^{(H_k^0-H_l^0)}&=\frac{C_{H_k^0 Z Z} C_{H_l^0 Z Z}^*}{(S-m_{H_k^0}^2)\,(S-m_{H_l^0}^2)} \frac{1}{8\, M_Z^4}(12\, M_Z^4-4\, M_Z^2 S+S^2)\nn&(f_{\ki \kj}^{(H_k^0-H_l^0)} d-g_{\ki \kj}^{(H_k^0-H_l^0)}\, m_i\, m_j)
	\end{align}
	\item  \underline{Chargino exchange}:
	\begin{align}
	\tilde w_{Z Z}^{t(\knk-\knl)}&=\Gamma_{Z Z}^{1 (\knk-\knl)}\,\mathcal{C}_{12}(\knk,\knl,Z,Z)+\Gamma_{Z Z}^{2 (\knk-\knl)}\,\mathcal{C}_{13}(\knk,\knl,Z,Z)\nn&+\Gamma_{Z Z}^{3 (\knk-\knl)}\,\mathcal{C}_{14}(\knk,\knl,Z,Z)+\Gamma_{Z Z}^{4 (\knk-\knl)}\,\mathcal{C}_{15}(\knk,\knl,Z,Z)\nn
	&+\left(m_{\knl}\,\Gamma_{Z Z}^{5 (\knk-\knl)}+m_{\knk}\,\Gamma_{Z Z}^{6 (\knk-\knl)}\right)\,\mathcal{C}_{16}(\knk,\knl,Z,Z)\nn&+\left(m_{\knl}\,\Gamma_{Z Z}^{7 (\knk-\knl)}+m_{\knk}\,\Gamma_{Z Z}^{8 (\knk-\knl)}\right)\,\mathcal{C}_{17}(\knk,\knl,Z,Z)\\
	\tilde w_{Z Z}^{u(\knk-\knl)}&=\Gamma_{Z Z}^{1 (\knk-\knl)}\,\mathcal{C}_{12}(\knk,\knl,Z,Z)+\Gamma_{Z Z}^{2 (\knk-\knl)}\,\mathcal{C}_{13}(\knk,\knl,Z,Z)\nn&+\Gamma_{Z Z}^{3 (\knk-\knl)}\,\mathcal{C}_{14}(\knk,\knl,Z,Z)+\Gamma_{Z Z}^{4 (\knk-\knl)}\,\mathcal{C}_{15}(\knk,\knl,Z,Z)\nn
	&+\left(m_{\knl}\,\Gamma_{Z Z}^{5 (\knk-\knl)}+m_{\knk}\,\Gamma_{Z Z}^{6 (\knk-\knl)}\right)\,\mathcal{C}_{16}(\knk,\knl,Z,Z)\nn&+\left(m_{\knl}\,\Gamma_{Z Z}^{7 (\knk-\knl)}+m_{\knk}\,\Gamma_{Z Z}^{8 (\knk-\knl)}\right)\,\mathcal{C}_{17}(\knk,\knl,Z,Z)\\
	\tilde w_{Z Z}^{tu(\knk-\knl)}&=I_{Z Z}^{1 (\knk-\knl)}\,J_{14}(\knk,\knl,Z,Z)+I_{Z Z}^{2 (\knk-\knl)}\,J_{15}(\knk,\knl,Z,Z)\nn&+I_{Z Z}^{3 (\knk-\knl)}\,J_{16}(\knk,\knl,Z,Z)+I_{Z Z}^{4 (\knk-\knl)}\,J_{17}(\knk,\knl,Z,Z)\nn
	&+I_{Z Z}^{5 (\knk-\knl)}\,J_{18}(\knk,\knl,Z,Z)+I_{Z Z}^{6 (\knk-\knl)}\,J_{19}(\knk,\knl,Z,Z)\nn&+I_{Z Z}^{7 (\knk-\knl)}\,J_{20}(\knk,\knl,Z,Z)+I_{Z Z}^{8 (\knk-\knl)}\,J_{21}(\knk,\knl,Z,Z)
	\end{align}
		\item  \underline{CP-even Higgs boson ($H_k^0$) interference with neutralino ($\knl$) }:
		\begin{align}
		\tilde w_{Z Z}^{st(H_k^0-\knl)}&=\frac{C_{H_k^0ZZ}}{s-m_{H_k^0}^2} \,\left( J_{ZZ}^{(5)(H_k^0 - \knl)}\,\mathcal{B}_{20}(H_k^0,\knl,Z,Z)+J_{ZZ}^{(6)(H_k^0 - \knl)}\right.\nn&\left.\mathcal{B}_{21}(H_k^0,\knl,Z,Z)+J_{ZZ}^{(7)(H_k^0 - \knl)}\,\mathcal{B}_{22}(H_k^0,\knl,Z,Z)+J_{ZZ}^{(8)(H_k^0 - \knl)}\right.\nn&\left.\mathcal{B}_{23}(H_k^0,\knl,Z,Z)\right)\\	
		\tilde w_{Z Z}^{su(H_k^0-\knl)}&=\frac{C_{H_k^0 Z Z}}{s-m_{H_k^0}^2} \,\left( J_{ZZ}^{(5)(H_k^0 - \knl)}\,\mathcal{B}_{20}(H_k^0,\knl,Z,Z)+J_{ZZ}^{(6)(H_k^0 - \knl)}\right.\nn&\left.\mathcal{B}_{21}(H_k^0,\knl,Z,Z)+J_{ZZ}^{(7)(H_k^0 - \knl)}\,\mathcal{B}_{22}(H_k^0,\knl,Z,Z)+J_{ZZ}^{(8)(H_k^0 - \knl)}\right.\nn&\left.\mathcal{B}_{23}(H_k^0,\knl,Z,W)\right)\\
		\tilde{w}_{ZZ}^{(H_k^0,\knl)}&=\left(\tilde{w}_{ZZ}^{st(H_k^0,\knl)}+\tilde{w}_{ZZ}^{su(H_k^0,\knl)}\right)+c.c. 
		\end{align}		
\end{itemize}			
\subsection{$\chi_i^0\chi_j^0\rightarrow f_m \bar f_n$}
This process is sum of s channel diagrams through CP even Higgs boson, CP odd Higgs boson and Z boson and sfermion t and u channel. This is the only process which involves flavour violatio effects from SUSY soft sector. The resultant cross section is given by:
\begin{align}
	\tilde{w}_{f_m \bar f_n}&=\sum_{k,l=1,2}	\tilde{w}_{f_m \bar f_n}^{(H_k^0,H_l^0)}+\tilde{w}_{f_m \bar f_n}^{(A_1^0,A_1^0)}+	\tilde{w}_{f_m \bar f_n}^{(Z,Z)}+\sum_{k=1,2}\tilde{w}_{f_m \bar f_n}^{(H_k^0,Z)}+\tilde{w}_{f_m \bar f_n}^{(A_1^0,Z)}+\sum_{k,l=1,p}\tilde{w}_{f_m \bar f_n}^{({\tilde f_k}-{\tilde f_l})}\nn&+\sum_{k=1,2}\sum_{l=1,p}\tilde{w}_{f_m \bar f_n}^{(H_k^0-\tilde f_l)}+\sum_{l=1,p}\tilde{w}_{f_m \bar f_n}^{(A_1^0-\tilde f_l)}+\sum_{l=1,p}\tilde{w}_{f_m \bar f_n}^{(Z-\tilde f_l)}
\end{align}
\begin{itemize}
\item  \underline{$H_k^0-H_l^0$, $A_1^0$ and Z boson exchange  exchange}:
\begin{align}
	\tilde{w}_{f_m \bar f_n}^{(H_k^0,H_l^0)}&=2\,\frac{C_{H_k^0f_m\bar f_n}\,C_{H_l^0f_m\bar f_n}^*}{(s-m_{H_k^0}^2)\,(s-m_{H_l^0}^2)}\left(d\, f_{\ki \kj}^{(H_k^0-H_l^0)}- g_{\ki \kj}^{(H_k^0-H_l^0)} m_i\, m_j\right)\nn&\left(d_{f_m\bar{f_n}}-m_{f_m}\,m_{\bar f_n}\right)\\
	\tilde{w}_{f_m \bar f_n}^{(A_1^0,A_1^0)}&=-2\,\frac{C_{A_1^0f_m\bar f_n}\,C_{A_1^0f_m\bar f_n}^*}{(s-m_{A_1^0}^2)\,(s-m_{A_1^0}^2)}\left(d\, f_{\ki \kj}^{(A_1^0-A_1^0)}- g_{\ki \kj}^{(A_1^0-A_1^0)} m_i\, m_j\right)\nn&\left(d_{f_m\bar{f_n}}+m_{f_m}\,m_{\bar f_n}\right)
	\end{align}
	\begin{align}
	\tilde{w}_{f_m \bar f_n}^{(Z,Z)}&=\frac{|C_{Zf_m\bar{f_n}}^V|^2+|C_{Zf_m\bar{f_n}}^A|^2}{12\,M_Z^4\,s^2\,(s-M_Z^2)^2}\,\left(12\,s^2\,(2\, M_Z^2 - s)\, (l_{f_m\bar{f_n}}\,l_{\bar{f_n}f_m} - 2\, d_{f_m\bar{f_n}}\, s)\right.\nn&\left.(d\, f_{\ki \kj}^{(Z-Z)}+ g_{\ki \kj}^{(Z-Z)} m_i\, m_j)+f_{\ki \kj}^{(Z-Z)}\,(2\, b_{f_m\bar{f_n}}^2\, M_Z^4 + 6\,l_i\,l_{ia} \,s^2\, (l_{f_m\bar{f_n}}\,l_{\bar{f_n}f_m}\right.\nn&+\left. 4\, d_{f_m\bar{f_n}}\, M_Z^2 - 2\, d_{f_m\bar{f_n}}\, s) + 
	3\,(\tau_{\bar{f_n}f_m}\,\omega_{f_m\bar{f_n}}+\omega_{\bar{f_n}f_m}\,\tau_{f_m\bar{f_n}})\,  (M_Z^4 - 2\, M_Z^2\, s))\right.\nn&+\left.48\,g_{\ki \kj}^{(Z-Z)}\,s^2\,M_Z^4\,d_{f_m\bar{f_n}}\,m_i\,m_j\right)+\frac{m_{f_m}\,m_{\bar{f_n}}\,(|C_{Zf_m\bar{f_n}}^V|^2-|C_{Zf_m\bar{f_n}}^A|^2)}{M_Z^4\,(s-M_Z^2)^2}\nn&\left((d\, f_{\ki \kj}^{(Z-Z)}+ g_{\ki \kj}^{(Z-Z)} m_i\, m_j)\,(-4\, M_Z^2\, s +2\, s^2)+4\,M_Z^4\,(d\, f_{\ki \kj}^{(Z-Z)}\right.\nn&+\left.2\, g_{\ki \kj}^{(Z-Z)} m_i\, m_j)+f_{\ki \kj}^{(Z-Z)}\,l_i\,l_{ia}\,(2\, M_Z^2 + s)\right)\nn&-\frac{(\tau_{f_m\bar{f_n}}\,\omega_{\bar{f_n}f_m}-\tau_{\bar{f_n}f_m}\,\omega_{f_m\bar{f_n}})\,(s+M_Z^2)}{4\,s^2\,(s-M_Z^2)^2\,M_Z^2}\nn
	&\left( C_{\ki \kj Z}^L C_{\ki \kj Z}^{L*} -C_{\ki \kj Z}^R C_{\ki \kj Z}^{R*}\right)\,\left(C_{Zf_m\bar{f_n}}^{A*}\,C_{Zf_m\bar{f_n}}^V+C_{Zf_m\bar{f_n}}^A\,C_{Zf_m\bar{f_n}}^{V*}\right)
\end{align}
\begin{align}
\tilde{w}_{f_m \bar f_n}^{(H_k^0,Z)}&=\frac{(l_{\bar{f_n}f_m}\,m_{f_m}-l_{f_m\bar{f_n}}\,m_{\bar{f_n}})}{2\,M_Z^2\,s\,(s-m_{H_k^0}^2)}\,C_{Zf_m\bar{f_n}}^{V*}\,\left( f_{\ki \kj}^{(Z-Z)}\,l_i\,m_j- g_{\ki \kj}^{(Z-Z)}\,l_{ia}\, m_i\right)\\
\tilde{w}_{f_m \bar f_n}^{(A_1^0,Z)}&=-\frac{(l_{\bar{f_n}f_m}\,m_{f_m}+l_{f_m\bar{f_n}}\,m_{\bar{f_n}})}{2\,M_Z^2\,s\,(s-m_{H_k^0}^2)}\,C_{Zf_m\bar{f_n}}^{A*}\,\left( f_{\ki \kj}^{(Z-Z)}\,l_i\,m_j- g_{\ki \kj}^{(Z-Z)}\,l_{ia}\, m_i\right)
\end{align}
	\item \underline{ Sfermion ($\tilde f_k-\tilde f_l$) exchange}:
\begin{align}	
	\tilde{w}_{f_m \bar f_n}^{t({\tilde f_k}-\tilde f_l)}&=\frac{1}{8\,b_{f_m\bar{f_n}}\,(R_{f_m\tilde f_k}-R_{f_m\tilde f_l})}\left(2\,b_{f_m\bar{f_n}}\,(R_{f_m\tilde f_k}-R_{f_m\tilde f_l})\right.\nn&\left.f_{\ki f_m}^{(\tilde f_k-\tilde f_l)}\,f_{\kj \bar{f_n}}^{(\tilde f_k-\tilde f_l)}\right)-\frac{\mathcal{X}_1(\tilde f_k,\tilde f_l,f_m,\bar{f_n})}{2\,b_{f_m\bar{f_n}}\,(R_{f_m\tilde f_k}-R_{f_m\tilde f_l})}\left(g_{\ki f_m}^{(\tilde f_k-\tilde f_l)}\, m_i\, m_{f_m}\, s\right.\nn&\left. (f_{\kj \bar{f_n}}^{(\tilde f_k-\tilde f_l)} (\omega_{\bar{f_n}f_m}-\tau_{f_m\bar{f_n}})+ 4\,g_{\kj \bar{f_n}}^{(\tilde f_k-\tilde f_l)}\, m_j\, m_{\bar{f_n}}\, s)\right)\nn&+\frac{\mathcal{Y}_1(\tilde f_k,\tilde f_l,f_m,\bar{f_n})}{8\,b_{f_m\bar{f_n}}\,(R_{f_m\tilde f_k}-R_{f_m\tilde f_l})}\,\left(f_{\ki f_m}^{(\tilde f_k-\tilde f_l)}\,f_{\kj \bar{f_n}}^{(\tilde f_k-\tilde f_l)}\right.\nn&\left. (\tau_{f_m\bar{f_n}}-\omega_{\bar{f_n}f_m})-4\, f_{\kj \bar{f_n}}^{(\tilde f_k-\tilde f_l)}\,g_{\ki f_m}^{(\tilde f_k-\tilde f_l)}\, m_i\, m_{f_m}\, s-4\,f_{\ki f_m}^{(\tilde f_k-\tilde f_l)}\, g_{\kj \bar{f_n}}^{(\tilde f_k-\tilde f_l)} \right.\nn&\left.m_j\, m_{\bar f_n}\, s\right)-\frac{\mathcal{Z}_1(\tilde f_k,\tilde f_l,f_m,\bar{f_n})}{8\,b_{f_m\bar{f_n}}\,(R_{f_m\tilde f_k}-R_{f_m\tilde f_l})}\left(f_{\ki f_m}^{(\tilde f_k-\tilde f_l)}\,f_{\kj \bar{f_n}}^{(\tilde f_k-\tilde f_l)}\right)
	\end{align}
	\begin{align}
\tilde{w}_{f_m \bar f_n}^{u({\tilde f_k}-\tilde f_l)}&=\frac{1}{8\,b_{f_m\bar{f_n}}\,(R_{\bar f_n\tilde f_k}-R_{\bar f_n\tilde f_l})}\left(2\,b_{f_m\bar{f_n}}\,(R_{\bar f_n\tilde f_k}-R_{\bar f_n\tilde f_l})\,f_{\ki \bar f_n}^{(\tilde f_k-\tilde f_l)}\,f_{\kj f_m}^{(\tilde f_k-\tilde f_l)}\right)\nn&+\frac{\mathcal{X}_1(\tilde f_k,\tilde f_l,\bar{f_n},f_m)}{2\,b_{f_m\bar{f_n}}\,(R_{\bar f_n\tilde f_k}-R_{\bar f_n\tilde f_l})}\left(g_{\ki \bar f_n}^{(\tilde f_k-\tilde f_l)}\, m_i\, m_{\bar f_n}\, s\, (f_{\kj f_m}^{(\tilde f_k-\tilde f_l)} (\tau_{\bar{f_n}f_m}-\omega_{f_m\bar{f_n}})\right.\nn&+\left. 4\,g_{\kj f_m}^{(\tilde f_k-\tilde f_l)}\, m_j\, m_{f_m}\, s)\right)+\frac{\mathcal{Y}_1(\tilde f_k,\tilde f_l,\bar{f_n},f_m)}{8\,b_{f_m\bar{f_n}}\,(R_{\bar f_n\tilde f_k}-R_{\bar f_n\tilde f_l})}\,\left(f_{\ki \bar f_n}^{(\tilde f_k-\tilde f_l)}\,f_{\kj f_m}^{(\tilde f_k-\tilde f_l)}\right.\nn&\left. (\omega_{f_m\bar{f_n}}-\tau_{\bar{f_n}f_m})-4\, f_{\kj f_m}^{(\tilde f_k-\tilde f_l)}\,g_{\ki \bar f_n}^{(\tilde f_k-\tilde f_l)}\, m_i\, m_{\bar f_n}\, s+4\,f_{\ki \bar f_n}^{(\tilde f_k-\tilde f_l)}\, g_{\kj f_m}^{(\tilde f_k-\tilde f_l)} \right.\nn&\left.m_j\, m_{f_m}\, s\right)-\frac{\mathcal{Z}_1(\tilde f_k,\tilde f_l,\bar{f_n},f_m)}{8\,b_{f_m\bar{f_n}}\,(R_{\bar f_n\tilde f_k}-R_{\bar f_n\tilde f_l})}\left(f_{\ki \bar f_n}^{(\tilde f_k-\tilde f_l)}\,f_{\kj f_m}^{(\tilde f_k-\tilde f_l)}\right)
\end{align}
\begin{align}
\tilde{w}_{f_m \bar f_n}^{tu({\tilde f_k}-\tilde f_l)}&=-\frac{1}{4}\,J_{f_m\bar f_n}^{(1)(\tilde f_k - \tilde f_l)} +\frac{J_{f_m\bar f_n}^{(1)(\tilde f_k - \tilde f_l)}}{16\,b_{f_m \bar f_n}\,(R_{ f_m\tilde f_k}+R_{\bar f_n\tilde f_l}-2\,s\,l_i)}\,\left(\mathcal{Y}_2(\tilde f_k,\tilde f_l,f_m,\bar{f_n})\right.\nn&\left.(\omega_{\bar f_n f_m}-\omega_{ f_m\bar f_n}-(3\,R_{ f_m\tilde f_k}\,L(\tilde f_k,f_m,\bar f_n)+R_{\bar f_n\tilde f_l}\,L(\tilde f_l,\bar f_n, f_m))\,\tau_{f_m\bar f_n})\right.\nn&-\left.(R_{ f_m\tilde f_k}\,L(\tilde f_k,f_m,\bar f_n)+3\,R_{\bar f_n\tilde f_l}\,L(\tilde f_l,\bar f_n, f_m))\,\tau_{\bar f_nf_m})+2\,(R_{ f_m\tilde f_k}^2\right.\nn&\left.L(\tilde f_k,f_m,\bar f_n)+R_{\bar f_n\tilde f_l}^2\,L(\tilde f_l,\bar f_n, f_m))+4\,l_i\,s^2\,(l_{f_m\bar f_n}\,L(\tilde f_k,f_m,\bar f_n)\right.\nn&+\left.l_{\bar f_nf_m}\,L(\tilde f_l,\bar f_n, f_m))\right)+\frac{s\,(L(\tilde f_k,f_m,\bar f_n)+L(\tilde f_l,\bar f_n, f_m))}{4\,b_{f_m \bar f_n}\,(R_{ f_m\tilde f_k}+R_{\bar f_n\tilde f_l}-2\,s\,l_i)}\,\left(4\,s\,(-d\, d_{f_m \bar f_n}\right.\nn &\left.J_{f_m\bar f_n}^{(1)(\tilde f_k - \tilde f_l)} + d_{f_m \bar f_n}\,J_{f_m\bar f_n}^{(4)(\tilde f_k - \tilde f_l)}\,  m_i\, m_j + d\,J_{f_m\bar f_n}^{(2)(\tilde f_k - \tilde f_l)}\, m_{f_m}\, m_{\bar f_n} +J_{f_m\bar f_n}^{(3)(\tilde f_k - \tilde f_l)}\right.\nn&\left. m_i\,m_j\,m_{f_m}\, m_{\bar f_n} )-\frac{s}{4\,b_{f_m \bar f_n}\,(R_{ f_m\tilde f_k}+R_{\bar f_n\tilde f_l}-2\,s\,l_i)}\,\left(\mathcal{Y}_2(\tilde f_k,\tilde f_l,f_m,\bar{f_n})\right.\right.\nn&-\left.\left.\tau_{f_m\bar f_n}\,L(\tilde f_k,f_m,\bar f_n)+\tau_{\bar f_nf_m}\,L(\tilde f_l,\bar f_n, f_m)\right)\,\left(-J_{f_m\bar f_n}^{(8)(\tilde f_k - \tilde f_l)}\, m_i\, m_{f_m} \right.\right.\nn&+\left.\left.J_{f_m\bar f_n}^{(5)(\tilde f_k - \tilde f_l)}\, m_j\, m_{f_m} + J_{f_m\bar f_n}^{(7)(\tilde f_k - \tilde f_l)}\, m_i\, m_{\bar f_n} -J_{f_m\bar f_n}^{(6)(\tilde f_k - \tilde f_l)}\, m_j\, m_{\bar f_n}\right)\right)
\end{align}
\begin{align}
\tilde{w}_{f_m \bar f_n}^{({\tilde f_k}-\tilde f_l)}&=\tilde{w}_{f_m \bar f_n}^{t({\tilde f_k}-\tilde f_l)}+\tilde{w}_{f_m \bar f_n}^{u({\tilde f_k}-\tilde f_l)}-(\tilde{w}_{f_m \bar f_n}^{tu({\tilde f_k}-\tilde f_l)}+c.c)	
\end{align}
\item \underline{Higgs boson ($H_k^0$, $A_1^0$), Z boson and sfermion ($\tilde f_l$) exchange}:
	\begin{align}
		\tilde{w}_{f_m \bar f_n}^{st(H_k^0-\tilde f_l)}&=-\frac{C_{H_k^0f_m \bar f_n}}{16\,b_{f_m \bar f_n}\,(s-m_{H_k^0}^2)}\,\left(J_{f_m \bar f_n}^{(5)(H_k^0 - \tilde f_l)}\left(4\, b_{f_m \bar f_n}\, s + (\tau_{f_m \bar f_n} +\omega{f_m \bar f_n})\, l_i\right.\right.\nn&\left.\left.L(\tilde f_l,f_m,\bar f_n) - 2\,s\, (4\, d\, (d_{f_m \bar f_n} - m_{f_m}\, m_{\bar f_n}) +R_{f_m\tilde f_l})\,L(\tilde f_l,f_m,\bar f_n)\right)\right.\nn&-\left.8\,s\,m_i\,m_j\,L(\tilde f_l,f_m,\bar f_n)\,J_{f_m \bar f_n}^{(6)(H_k^0 - \tilde f_l)}\,\left(d_{f_m \bar f_n} - m_{f_m}\, m_{\bar f_n}\right)-2\,m_j\right.\nn&\left.J_{f_m \bar f_n}^{(7)(H_k^0 - \tilde f_l)}\,\left(2\, b_{f_m \bar f_n}\, (m_{f_m} + m_{\bar f_n}) - ((m_{f_m} + m_{\bar f_n})\,R_{f_m\tilde f_l} - 2\,s\, l_i\, m_{f_m})\right.\right.\nn&\left.\left.L(\tilde f_l,f_m,\bar f_n)\right)-2\,m_i\,J_{f_m \bar f_n}^{(8)(H_k^0 - \tilde f_l)}\,\left(2\, b_{f_m \bar f_n}\, (m_{f_m} + m_{\bar f_n}) - ((m_{f_m} + m_{\bar f_n})\right.\right.\nn&\left.\left.+(m_{f_m}\,(\tau_{f_m \bar f_n}-\omega_{\bar f_n f_m}) - (m_{f_m} + m_{\bar f_n})\,R_{f_m\tilde f_l}) + 2\,s\,m_{\bar f_n}\,l_{f_m \bar f_n})\right.\right.\nn&\left.\left.L(\tilde f_l,f_m,\bar f_n)\right)\right)
		\end{align}
		\begin{align}
			\tilde{w}_{f_m \bar f_n}^{su(H_k^0-\tilde f_l)}&=-\frac{C_{H_k^0f_m\bar f_n}}{16\,b_{f_m \bar f_n}\,(s-m_{H_k^0}^2)}\,\left(J_{\bar f_nf_m}^{(5)(H_k^0 - \tilde f_l)}\left(4\, b_{f_m \bar f_n}\, s + (\tau_{\bar f_nf_m} +\omega{\bar f_nf_m})\, l_i\right.\right.\nn&\left.\left.L(\tilde f_l,\bar f_n,f_m) - 2\,s\, (4\, d\, (d_{f_m \bar f_n} - m_{f_m}\, m_{\bar f_n}) +R_{\bar f_n\tilde f_l})\,L(\tilde f_l,\bar f_n,f_m)\right)\right.\nn&-\left.8\,s\,m_i\,m_j\,L(\tilde f_l,\bar f_n,f_m)\,J_{\bar f_n,f_m}^{(6)(H_k^0 - \tilde f_l)}\,\left(d_{f_m \bar f_n} - m_{f_m}\, m_{\bar f_n}\right)-2\,m_j\right.\nn&\left.J_{\bar f_nf_m}^{(7)(H_k^0 - \tilde f_l)}\,\left(2\, b_{f_m \bar f_n}\, (m_{f_m} + m_{\bar f_n}) - ((m_{f_m} + m_{\bar f_n})\,R_{\bar{f_n}\tilde f_l} - 2\,s\, l_i\, m_{\bar f_n})\right.\right.\nn&\left.\left.L(\tilde f_l,\bar f_n,f_m)\right)-2\,m_i\,J_{\bar f_nf_m}^{(8)(H_k^0 - \tilde f_l)}\,\left(2\, b_{f_m \bar f_n}\, (m_{f_m} + m_{\bar f_n}) - ((m_{f_m} + m_{\bar f_n})\right.\right.\nn&\left.\left.+(m_{\bar f_n}\,(\tau_{\bar f_n f_m}-\omega_{f_m\bar f_n}) - (m_{f_m} + m_{\bar f_n})\,R_{\bar f_n\tilde f_l}) + 2\,s\,m_{f_m}\,l_{\bar f_n f_m})\right.\right.\nn&\left.\left.L(\tilde f_l,\bar f_n,f_m)\right)\right)
			\end{align}
			\begin{align}
	\tilde{w}_{f_m \bar f_n}^{st(A_1^0-\tilde f_l)}&=\frac{C_{A_1^0f_m \bar f_n}}{8\,b_{f_m \bar f_n}\,(s-m_{A_1^0}^2)}\,\left(-s\,J_{f_m\bar f_n}^{(9)(A_1^0 - \tilde f_l)}\,\left(-2\, b_{f_m\bar f_n} + (-l_{f_m\bar f_n}\, l_i \right.\right.\nn &+\left.\left.  4\, d \,(d_{f_m\bar f_n} + m_{f_m}\, m_{\bar f_n})+R_{f_m\tilde f_l})\, L(\tilde f_l,f_m,\bar f_n)\right)+4\,m_i\,m_j\,s\right.\nn&\left.J_{f_m\bar f_n}^{(10)(A_1^0 - \tilde f_l)}\, L(\tilde f_l,f_m,\bar f_n)\,(d_{f_m\bar f_n} + m_{f_m}\, m_{\bar f_n})-m_j\,J_{f_m\bar f_n}^{(11)(A_1^0 - \tilde f_l)}\right.\nn&\left.\left((m_{f_m} - m_{\bar f_n})\,(2\,b_{f_m\bar f_n}-R_{f_m\tilde f_l} \,L(\tilde f_l,f_m,\bar f_n))+ 2\,l_i\, m_{f_m}\,s\, L(\tilde f_l,f_m,\bar f_n)\right)\right.\nn&-\left.m_i\,J_{f_m\bar f_n}^{(12)(A_1^0 - \tilde f_l)}\,\left(2\, b_{f_m\bar f_n}\, (m_{f_m} - m_{\bar f_n})+ L(\tilde f_l,f_m,\bar f_n)\,(m_{f_m}\right.\right.\nn&\left.\left.(\tau_{f_m\bar f_n}-\omega_{\bar f_n f_m}) -R_{f_m\tilde f_l}\,(m_{f_m} - m_{\bar f_n}) - 2\, l_{f_m\bar f_n} m_{\bar f_n}\,s)\right)\right)
	\end{align}
	\begin{align}
		\tilde{w}_{f_m \bar f_n}^{su(A_1^0-\tilde f_l)}&=\frac{C_{A_1^0f_m \bar f_n}}{8\,b_{f_m \bar f_n}\,(s-m_{A_1^0}^2)}\,\left(-s\,J_{\bar f_nf_m}^{(9)(A_1^0 - \tilde f_l)}\,\left(-2\, b_{f_m\bar f_n} + (-l_{\bar f_nf_m}\, l_i \right.\right.\nn &+\left.\left.  4\, d \,(d_{f_m\bar f_n} + m_{f_m}\, m_{\bar f_n})+R_{\bar f_n\tilde f_l})\, L(\tilde f_l,\bar f_n,f_m)\right)+4\,m_i\,m_j\,s\right.\nn&\left.J_{\bar f_nf_m}^{(10)(A_1^0 - \tilde f_l)}\, L(\tilde f_l,\bar f_n,f_m)\,(d_{f_m\bar f_n} + m_{f_m}\, m_{\bar f_n})-m_j\,J_{\bar f_nf_m}^{(11)(A_1^0 - \tilde f_l)}\right.\nn&\left.\left((m_{\bar f_n} - m_{ f_m})\,(2\,b_{f_m\bar f_n}-R_{\bar f_n\tilde f_l} \,L(\tilde f_l,\bar f_n,f_m))+ 2\,l_i\, m_{\bar f_n}\,s\, L(\tilde f_l,\bar f_n,f_m)\right)\right.\nn&-\left.m_i\,J_{\bar f_nf_m}^{(12)(A_1^0 - \tilde f_l)}\,\left(2\, b_{f_m\bar f_n}\, (m_{\bar f_n} - m_{f_m})+ L(\tilde f_l,\bar f_n,f_m)\,(m_{\bar f_n}\right.\right.\nn&\left.\left.(\tau_{\bar f_n f_m}-\omega_{f_m\bar f_n}) -R_{\bar f_n\tilde f_l}\,(m_{\bar f_m} - m_{f_m}) - 2\,s\, l_{\bar f_nf_m} m_{f_m})\right)\right)
		\end{align}
		\begin{align}
	\tilde{w}_{f_m \bar f_n}^{st(Z-\tilde f_l)}&=\frac{C_{Zf_m\bar f_n}^V\,J_{f_m\bar f_n}^{(5)(Z - \tilde f_l)}-C_{Zf_m\bar f_n}^A\,J_{f_m\bar f_n}^{(9)(Z - \tilde f_l)}}{8\,b_{f_m\bar f_n}\,M_Z^2\,(s-M_Z^2)}\,m_j\, m_{\bar f_n}\,(s-4\,M_Z^2)\left(2\, b_{f_m \bar f_n}\right.\nn&-\left.R_{f_m\tilde f_l}\,L(\tilde f_l,f_m,\bar f_n)\right)+\frac{C_{Zf_m\bar f_n}^V\,J_{f_m\bar f_n}^{(5)(Z - \tilde f_l)}+C_{Zf_m\bar f_n}^A\,J_{f_m\bar f_n}^{(9)(Z - \tilde f_l)}}{8\,b_{f_m\bar f_n}\,M_Z^2\,(s-M_Z^2)}\nn& m_j\, m_{f_m}\,\left((s-2\,M_Z^2)\,((R_{f_m\tilde f_l} - 2\, l_i\,s)\,L(\tilde f_l,f_m,\bar f_n)-2\, b_{f_m \bar f_n})\right.\nn&+\left.2\,s\,\tau_{\bar f_nf_m}\,L(\tilde f_l,f_m,\bar f_n)\right)+\frac{C_{Zf_m\bar f_n}^V\,J_{f_m\bar f_n}^{(6)(Z - \tilde f_l)}-C_{Zf_m\bar f_n}^A\,J_{f_m\bar f_n}^{(10)(Z - \tilde f_l)}}{8\,b_{f_m\bar f_n}\,M_Z^2\,(s-M_Z^2)}\nn& m_i\,m_{\bar f_n}\,\left((s-2\,M_Z^2)\,((R_{f_m\tilde f_l} - 2\, l_{f_m\bar f_n}\,s)\,L(\tilde f_l,f_m,\bar f_n)-2\, b_{f_m \bar f_n})\right.\nn&+\left.2\,s\,\omega_{f_m\bar f_n}\,L(\tilde f_l,f_m,\bar f_n)\right)+\frac{C_{Zf_m\bar f_n}^V\,J_{f_m\bar f_n}^{(6)(Z - \tilde f_l)}+C_{Zf_m\bar f_n}^A\,J_{f_m\bar f_n}^{(10)(Z - \tilde f_l)}}{8\,b_{f_m\bar f_n}\,M_Z^2\,(s-M_Z^2)}\nn&m_i\,m_{f_m}\,(s-4\,M_Z^2)\left((\tau_{f_m\bar f_n} -\omega_{\bar f_nf_m}  -R_{f_m\tilde f_l})\,L(\tilde f_l,f_m,\bar f_n)+2\,b_{f_m\bar f_n}\right)\nn&-\frac{C_{Zf_m\bar f_n}^V\,J_{f_m\bar f_n}^{(7)(Z - \tilde f_l)}-C_{Zf_m\bar f_n}^A\,J_{f_m\bar f_n}^{(11)(Z - \tilde f_l)}}{32\,s\,b_{f_m\bar f_n}\,M_Z^2\,(s-M_Z^2)}\,\left(8\, b_{f_m\bar f_n}\, M_Z^2\right.\nn&\left. (\omega_{\bar f_nf_m} + R_{f_m\tilde f_l})+ 
	2\, b_{f_m\bar f_n}\,s\, (\tau_{\bar f_nf_m}-\tau_{f_m\bar f_n} + \omega_{f_m\bar f_n} - \omega_{\bar f_nf_m})\right.\nn&-\left. (4\,M_Z^2\,R_{f_m\tilde f_l}\,(R_{f_m\tilde f_l}-\tau_{f_m\bar f_n})+ 
	\tau_{f_m\bar f_n}^2\,s - 3\,\tau_{\bar f_nf_m}\,\omega_{f_m\bar f_n}\,s \right.\nn&+\left.s\,\tau_{\bar f_nf_m}\,R_{f_m\tilde f_l} +\tau_{f_m\bar f_n}\,R_{f_m\tilde f_l}\,s - 
	\tau_{f_m\bar f_n}\, (\tau_{\bar f_nf_m} +\omega_{f_m\bar f_n} +R_{f_m\tilde f_l} )\, s\right.\nn&-\left. 16\, d\, d_{f_m\bar f_n}\,s^3 - \omega_{\bar f_nf_m}\,R_{f_m\tilde f_l}\,(s-4\,M_Z^2) + 
	8\, d_{f_m\bar f_n}\,s^2\,l_i\,l_{ia}\right.\nn &+\left. 8\, d\, s^2\, l_{f_m\bar f_n}\,l_{\bar f_nf_m})\,L(\tilde f_l,f_m,\bar f_n)\right)\nn&+\frac{C_{Zf_m\bar f_n}^V\,J_{f_m\bar f_n}^{(7)(Z - \tilde f_l)}+C_{Zf_m\bar f_n}^A\,J_{f_m\bar f_n}^{(11)(Z - \tilde f_l)}}{4\,b_{f_m\bar f_n}\,M_Z^2}\,m_{f_m}\, m_{\bar f_n}\,s\,L(\tilde f_l,f_m,\bar f_n)\nn&\left(l_i\,l_{ia}-2\, d\,(s-2\,M_Z^2)\right)+\frac{C_{Zf_m\bar f_n}^V\,J_{f_m\bar f_n}^{(8)(Z - \tilde f_l)}-C_{Zf_m\bar f_n}^A\,J_{f_m\bar f_n}^{(12)(Z - \tilde f_l)}}{4\,b_{f_m\bar f_n}\,M_Z^2}\nn&\left(l_{f_m\bar f_n}\,l_{\bar f_nf_m}-2\, d_{f_m\bar f_n}\, (s-2\,M_Z^2)\right)\nn&+\frac{C_{Zf_m\bar f_n}^V\,J_{f_m\bar f_n}^{(8)(Z - \tilde f_l)}+C_{Zf_m\bar f_n}^A\,J_{f_m\bar f_n}^{(12)(Z - \tilde f_l)}}{2\,b_{f_m\bar f_n}\,M_Z^2}\,m_i\, m_j\,m_{f_m}\, m_{\bar f_n}\,(s-4\,M_Z^2)\nn&L(\tilde f_l,\bar f_n,f_m)
	\end{align}
	\begin{align}
	\tilde{w}_{f_m \bar f_n}^{su(Z-\tilde f_l)}&=\frac{C_{Zf_m\bar f_n}^V\,J_{\bar f_nf_m}^{(5)(Z - \tilde f_l)}+C_{Zf_m\bar f_n}^A\,J_{\bar f_nf_m}^{(9)(Z - \tilde f_l)}}{8\,b_{f_m\bar f_n}\,M_Z^2\,(s-M_Z^2)}\,m_j\, m_{ f_m}\,(s-4\,M_Z^2)\,\left(2\, b_{f_m \bar f_n}\right.\nn&-\left.R_{\bar f_n\tilde f_l}\,L(\tilde f_l,\bar f_n,f_m)\right)+\frac{C_{Zf_m\bar f_n}^V\,J_{\bar f_nf_m}^{(5)(Z - \tilde f_l)}-C_{Zf_m\bar f_n}^A\,J_{\bar f_nf_m}^{(9)(Z - \tilde f_l)}}{8\,b_{f_m\bar f_n}\,M_Z^2\,(s-M_Z^2)}\nn& m_j\, m_{\bar f_n}\,\left((s-2\,M_Z^2)\,((R_{\bar f_n\tilde f_l} - 2\, l_i\,s)\,L(\tilde f_l,\bar f_n,f_m)-2\, b_{f_m \bar f_n})\right.\nn&+\left.2\,s\,\tau_{f_m\bar f_n}\,L(\tilde f_l,\bar f_n,f_m)\right)+\frac{C_{Zf_m\bar f_n}^V\,J_{\bar f_nf_m}^{(6)(Z - \tilde f_l)}+C_{Zf_m\bar f_n}^A\,J_{\bar f_nf_m}^{(10)(Z - \tilde f_l)}}{8\,b_{f_m\bar f_n}\,M_Z^2\,(s-M_Z^2)}\nn&m_i\,m_{ f_m}\,\left((s-2\,M_Z^2)\,((R_{\bar f_n\tilde f_l} - 2\, l_{\bar f_nf_m}\,s)\,L(\tilde f_l,\bar f_n,f_m)-2\, b_{f_m \bar f_n})\right.\nn&+\left.2\,s\,\omega_{\bar f_nf_m}\,L(\tilde f_l,\bar f_n,f_m)\right)+\frac{C_{Zf_m\bar f_n}^V\,J_{\bar f_nf_m}^{(6)(Z - \tilde f_l)}-C_{Zf_m\bar f_n}^A\,J_{\bar f_nf_m}^{(10)(Z - \tilde f_l)}}{8\,b_{f_m\bar f_n}\,M_Z^2\,(s-M_Z^2)}\nn&m_i\,m_{\bar f_n}\,(s-4\,M_Z^2)\left((\tau_{\bar f_nf_m} -\omega_{f_m\bar f_n}  -R_{\bar f_n\tilde f_l})\,L(\tilde f_l,\bar f_n,f_m)+2\,b_{f_m\bar f_n}\right)\nn&-\frac{C_{Zf_m\bar f_n}^V\,J_{\bar f_nf_m}^{(7)(Z - \tilde f_l)}+C_{Zf_m\bar f_n}^A\,J_{\bar f_nf_m}^{(11)(Z - \tilde f_l)}}{32\,s\,b_{f_m\bar f_n}\,M_Z^2\,(s-M_Z^2)}\,\left(8\, b_{f_m\bar f_n}\, M_Z^2\right.\nn&\left. (\omega_{f_m\bar f_n} + R_{\bar f_n\tilde f_l})+ 
	2\, b_{f_m\bar f_n}\,s\, (\tau_{f_m\bar f_n}-\tau_{\bar f_nf_m} + \omega_{\bar f_nf_m} - \omega_{f_m\bar f_n})\right.\nn&-\left. (4\,M_Z^2\,R_{\bar f_n\tilde f_l}\,(R_{\bar f_n\tilde f_l}-\tau_{\bar f_nf_m})+ 
	\tau_{\bar f_nf_m}^2\,s - 3\,\tau_{f_m\bar f_n}\,\omega_{\bar f_nf_m}\,s \right.\nn&+\left.s\,\tau_{f_m\bar f_n}\,R_{\bar f_n\tilde f_l} +\tau_{\bar f_nf_m}\,R_{\bar f_n\tilde f_l}\,s - 
	\tau_{\bar f_nf_m}\, (\tau_{f_m\bar f_n} +\omega_{\bar f_nf_m} +R_{\bar f_n\tilde f_l} )\, s\right.\nn&-\left. 16\, d\, d_{f_m\bar f_n}\,s^3 - \omega_{f_m\bar f_n}\,R_{\bar f_n\tilde f_l}\,(s-4\,M_Z^2) + 
	8\, d_{f_m\bar f_n}\,s^2\,l_i\,l_{ia}\right.\nn &+\left. 8\, d\, s^2\, l_{f_m\bar f_n}\,l_{\bar f_nf_m})\,L(\tilde f_l,\bar f_n,f_m)\right)\nn&+\frac{C_{Zf_m\bar f_n}^V\,J_{\bar f_nf_m}^{(7)(Z - \tilde f_l)}-C_{Zf_m\bar f_n}^A\,J_{\bar f_nf_m}^{(11)(Z - \tilde f_l)}}{4\,b_{f_m\bar f_n}\,M_Z^2}\,m_{f_m}\, m_{\bar f_n}\,s\,L(\tilde f_l,\bar f_n,f_m)\nn&\left(l_i\,l_{ia}-2\, d\,(s-2\,M_Z^2)\right)+\frac{C_{Zf_m\bar f_n}^V\,J_{\bar f_nf_m}^{(8)(Z - \tilde f_l)}+C_{Zf_m\bar f_n}^A\,J_{\bar f_nf_m}^{(12)(Z - \tilde f_l)}}{4\,b_{f_m\bar f_n}\,M_Z^2}\nn&\left(l_{f_m\bar f_n}\,l_{\bar f_nf_m}-2\, d_{f_m\bar f_n}\, (s-2\,M_Z^2)\right)\nn&+\frac{C_{Zf_m\bar f_n}^V\,J_{\bar f_nf_m}^{(8)(Z - \tilde f_l)}-C_{Zf_m\bar f_n}^A\,J_{\bar f_nf_m}^{(12)(Z - \tilde f_l)}}{2\,b_{f_m\bar f_n}\,M_Z^2}\,m_i\, m_j\,m_{f_m}\, m_{\bar f_n}\,(s-4\,M_Z^2)\nn&L(\tilde f_l,\bar f_n,f_m)\\
	\tilde{w}_{f_m \bar f_n}^{(H_k^0-\tilde f_l)}&=(\tilde{w}_{f_m \bar f_n}^{st(H_k^0-\tilde f_l)}-\tilde{w}_{f_m \bar f_n}^{su(H_k^0-\tilde f_l)})+c.c.\\
	\tilde{w}_{f_m \bar f_n}^{(A_1^0-\tilde f_l)}&=(\tilde{w}_{f_m \bar f_n}^{st(A_1^0-\tilde f_l)}-\tilde{w}_{f_m \bar f_n}^{su(A_1^0-\tilde f_l)})+c.c.\\
	\tilde{w}_{f_m \bar f_n}^{(Z-\tilde f_l)}&=	(\tilde{w}_{f_m \bar f_n}^{s(Z-\tilde f_l)}-	\tilde{w}_{f_m \bar f_n}^{su(Z-\tilde f_l)})+c.c						
	\end{align}
	\end{itemize}
where $\tilde f_k$ is a sfermion mass eigen stat with k=1,..,p; for squarks and charged leptons p=6 andfor s neutrinos p=3
\section{Summary}
We have presented, for the first time the full neutralino neutralino annihilation cross section with flavour violation. And we compared our results with the ones given in Nihei etal. We hope to extend this analysis to neutralino-slepton co-annihilation where we expect large changes and also to Sommerfeld enhancement.
\appendix
\section{Mixing matrices and Coupling constants}
In this appendix , we collect various couplings involving mixing matrices are listed.
Feynman diagrams corresponding to all the annihilation channels
are given in \cite{Rosiek:1989rs}. We closely follow Ref.\cite{Rosiek:1989rs} for notational consistency. For the sake of ease and completeness we present in this section relevant spectrum which is used in calculations in this paper. And for more details we urge readers to refer to \cite{Rosiek:1989rs}
\begin{itemize}
	\item  Neutral Higgs scalars. Two CP even and one CP odd Higgs boson eigen values are given in terms of their mixing matrices.
	\begin{itemize}
		\item[i)] CP even Higgs boson  $H_i^0$ , $i=1,2 $, defined as:
		\begin{center}
			$\sqrt{2} \Re H_i^i = Z_R^{ij} H_j^0 + v_i$ (no sum over $i$)
		\end{center}
		\item[ii)] CP odd Higgs boson  $A_i^0$ , $i=1,2$:
		\begin{center}
			$ \sqrt{2} \Im H_i^i = Z_H^{ij} A_j^0$ 
		\end{center}
		$A_1^0$ is the CP odd Higgs boson and $A_2^0$ is the mass less Goldstone boson which is not a physical degree of freedom.
		\bea
		Z_H =  \left(
		\begin{array}{cc}
			{\sin}{\beta}&-  {\cos}{\beta}\\
			{\cos}{\beta}&  {\sin}{\beta}
		\end{array}
		\right) 
		\hskip 1cm
		Z_R =  \left(
		\begin{array}{cc}
			{\cos}{\alpha}&-  {\sin}{\alpha}\\
			{\sin}{\alpha}&  {\cos}{\alpha}
		\end{array}
		\right)
		\eea
		where:
		\bea
		\tan\beta &=& {v_2\over v_1}\\
		\tan 2\alpha &=& \tan 2\beta\, {m^2_A + M_Z^2 \over m_A^2 - M_Z^2}\\
		m_A^2&=& m_{H_1}^2+m_{H_2}^2+2 \mu^2
		\eea
	\end{itemize}
	\begin{eqnarray}
	A_M^{ij} &=& Z_R^{1i} Z_H^{1j} - Z_R^{2i} Z_H^{2j} \hspace{1cm} A_R^{ij} = Z_R^{1i} Z_R^{1j} - Z_R^{2i} Z_H^{2j}\\
	A_H^{ij} &=& Z_H^{1i} Z_H^{1j} - Z_H^{2i} Z_H^{2j}\hspace{1cm} B^i_R=v_1Z_R^{1i}-v_2Z_R^{2i}\\
	A_P^{ij} &=& Z_R^{1i} Z_H^{2j} + Z_R^{2i} Z_H^{1j}
	\end{eqnarray}
	\item  Charged Higgs scalars: There exist two charged physical Higgs bsons and two charged Goldstone bosons in MSSM. Masses of the physical charged Higgses are given by
	
	\begin{equation}
	M_{H_1^{\pm}}^2 = M_W^{2} + m_{H_1}^2 + m_{H_2}^2 + 2|\mu|^2
	\end{equation}
	
	and the charged Goldstone bosons are mass less. In the unitary gauge
	$H_2^{\pm} (\equiv G^{\pm})$ disappear as they are not physical degrees of freedom. And $Z_H$ is matrix which mixes $H_1^+$ and $H_2^+$ with the initial Higgs fields.
	\end{itemize} 
\begin{align}
C_{A_i^0A_j^0 H_k^0}&=-\frac{ie^2}{4 s_W^2 c_W^2}A^{ij}_R B^k_R\\
C_{H_i^0H_j^0 H_k^0}&=-\frac{ie^2}{4 s_W^2 c_W^2}(A^{ij}_RB^k_R+A^{jk}_RB^i_R+A^{ki}_RB^j_R)\\
C_{H_i^0H_1^+ H_1^-}&=-i\left(\frac{e^2}{4 s_W^2 c_W^2}\,A^{11}_HB^i_R+\frac{e\,M_W}{s_W}A^{i1}_P\right)\\
C_{H_i^0 A^0_j Z}&={e \over 2s_Wc_W}A^{ij}_M\\
C_{H_1^+H_1^- Z}&=ie{c_W^2-s_W^2\over2 s_W c_W}\\
C_{A_1^0W^+H^-}&=-\frac{e}{2s_W}\\
C_{H_i^0W^+H^-}&=-\frac{e}{2s_W}A^{i1}_M\\
C_{H_i^0ZZ}&=\frac{ie^2}{2\,s_W^2\,c_W^2}\,g^{\mu\nu}\,C_R^i\\
C_{H_k^0 ZZ}&={ie^2 \over 2s_W^2c_W^2}g^{\mu\nu} \left[v_1 Z_R^{1k}+v_2 Z_R^{2k}\right]\\
C_{H_k^0 WW}&={ie^2 \over 2s_W^2}g^{\mu\nu} \left[v_1 Z_R^{1k}+v_2 Z_R^{2k}\right]\\
C_{Zf_m\bar f_n}^V&={ie^2 \over 4\,s_W^2c_W^2}\,(1-2\,s_W^2)\\
C_{Zf_m\bar f_n}^A&={ie^2 \over 4\,s_W^2c_W^2}\\
C_{\ki \knl H_m^0}^{L}&= {e \over 2s_Wc_W}  \left[(Z_R^{1m} Z_N^{3i} -Z_R^{2m} Z_N^{4i})(Z_N^{1l} s_W - Z_N^{2l} c_W)\right]
\end{align}
\begin{align}
C_{\ki \knl H_m^0}^{R}&=  {e \over 2s_Wc_W}\left[(Z_R^{1m} Z_N^{3l\star} - Z_R^{2m}
Z_N^{4l\star}) (Z_N^{1i\star} s_W - Z_N^{2i\star} c_W) \right]\\
C_{\knl \kj  A_n^0}^{L}&={ie \over 2s_Wc_W} \left[(Z_H^{1n} Z_N^{3l} -
Z_H^{2n} Z_N^{4l}) (Z_N^{1j} s_W - Z_N^{2j} c_W) \right]\\
C_{\knl \kj  A_n^0}^{R}&={-ie \over 2s_Wc_W} \left[ (Z_H^{1n} Z_N^{3j\star} - Z_H^{2n}
Z_N^{4j\star}) (Z_N^{1l\star} s_W - Z_N^{2l\star} c_W) \right]
\end{align}
\begin{align}
C_{\ki \kj Z}^L&={e \over 4 s_W c_W}(Z_N^{4j\star} Z_N^{4i} - Z_N^{3j\star} Z_N^{3i}) P_L\\
C_{\ki \kj Z}^R&={-e \over 4 s_W c_W}(Z_N^{4j}Z_N^{4i\star} - Z_N^{3j} Z_N^{3i\star}) P_R\\
C_{\ki \knl Z}^{L}&= {ie \over 2s_Wc_W} \left[(Z_N^{4i*}Z_N^{4l}-Z_N^{3i*}Z_N^{3l})\right]\\
C_{\ki \knl Z}^{R}&= {ie \over 2s_Wc_W} \left[(Z_N^{3i}Z_N^{3l*}-Z_N^{4i}Z_N^{4l*})\right]\\
C_{\ki \chi^j H^+_k}^{L}&={ie \over s_W c_W}\left[Z_H^{1k}\left(\frac{1}{\sqrt 2}Z_-^{2j}(Z_N^{1i}s_W+Z_N^{2i}c_W)-Z_-^{1j}Z_N^{3i}c_W\right)\right]\\
C_{\ki \chi^j H^+_k}^{R}&=-{ie \over s_W c_W}\left[Z_H^{2k}\left(\frac{1}{\sqrt 2}Z_+^{2j*}(Z_N^{1i*}s_W+Z_N^{2i*}c_W)+Z_+^{1j*}Z_N^{4i*}c_W\right)\right]\\
C_{\ki \kl W}^{L}&={ie \over s_W}\left[(Z_N^{2l}Z_+^{1i*}-\frac{1}{\sqrt{2}}Z_N^{4l}Z_+^{2i*})\right]\\
C_{\ki \kl W}^{R}&={ie \over s_W}\left[(Z_N^{2l*}Z_-^{1i}+\frac{1}{\sqrt{2}}Z_N^{4l*}Z_-^{2i})\right]
\end{align}

\begin{eqnarray}
f_{\ki \kj}^{A-B}&=& C_{\ki \kj A}^L C_{\ki \kj B}^{L*} +C_{\ki \kj A}^R C_{\ki \kj B}^{R*} \nn
g_{\ki \kj}^{A_B}&=& C_{\ki \kj A}^L C_{\ki \kj B}^{R*} +C_{\ki \kj A}^R C_{\ki \kj B}^{L*} \nn
\Gamma_{\phi_1 \phi_2}^{1 (A-B)}&=& C_{\ki A\, \phi_1}^{L} C_{\ki B\, \phi_1}^{L*} C_{\kj A\, \phi_2}^{L}  C_{\kj B\, \phi_2}^{L*} +C_{\ki A \phi_1}^{R} C_{\ki B \phi_1}^{R*}    C_{\kj A\, \phi_2}^{R}  C_{\kj B\, \phi_2}^{R*}\nn
\Gamma_{\phi_1 \phi_2}^{2 (A-B)}&=& C_{\ki A\, \phi_1}^{L} C_{\ki B\, \phi_1}^{L*} C_{\kj A\, \phi_2}^{R}  C_{\kj B\, \phi_2}^{R*} +C_{\ki A \phi_1}^{R} C_{\ki B \phi_1}^{R*}    C_{\kj A\, \phi_2}^{L}  C_{\kj B\, \phi_2}^{L*}\nn
\Gamma_{\phi_1 \phi_2}^{3 (A-B)}&=& C_{\ki A\, \phi_1}^{L} C_{\ki B\, \phi_1}^{R*} C_{\kj A\, \phi_2}^{L}  C_{\kj B\, \phi_2}^{R*} +C_{\ki A \phi_1}^{R} C_{\ki B \phi_1}^{L*}    C_{\kj A\, \phi_2}^{R}  C_{\kj B\, \phi_2}^{L*}\nn
\Gamma_{\phi_1 \phi_2}^{4 (A-B)}&=& C_{\ki A\, \phi_1}^{L} C_{\ki B\, \phi_1}^{R*} C_{\kj A\, \phi_2}^{R}  C_{\kj B\, \phi_2}^{L*} +C_{\ki A \phi_1}^{R} C_{\ki B \phi_1}^{L*}    C_{\kj A\, \phi_2}^{L}  C_{\kj B\, \phi_2}^{R*}\nonumber
\end{eqnarray}
\begin{eqnarray}
\Gamma_{\phi_1 \phi_2}^{5 (A-B)}&=& C_{\ki A\, \phi_1}^{L} C_{\ki B\, \phi_1}^{L*} C_{\kj A\, \phi_2}^{L}  C_{\kj B\, \phi_2}^{R*} +C_{\ki A \phi_1}^{R} C_{\ki B \phi_1}^{R*}    C_{\kj A\, \phi_2}^{R}  C_{\kj B\, \phi_2}^{L*}\nn
\Gamma_{\phi_1 \phi_2}^{6 (A-B)}&=& C_{\ki A\, \phi_1}^{L} C_{\ki B\, \phi_1}^{L*} C_{\kj A\, \phi_2}^{R}  C_{\kj B\, \phi_2}^{L*} +C_{\ki A \phi_1}^{R} C_{\ki B \phi_1}^{R*}    C_{\kj A\, \phi_2}^{L}  C_{\kj B\, \phi_2}^{R*}\nn
\Gamma_{\phi_1 \phi_2}^{7 (A-B)}&=& C_{\ki A\, \phi_1}^{L} C_{\ki B\, \phi_1}^{R*} C_{\kj A\, \phi_2}^{L}  C_{\kj B\, \phi_2}^{L*} +C_{\ki A \phi_1}^{R} C_{\ki B \phi_1}^{L*}    C_{\kj A\, \phi_2}^{R}  C_{\kj B\, \phi_2}^{R*}\nn
\Gamma_{\phi_1 \phi_2}^{8 (A-B)}&=& C_{\ki A\, \phi_1}^{L} C_{\ki B\, \phi_1}^{R*} C_{\kj A\, \phi_2}^{R}  C_{\kj B\, \phi_2}^{R*} +C_{\ki A \phi_1}^{R} C_{\ki B \phi_1}^{L*}    C_{\kj A\, \phi_2}^{L}  C_{\kj B\, \phi_2}^{L*}\nonumber
\end{eqnarray}
\begin{eqnarray}
I_{\phi_1 \phi_2}^{1 (A-B)}&=& C_{\ki A\, \phi_1}^{L} C_{\ki B\, \phi_2}^{L*} C_{\kj A\, \phi_2}^{L}  C_{\kj B\, \phi_1}^{L*} +C_{\ki A \phi_1}^{R} C_{\ki B \phi_2}^{R*}    C_{\kj A\, \phi_2}^{R}  C_{\kj B\, \phi_1}^{R*}\nn
I_{\phi_1 \phi_2}^{2 (A-B)}&=& C_{\ki A\, \phi_1}^{L} C_{\ki B\, \phi_2}^{L*} C_{\kj A\, \phi_2}^{R}  C_{\kj B\, \phi_1}^{R*} +C_{\ki A \phi_1}^{R} C_{\ki B \phi_2}^{R*}    C_{\kj A\, \phi_2}^{L}  C_{\kj B\, \phi_1}^{L*}\nn
I_{\phi_1 \phi_2}^{3 (A-B)}&=& C_{\ki A\, \phi_1}^{L} C_{\ki B\, \phi_2}^{R*} C_{\kj A\, \phi_2}^{L}  C_{\kj B\, \phi_1}^{R*} +C_{\ki A \phi_1}^{R} C_{\ki B \phi_2}^{L*}    C_{\kj A\, \phi_2}^{R}  C_{\kj B\, \phi_1}^{L*}\nonumber
\end{eqnarray}
\begin{eqnarray}
I_{\phi_1 \phi_2}^{4 (A-B)}&=& C_{\ki A\, \phi_1}^{L} C_{\ki B\, \phi_2}^{R*} C_{\kj A\, \phi_2}^{R}  C_{\kj B\, \phi_1}^{L*} +C_{\ki A \phi_1}^{R} C_{\ki B \phi_2}^{L*}    C_{\kj A\, \phi_2}^{L}  C_{\kj B\, \phi_1}^{R*}\nn
I_{\phi_1 \phi_2}^{5 (A-B)}&=& C_{\ki A\, \phi_1}^{L} C_{\ki B\, \phi_2}^{L*} C_{\kj A\, \phi_2}^{L}  C_{\kj B\, \phi_1}^{R*} +C_{\ki A \phi_1}^{R} C_{\ki B \phi_2}^{R*}    C_{\kj A\, \phi_2}^{R}  C_{\kj B\, \phi_1}^{L*}\nn
I_{\phi_1 \phi_2}^{6 (A-B)}&=& C_{\ki A\, \phi_1}^{L} C_{\ki B\, \phi_2}^{L*} C_{\kj A\, \phi_2}^{R}  C_{\kj B\, \phi_1}^{L*} +C_{\ki A \phi_1}^{R} C_{\ki B \phi_2}^{R*}    C_{\kj A\, \phi_2}^{L}  C_{\kj B\, \phi_1}^{R*}\nn
I_{\phi_1 \phi_2}^{7 (A-B)}&=& C_{\ki A\, \phi_1}^{L} C_{\ki B\, \phi_2}^{R*} C_{\kj A\, \phi_2}^{L}  C_{\kj B\, \phi_1}^{L*} +C_{\ki A \phi_1}^{R} C_{\ki B \phi_2}^{L*}    C_{\kj A\, \phi_2}^{R}  C_{\kj B\, \phi_1}^{R*}\nn
I_{\phi_1 \phi_2}^{8 (A-B)}&=& C_{\ki A\, \phi_1}^{L} C_{\ki B\, \phi_2}^{R*} C_{\kj A\, \phi_2}^{R}  C_{\kj B\, \phi_1}^{R*} +C_{\ki A \phi_1}^{R} C_{\ki B \phi_2}^{L*}    C_{\kj A\, \phi_2}^{L}  C_{\kj B\, \phi_1}^{L*}\nn
J_{\phi_1 \phi_2}^{(1)(\phi - \chi)}&=& C_{\ki \chi \phi_1}^{L*} C_{\kj \chi \phi_2}^{L*} C_{\ki \kj \phi}^{R} + C_{\ki \chi \phi_1}^{L*} C_{\kj \chi \phi_2}^{R*} C_{\ki \kj \phi}^{R} \nn
J_{\phi_1 \phi_2}^{(2)(\phi - \chi)}&=& C_{\ki \chi \phi_1}^{R*} C_{\kj \chi \phi_2}^{R*} C_{\ki \kj \phi}^{R} + C_{\ki \chi \phi_1}^{R*} C_{\kj \chi \phi_2}^{L*} C_{\ki \kj \phi}^{R} \nn
J_{\phi_1 \phi_2}^{(3)(\phi - \chi)}&=& C_{\ki \chi \phi_1}^{L*} C_{\kj \chi \phi_2}^{L*} C_{\ki \kj \phi}^{L} + C_{\ki \chi \phi_1}^{L*} C_{\kj \chi \phi_2}^{R*} C_{\ki \kj \phi}^{L} \nn
J_{\phi_1 \phi_2}^{(4)(\phi - \chi)}&=& C_{\ki \chi \phi_1}^{R*} C_{\kj \chi \phi_2}^{R*} C_{\ki \kj \phi}^{R} + C_{\ki \chi \phi_1}^{R*} C_{\kj \chi \phi_2}^{L*} C_{\ki \kj \phi}^{L} \nn
J_{\phi_1 \phi_2}^{(5)(\phi - \chi)}&=& C_{\ki \chi \phi_1}^{L*} C_{\kj \chi \phi_2}^{L*} C_{\ki \kj \phi}^{L} + C_{\ki \chi \phi_1}^{R*} C_{\kj \chi \phi_2}^{R*} C_{\ki \kj \phi}^{R} \nn
J_{\phi_1 \phi_2}^{(6)(\phi - \chi)}&=& C_{\ki \chi \phi_1}^{L*} C_{\kj \chi \phi_2}^{L*} C_{\ki \kj \phi}^{R} + C_{\ki \chi \phi_1}^{R*} C_{\kj \chi \phi_2}^{R*} C_{\ki \kj \phi}^{L} \nn
J_{\phi_1 \phi_2}^{(7)(\phi - \chi)}&=& C_{\ki \chi \phi_1}^{L*} C_{\kj \chi \phi_2}^{R*} C_{\ki \kj \phi}^{L} + C_{\ki \chi \phi_1}^{R*} C_{\kj \chi \phi_2}^{L*} C_{\ki \kj \phi}^{R} \nn
J_{\phi_1 \phi_2}^{(8)(\phi - \chi)}&=& C_{\ki \chi \phi_1}^{L*} C_{\kj \chi \phi_2}^{R*} C_{\ki \kj \phi}^{R} + C_{\ki \chi \phi_1}^{R*} C_{\kj \chi \phi_2}^{L*} C_{\ki \kj \phi}^{L} \nn
J_{\phi_1 \phi_2}^{(9)(\phi - \chi)}&=& C_{\ki \chi \phi_1}^{L*} C_{\kj \chi \phi_2}^{L*} C_{\ki \kj \phi}^{L}- C_{\ki \chi \phi_1}^{R*} C_{\kj \chi \phi_2}^{R*} C_{\ki \kj \phi}^{R} \nn
J_{\phi_1 \phi_2}^{(10)(\phi - \chi)}&=& C_{\ki \chi \phi_1}^{L*} C_{\kj \chi \phi_2}^{L*} C_{\ki \kj \phi}^{R} - C_{\ki \chi \phi_1}^{R*} C_{\kj \chi \phi_2}^{R*} C_{\ki \kj \phi}^{L} \nn
J_{\phi_1 \phi_2}^{(11)(\phi - \chi)}&=& C_{\ki \chi \phi_1}^{L*} C_{\kj \chi \phi_2}^{R*} C_{\ki \kj \phi}^{L} - C_{\ki \chi \phi_1}^{R*} C_{\kj \chi \phi_2}^{L*} C_{\ki \kj \phi}^{R} \nn
J_{\phi_1 \phi_2}^{(12)(\phi - \chi)}&=& C_{\ki \chi \phi_1}^{L*} C_{\kj \chi \phi_2}^{R*} C_{\ki \kj \phi}^{R} - C_{\ki \chi \phi_1}^{R*} C_{\kj \chi \phi_2}^{L*} C_{\ki \kj \phi}^{L} \nonumber
\end{eqnarray}

\section{Auxilary functions}
\label{aux}
In this appendix, we collected various kinetic functions and auxilary functions used in the main text.
\begin{eqnarray}
l_i&=& m_i^2 - m_j^2 + s \nn
l_{ia}&=& m_j^2 - m_i^2 + s\nn
l_{\phi_1\phi_2}&=&m_1^2-m_2^2+s\nn
\tau_{\phi_1\phi_2}&=& l_i\, (m_1^2 - m_2^2 + s)\nn
\omega_{\phi_1\phi_2}&=& l_{ia}\, (m_1^2 - m_2^2 + s)\nn
R_{\phi_1\phi_2}&=& 2 (m_1^2 - m_{2}^2 + m_i^2) s\nn
b_{\phi_1 \phi_2} &=& \sqrt{s^2 + m_i^4 + m_j^4 - 2\,s\,(m_i^2 + m_j^2) - 2\,m_i^2\,m_j^2}\nn
&& \,\sqrt{s^2 + m_1^4 + m_2^4 - 2\,s\,(m_1^2 + m_2^2) - 2\,m_1^2\,m_2^2}\nn
d&=&\frac{1}{2}\left(s-(m_{i}^2+m_j^2)\right)\nn
d_{\phi_1\phi_2}&=&\frac{1}{2}\left(s-(m_{1}^2+m_2^2)\right)\nonumber
\end{eqnarray}
\begin{eqnarray}
L(\chi,\phi_1,\phi_2)&=&\log\frac{R_{\phi_1\chi}+b_{\phi_1\phi_2}-\tau_{\phi_1\phi_2}}{R_{\phi_1\chi}-b_{\phi_1\phi_2}-\tau_{\phi_1\phi_2}}\nn
\mathcal{A}(\chi,\phi_1,\phi_2)&=&\arctanh {\frac{b_{\phi_1\phi_2}}{R_{\phi_1\chi}-\tau_{\phi_1\phi_2}}}\nn
\mathcal{B}(\chi,\phi_1,\phi_2)&=&R_{\phi_1\chi}-\tau_{\phi_1\phi_2}-\omega_{\phi_1\phi_2}\nn
\mathcal{C}(\chi,\phi_1,\phi_2)&=&R_{\phi_1\chi}-\tau_{\phi_1\phi_2}+\omega_{\phi_2\phi_1}\nn
\rho(\chi,\phi_1,\phi_2)&=&R_{\phi_1\chi}-\omega_{\phi_1\phi_2}\nn
\mathcal{X}(\phi_1,\phi_2,\chi_1)&=&-8\,s^2\,(2\,d\,d_{\phi_1\phi_2}-(2\,d+l_{ia}\,m_i^2)+2\,R_{\phi_1\chi_1}^2-R_{\phi_1\chi_1}\,(3\,\tau_{\phi_1\phi_2}+\tau_{\phi_2\phi_1}\nn
&+&\omega_{\phi_1\phi_2}-\omega_{\phi_2\phi_1})+\tau_{\phi_1\phi_2}\,(\tau_{\phi_1\phi_2}+\tau_{\phi_2\phi_1}+\omega_{\phi_1\phi_2}+\omega_{\phi_2\phi_1})\nn
\mathcal{Q}_1(\chi_1,\chi_2,\phi_1,\phi_2)&=&R_{\phi_1\chi_1}^4\,L(\chi_1,\phi_1,\phi_2)-R_{\phi_1\chi_2}^4\,L(\chi_2,\phi_1,\phi_2)\nn
\mathcal{R}_1(\chi_1,\chi_2,\phi_1,\phi_2)&=&R_{\phi_1\chi_1}^3\,L(\chi_1,\phi_1,\phi_2)-R_{\phi_1\chi_2}^3\,L(\chi_2,\phi_1,\phi_2)\nn
\mathcal{Z}_1(\chi_1,\chi_2,\phi_1,\phi_2)&=&R_{\phi_1\chi_1}^2\,L(\chi_1,\phi_1,\phi_2)-R_{\phi_1\chi_2}^2\,L(\chi_2,\phi_1,\phi_2)\nn
\mathcal{Y}_1(\chi_1,\chi_2,\phi_1,\phi_2)&=&R_{\phi_1\chi_1}\,L(\chi_1,\phi_1,\phi_2)-R_{\phi_1\chi_2}\,L(\chi_2,\phi_1,\phi_2)\nn
\mathcal{X}_1(\chi_1,\chi_2,\phi_1,\phi_2)&=&L(\chi_1,\phi_1,\phi_2)-L(\chi_2,\phi_1,\phi_2)\nn
\mathcal{Q}_2(\chi_1,\chi_2,\phi_1,\phi_2)&=&R_{\phi_1\chi_1}^4\,L(\chi_1,\phi_1,\phi_2)-R_{\phi_2\chi_2}^4\,L(\chi_2,\phi_2,\phi_1)\nn
\mathcal{R}_2(\chi_1,\chi_2,\phi_1,\phi_2)&=&R_{\phi_1\chi_1}^3\,L(\chi_1,\phi_1,\phi_2)-R_{\phi_2\chi_2}^3\,L(\chi_2,\phi_2,\phi_1)\nn
\mathcal{Z}_2(\chi_1,\chi_2,\phi_1,\phi_2)&=&R_{\phi_1\chi_1}^2\,L(\chi_1,\phi_1,\phi_2)-R_{\phi_2\chi_2}^2\,L(\chi_2,\phi_2,\phi_1)\nn
\mathcal{Y}_2(\chi_1,\chi_2,\phi_1,\phi_2)&=&R_{\phi_1\chi_1}\,L(\chi_1,\phi_1,\phi_2)-R_{\phi_2\chi_2}\,L(\chi_2,\phi_2,\phi_1)\nn
\mathcal{X}_2(\chi_1,\chi_2,\phi_1,\phi_2)&=&L(\chi_1,\phi_1,\phi_2)-L(\chi_2,\phi_2,\phi_1)\nonumber
\end{eqnarray}
\begin{eqnarray}
\beta_1(\phi_1,\phi_2,\chi_1,\chi_2)&=&-4\, d\,(m_1^2-m_i^2)-\frac{1}{4\,s}\,(s + 8\, m_i^2\,\omega_{\phi_1\phi_2})-\frac{1}{4\,b_{\phi_1 \phi_2}\,(R_{\phi_1\chi_1}-R_{\phi_1\chi_2})}\nn
&&\left[R_{\phi_1\chi_2}\,\mathcal{B}(\chi_2,\phi_1,\phi_2)\,\mathcal{A}(\chi_2,\phi_1,\phi_2)-\{\chi_2\rightarrow\chi_1\}\right]\nn
\beta_2(\phi_1,\phi_2,\chi_1,\chi_2)&=&4\,(m_1^2+m_i^2)+\frac{4\,s}{b_{\phi_1 \phi_2}\,(R_{\phi_1\chi_1}-R_{\phi_1\chi_2})}\nn
&&\left[R_{\phi_1\chi_2}\,\mathcal{B}(\chi_1,\phi_1,\phi_2)\,\mathcal{A}(\chi_1,\phi_1,\phi_2)-\{\chi_1\rightarrow\chi_2\}\right]\nn
\beta_3(\phi_1,\phi_2,\chi_1,\chi_2)&=&4\,m_i^2+\frac{s}{2\,b_{\phi_1 \phi_2}\,(R_{\phi_1\chi_1}-R_{\phi_1\chi_2})}\nn
&&\left[R_{\phi_1\chi_2}\,\mathcal{B}(\chi_1,\phi_1,\phi_2)\,\mathcal{A}(\chi_1,\phi_1,\phi_2)-\{\chi_1\rightarrow\chi_2\}\right]\nonumber
\end{eqnarray}
\begin{eqnarray}
J_1(\phi_1,\phi_2,\chi_1,\chi_2)&=&\frac{L(\chi_1,\phi_1,\phi_2)+L(\chi_2,\phi_2,\phi_1)}{b_{\phi_1\phi_2}\,(R_{\phi_1\chi_1}+R_{\phi_2\chi_2}-2 s)}\nn
J_2(\phi_1,\phi_2,\chi_1,\chi_2)&=&-\frac{1}{4}+\frac{\mathcal{X}(\phi_1,\phi_2,\chi_1)\,L(\chi_1,\phi_1,\phi_2)+\mathcal{X}(\phi_2,\phi_1,\chi_2)\,L(\chi_2,\phi_2,\phi_1)}{16 b_{\phi_1\phi_2}\,(R_{\phi_1\chi_1}+R_{\phi_2\chi_2}-2 s)}\nn
J_3(\phi_1,\phi_2,\chi_1,\chi_2)&=&\frac{1}{b_{\phi_1\phi_2}\,(R_{\phi_1\chi_1}+R_{\phi_2\chi_2}-2 s)}(R_{\phi_2\chi_2}-4\,m_i^2\,s)\,L(\chi_2,\phi_2,\phi_1)\nn
&&-(R_{\phi_1\chi_1}+4\,m_i^2\, s-2\,s\,l_i\,)L(\chi_1,\phi_1,\phi_2)\nn
J_4(\phi_1,\phi_2,\chi_1,\chi_2)&=&\frac{1}{b_{\phi_1\phi_2}\,(R_{\phi_1\chi_1}+R_{\phi_2\chi_2}-2 s)}\left((\mathcal{C}(\chi_1,\phi_1,\phi_2) - 4\,d\, s )\,L(\chi_1,\phi_1,\phi_2)-\right.\nn
&&\left.2\,( 4\,d\, s +\mathcal{B}(\chi_2,\phi_2,\phi_1))\,\mathcal{A}(\chi_2,\phi_2,\phi_1)\right)\nonumber
\end{eqnarray}
\begin{eqnarray}
J_5(\phi_1,\phi_2,\chi_1,\chi_2)&=&\frac{1}{b_{\phi_1\phi_2}\,(R_{\phi_1\chi_1}+R_{\phi_2\chi_2}-2 s)}\left((4\,d\, s +\mathcal{B}(\chi_1,\phi_1,\phi_2))\,L(\chi_1,\phi_1,\phi_2)\right.\nn
&&\left.-2\,(\mathcal{C}(\chi_2,\phi_2,\phi_1) - 4\,d\, s)\,\mathcal{A}(\chi_2,\phi_2,\phi_1)\right)\nn
J_6(\phi_1,\phi_2,\chi_1,\chi_2)&=&\frac{m_j\,m_{\chi_2}}{8\,b_{\phi_1\phi_2}\,m_1^2\,(\ruaa-\rubb)}\left(2\,\ba (\ruaa - \rubb) - 8\, m_i^2\,m_1^2\, s^2\right.\nn&&\left.\mathcal{X}_2(\chi_1,\chi_2,\phi_1,\phi_2) + 6\, m_1^2\, s\, \mathcal{Y}_2(\chi_1,\chi_2,\phi_1,\phi_2) - \mathcal{Z}_2(\chi_1,\chi_2,\phi_1,\phi_2)\right)\nn
J_7(\phi_1,\phi_2,\chi_1,\chi_2)&=&\frac{3\,m_j\,m_{\chi_1}\,s}{4\,b_{\phi_1\phi_2}\,(\ruaa-\rubb)}\left(-2\, l_i\, s  + 4\, m_i^2\, s)\,\mathcal{X}_2(\chi_1,\chi_2,\phi_1,\phi_2) \right.\nn&+&\left.\mathcal{Y}_2(\chi_1,\chi_2,\phi_1,\phi_2)\right)\nn
J_8(\phi_1,\phi_2,\chi_1,\chi_2)&=&-\frac{3\,m_i\,m_{\chi_2}\,s}{4\,b_{\phi_1\phi_2}\,(\ruaa-\rubb)}\left((4\, d\, s- 2\, l_{\phi_1\phi_2}\, s)\,\mathcal{X}_2(\chi_1,\chi_2,\phi_1,\phi_2) \right.\nn&+&\left.\mathcal{Y}_2(\chi_1,\chi_2,\phi_1,\phi_2)\right)\nn
J_9(\phi_1,\phi_2,\chi_1,\chi_2)&=&\frac{m_i\,m_{\chi_1}}{4\,m_1^2}+\frac{m_i\,m_{\chi_1}\,\mathcal{X}_2(\chi_1,\chi_2,\phi_1,\phi_2)}{8\,\ba\,m_1^2\,(\ruaa-\rubb)}\,\left(2\,m_1^2\, s(\taua-\taub)\right.\nn&-&\left. 8\,d_{\phi_1\phi_2}\, (l_{\phi_1\phi_2}- m_1^2)\, s^2\right)+\frac{m_i\,m_{\chi_1}\,\mathcal{Y}_2(\chi_1,\chi_2,\phi_1,\phi_2)}{8\,\ba\,m_1^2\,(\ruaa-\rubb)}\nn
&&\left(4\, d_{\phi_1\phi_2}\, s + 2\, l_{\phi_1\phi_2}\,s - 2\,m_1^2\, s\right)-\frac{m_i\,m_{\chi_1}\,\mathcal{Z}_2(\chi_1,\chi_2,\phi_1,\phi_2)}{8\,\ba\,m_1^2\,(\ruaa-\rubb)}\nonumber
\end{eqnarray}
\begin{eqnarray}
J_{10}(\phi_1,\phi_2,\chi_1,\chi_2)&=&\frac{1}{2}+\frac{\mathcal{X}_2(\chi_1,\chi_2,\phi_1,\phi_2)}{4\,\ba\,m_1^2\,(\ruaa-\rubb)}\,\left(-\ba + s^2\,(4\, \taua + 3\,\omegaa \right.\nn&-&\left.\omegab)\, m_i^2\,m_1^2 + (4\, d_{\phi_1\phi_2}\,l_{\phi_1\phi_2}\, m_i^2 + (-3\,\taua + 
4\, d\, (d_{\phi_1\phi_2} + 2\, l_i \right.\nn&-&\left. 3\, m_i^2))\, m_1^2)\, s\right)-\frac{\mathcal{Y}_2(\chi_1,\chi_2,\phi_1,\phi_2)}{32\,\ba\,m_1^2\,(\ruaa-\rubb)\,s}\,\left(2\,m_1^2\,s\,(-7\,\taua \right.\nn&-&\left. 3\, (\taub + \omegaa) + \omegab)\right.\nn&+&\left.16\, (d + m_i^2)\, (d_{\phi_1\phi_2} + 2\, m_1^2)\, s^2\right)+\frac{\mathcal{Z}_2(\chi_1,\chi_2,\phi_1,\phi_2)\,\left(4\, s - 8\,m_1^2\, s\right)}{32\,\ba\,m_1^2\,(\ruaa-\rubb)\,s}\nn
J_{11}(\phi_1,\phi_2,\chi_1,\chi_2)&=&\frac{m_{\chi_1}\,m_{\chi_2}}{4\,m_1^2}+\frac{d\,s^2\,m_{\chi_1}\,m_{\chi_2}\,\mathcal{X}_2(\chi_1,\chi_2,\phi_1,\phi_2)}{\ba\,(\ruaa-\rubb)}\nn&+&\frac{l_{\phi_1\phi_2}\,s\,m_{\chi_1}\,m_{\chi_2}\,\mathcal{Y}_2(\chi_1,\chi_2,\phi_1,\phi_2)}{4\,\ba\,m_1^2\,(\ruaa-\rubb)}-\frac{m_{\chi_1}\,m_{\chi_2}\,\mathcal{Z}_2(\chi_1,\chi_2,\phi_1,\phi_2)}{8\,\ba\,m_1^2\,(\ruaa-\rubb)}\nonumber
\end{eqnarray}
\begin{eqnarray}
J_{12}(\phi_1,\phi_2,\chi_1,\chi_2)&=&\frac{m_i\,m_j}{4\,m_1^2}+\frac{s^2\,m_i\,m_j\,\mathcal{X}_2(\chi_1,\chi_2,\phi_1,\phi_2)}{2\,\ba\,(\ruaa-\rubb)}\,\left(-6\, d_{\phi_1\phi_2} + l_i - 2\, m_i^2\right)\nn&+&\frac{s\,m_i\,m_j\,\mathcal{Y}_2(\chi_1,\chi_2,\phi_1,\phi_2)}{2\,\ba\,m_1^2\,(\ruaa-\rubb)}\,\left(d_{\phi_1\phi_2}+m_1^2\right)\nn&-&\frac{m_i\,m_j\,\mathcal{Z}_2(\chi_1,\chi_2,\phi_1,\phi_2)}{8\,\ba\,m_1^2\,(\ruaa-\rubb)}\nn
J_{13}(\phi_1,\phi_2,\chi_1,\chi_2)&=&\frac{3\,m_i\,m_j\,m_{\chi_1}\,m_{\chi_2}\,s^2\,\mathcal{X}_2(\chi_1,\chi_2,\phi_1,\phi_2)}{\ba\,(\ruaa-\rubb)}\nn
J_{14}(\phi_1,\phi_2,\chi_1,\chi_2)&=&-\frac{\ba^2}{48\,m_1^4\,s^2} -\frac{1}{16\,m_1^4\,s^2}\, \left(\ruaa^2 + \rubb^2+\ruaa\,\rubb- 3\,l_i\,s \right.\nn&&\left.(\ruaa+\rubb)\right)-\frac{1}{16\,m_1^4}\,\left(-8\, d\, d_{\phi_1\phi_2} + 2\, l_i^2 + l_i\, l_{ia} + 8\, d_{\phi_1\phi_2}\, m_i^2\right.\nn&-&\left. 4\, m_1^2\, (l_{ia} +m_1^2 + 2\, s)\right)
+\frac{s^2\,\mathcal{X}_2(\chi_1,\chi_2,\phi_1,\phi_2)}{8\,\ba\,m_1^4\,(\ruaa-\rubb)}\,\left(-2\, l_i^2 m_1^2\,s \right.\nn&+&\left. 4\, d_{\phi_1\phi_2}\, m_i^2\, (-4\, l_{ia}\, m_1^2 + l_i\,s) - 
m_1^4 (-12\, l_{ia}\, m_i^2 + 2\, l_i\, s)\right)\nn&-&\frac{d\,s^2\,\mathcal{X}_2(\chi_1,\chi_2,\phi_1,\phi_2)}{\ba\,m_1^4\,(\ruaa-\rubb)}\,\left(-2\, d_{\phi_1\phi_2}^2\, m_i^2 + d_{\phi_1\phi_2}\, l_i\,m_1^2 + (5\, d_{\phi_1\phi_2} \right.\nn&-&\left. 4\, l_i + 5\, m_i^2)\, m_1^4\right)-\frac{l_i\,s\,\mathcal{Y}_2(\chi_1,\chi_2,\phi_1,\phi_2)}{16\,\ba\,m_1^4\,(\ruaa-\rubb)}\nn
&&\left(-8\, d\, d_{\phi_1\phi_2} + l_i^2 + l_i\, l_{ia} + 8\, d_{\phi_1\phi_2}\, m_i^2 - 4\,m_1^2\, (l_{ia} + m_1^2 + 2\, s)\right)\nn
&+&\frac{\mathcal{Z}_2(\chi_1,\chi_2,\phi_1,\phi_2)}{16\,\ba\,m_1^4\,(\ruaa-\rubb)}\,\left(-4\, d\,d_{\phi_1\phi_2} + 2\, l_i^2 + 4\,d_{\phi_1\phi_2}\, m_i^2 + l_i\, s \right.\nn&-&\left. 2\, m_1^2\, (l_{ia} + m_1^2 + 2\, s)\right)-\frac{4\,l_i\,\mathcal{R}_1(\chi_1,\chi_2,\phi_1,\phi_2)-\mathcal{Q}_2(\chi_1,\chi_2,\phi_1,\phi_2)}{32\,\ba\,m_1^4\,(\ruaa-\rubb)\,s^2}\nonumber
\end{eqnarray}
\begin{eqnarray}
J_{15}(\phi_1,\phi_2,\chi_1,\chi_2)&=&\frac{m_{\chi_1}\,m_{\chi_2}\,(m_1^2-d_{\phi_1\phi_2})}{2\,m_1^2}-\frac{m_{\chi_1}\,m_{\chi_2}\,s^2\,\mathcal{X}_2(\chi_1,\chi_2,\phi_1,\phi_2)}{4\,\ba\,m_1^2\,(\ruaa-\rubb)}\,\left(l_i\, (l_i - l_{ia}) \right.\nn&&\left.m_1^2+ 12\, d\,m_1^4 + d_{\phi_1\phi_2}\, (8\, d\, d_{\phi_1\phi_2} - 2\, l_i\, s)\right)\nn&-&\frac{l_i\,m_{\chi_1}\,m_{\chi_2}\,s\,\mathcal{Y}_2(\chi_1,\chi_2,\phi_1,\phi_2)}{2\,\ba\,m_1^4\,(\ruaa-\rubb)}\,\left(d_{\phi_1\phi_2}-m_1^2\right)
\nn&+&\frac{m_{\chi_1}\,m_{\chi_2}\,\mathcal{Z}_2(\chi_1,\chi_2,\phi_1,\phi_2)}{4\,\ba\,m_1^4\,(\ruaa-\rubb)}\,\left(d_{\phi_1\phi_2}-m_1^2\right)\nonumber
\end{eqnarray}
\begin{eqnarray}
J_{16}(\phi_1,\phi_2,\chi_1,\chi_2)&=&\frac{m_i\,m_j\,(-d_{\phi_1\phi_2} +m_1^2)}{2\,m_1^4}+\frac{m_i\,m_j\,s^2\,\mathcal{X}_2(\chi_1,\chi_2,\phi_1,\phi_2)}{2\,\ba\,m_1^4\,(\ruaa-\rubb)}\,\left(4\, d_{\phi_1\phi_2}^2\, m_i^2\right.\nn&-&\left. l_i^2\,m_1^2- (6\, d_{\phi_1\phi_2} + l_i - 6\, m_i^2)\, m_1^4 + 4\, d_{\phi_1\phi_2}\,m_1^2\, s\right)\nn&+&\frac{m_i\,m_j\,(d_{\phi_1\phi_2}-m_1^2)}{4\,\ba\,m_1^4\,(\ruaa-\rubb)}\,\left(-2\,l_i\,s\,\mathcal{Y}_2(\chi_1,\chi_2,\phi_1,\phi_2)\right.\nn&+&\left.\mathcal{Z}_2(\chi_1,\chi_2,\phi_1,\phi_2)\right)\nn
J_{17}(\phi_1,\phi_2,\chi_1,\chi_2)&=&-\frac{m_i\,m_j\,m_{cji_1}\,m_{\chi_2}\,s^2\,\mathcal{X}_2(\chi_1,\chi_2,\phi_1,\phi_2)}{\ba\,m_1^4\,(\ruaa-\rubb)}\,\left(-2\, d_{\phi_1\phi_2}^2 + 5\,m_1^4\right)\nonumber
\end{eqnarray}
\begin{eqnarray}
J_{18}(\phi_1,\phi_2,\chi_1,\chi_2)&=&\frac{m_j\,m_{\chi_2}\,(2\,d_{\phi_1\phi_2}+m_1^2)}{4\,m_1^4}+\frac{m_j\,m_{\chi_2}\,s^2\,\mathcal{X}_2(\chi_1,\chi_2,\phi_1,\phi_2)}{\ba\,m_1^4\,(\ruaa-\rubb)}\nn&&\left(-2\, d_{\phi_1\phi_2}^2\, m_i^2 +(d_{\phi_1\phi_2} - l_i)\, l_i\,m_1^2 + 3\, m_i^2\, m_1^4\right)\nn&+&\frac{m_j\,m_{\chi_2}\,s\,\mathcal{Y}_2(\chi_1,\chi_2,\phi_1,\phi_2)}{4\,\ba\,m_1^4\,(\ruaa-\rubb)}\,\left(2\,d_{\phi_1\phi_2}\,l_i - 2\, (d_{\phi_1\phi_2} - 2\, l_i)\,m_1^2 - 5\,m_1^4\right)\nn&-&\frac{m_j\,m_{\chi_2}\,(2\,d_{\phi_1\phi_2}+m_1^2)}{8\,\ba\,m_1^4\,(\ruaa-\rubb)}\,\mathcal{Z}_2(\chi_1,\chi_2,\phi_1,\phi_2)\nn
J_{19}(\phi_1,\phi_2,\chi_1,\chi_2)&=&\frac{m_j\,m_{\chi_1}\,(2\,d_{\phi_1\phi_2}+m_1^2)}{4\,m_1^4}+\frac{m_j\,m_{\chi_1}\,s^2\,\mathcal{X}_2(\chi_1,\chi_2,\phi_1,\phi_2)}{2\,\ba\,m_1^4\,(\ruaa-\rubb)}\nn&&\left(-4\, d_{\phi_1\phi_2}^2\, m_i^2 + l_i^2\,m_1^2 + (-5\, l_i + 6\, m_i^2)\,m_1^4\right)\nn&+&\frac{m_j\,m_{\chi_1}\,s\,\mathcal{Y}_2(\chi_1,\chi_2,\phi_1,\phi_2)}{4\,\ba\,m_1^4\,(\ruaa-\rubb)}\,\left(-2\, l_i\,m_1^2 + 5\,m_1^4 \right.\nn&+&\left. 2\,d_{\phi_1\phi_2}\, (l_i + m_1^2)\right)-\frac{m_j\,m_{\chi_1}\,(2\,d_{\phi_1\phi_2}+m_1^2)}{8\,\ba\,m_1^4\,(\ruaa-\rubb)}\,\mathcal{Z}_2(\chi_1,\chi_2,\phi_1,\phi_2)\nn
J_{20}(\phi_1,\phi_2,\chi_1,\chi_2)&=&\frac{m_i\,m_{\chi_2}\,(2\,d_{\phi_1\phi_2}+m_1^2)}{4\,m_1^4}-\frac{m_i\,m_{\chi_2}\,s^2\,\mathcal{X}_2(\chi_1,\chi_2,\phi_1,\phi_2)}{4\,\ba\,m_1^4\,(\ruaa-\rubb)}\,\left((4\, d_{\phi_1\phi_2} - l_i)\right.\nn&&\left. (l_i - l_{ia})\,m_1^2 + d\, (-8\, d_{\phi_1\phi_2}^2 + 12\, m_1^4) + 
2\,(d_{\phi_1\phi_2}\, l_i +m_1^4)\, s\right)\nn&+&\frac{m_i\,m_{\chi_2}\,s\,\mathcal{Y}_2(\chi_1,\chi_2,\phi_1,\phi_2)}{8\,\ba\,m_1^4\,(\ruaa-\rubb)}\,\left(4\, d_{\phi_1\phi_2}\, l_i + (8\, d_{\phi_1\phi_2} - l_i + 3\, l_{ia})\,m_1^2 \right.\nn&+&\left. 2 m_1^4\right)-\frac{m_i\,m_{\chi_2}\,(2\,d_{\phi_1\phi_2}+m_1^2)}{8\,\ba\,m_1^4\,(\ruaa-\rubb)}\,\mathcal{Z}_2(\chi_1,\chi_2,\phi_1,\phi_2)\nonumber
\end{eqnarray}
\begin{eqnarray}
J_{21}(\phi_1,\phi_2,\chi_1,\chi_2)&=&\frac{m_i\,m_{\chi_1}\,(2\,d_{\phi_1\phi_2}+m_1^2)}{4\,m_1^4}+\frac{m_i\,m_{\chi_1}\,s^2\,\mathcal{X}_2(\chi_1,\chi_2,\phi_1,\phi_2)}{4\,\ba\,m_1^4\,(\ruaa-\rubb)}\,\left(-m_1^2\,(l_i - l_{ia})\right.\nn&&\left. (2\, l_i - m_1^2) + 4\, d\, (2\, d_{\phi_1\phi_2}^2 - 3\,m_1^4) - 
2\, d_{\phi_1\phi_2}\, (l_i - 4\,m_1^2)\, s\right)\nn
&&\frac{m_i\,m_{\chi_1}\,s\,\mathcal{Y}_2(\chi_1,\chi_2,\phi_1,\phi_2)}{8\,\ba\,m_1^4\,(\ruaa-\rubb)}\,\left(4\, d_{\phi_1\phi_2}\, l_i + (-8\, d_{\phi_1\phi_2} + 5\, l_i - 3\, l_{ia})\, m_1^2\right.\nn&-&\left. 2\,m_1^4\right)-\frac{m_i\,m_{\chi_1}\,(2\,d_{\phi_1\phi_2}+m_1^2)}{8\,\ba\,m_1^4\,(\ruaa-\rubb)}\,\mathcal{Z}_2(\chi_1,\chi_2,\phi_1,\phi_2)\nonumber
\end{eqnarray}
\begin{eqnarray}
\mathcal{B}_1(\chi,\phi_1,\phi_2)&=&-\frac{b_{\phi_1\phi_2}\,m_{\chi}^4}{6\,s^2}+{8\over s^2}\,(s-m_{\chi}^2)^2\,(m_1^2-m_2^2)^2\,\tau_{\phi_1\phi_2}\,\tau_{\phi_2\phi_1}-\frac{d}{6}\left((m_1^2+m_2^2\right.\nn
&-&\left.2\,d_{\phi_1\phi_2})\,m_{\chi}^4+(s-2\,m_{\chi}^2)\,(m_1^2-m_2^2)^2\right)\nn
\mathcal{B}_2(\chi,\phi_1,\phi_2)&=&m_i\,m_j\,\left((s-2\,m_{\chi}^2)\,(m_1^2-m_2^2)^2+(m_i^2+m_j^2-2\,d_{\phi_1\phi_2})\,m_{\chi}^4\right)\nn
\mathcal{B}_3(\phi,\chi,\phi_1,\phi_2)&=&b_{\phi_1\phi_2}\,m_{\phi}^2+\left(m_{\phi}^2\,(R_{\phi_1\chi}-2\,s\,l_i)-2\,s\,l_i\ka)\right)\nn
&&\mathcal{A}(\chi,\phi_1,\phi_2)\nonumber
\end{eqnarray}
\begin{eqnarray}
\mathcal{B}_4(\phi,\chi,\phi_1,\phi_2)&=&2\,b_{\phi_1\phi_2}\,m_{\phi}^2+\left(2\,\ka\,l_{ia}+m_{\phi}^2\,(\mathcal{C}(\chi,\phi_1,\phi_2)-2\,s\,l_{\phi_1\phi_2})\right)\nn
&&\mathcal{A}(\chi,\phi_1,\phi_2)\nonumber
\end{eqnarray}
\begin{eqnarray}
\mathcal{B}_5(\phi,\chi,\phi_1,\phi_2)&=&\frac{\rua^2}{8\,b_{\phi_1\phi_2}\,s}\,L(\chi,\phi_1,\phi_2)+\frac{\rua}{8\,b_{\phi_1\phi_2}\,m_\phi^2\,s}\left(-2\,b_{\phi_1\phi_2}\,m_\phi^2 -L(\chi,\phi_1,\phi_2)\right.\nn&&\left. (l_{\phi_2\phi_1}\, m_{\phi}^2\,(m_i^2-m_j^2)+(2\, m_i^2\, m_{\phi}^2 -(m_1^2-m_2^2)\,(m_i^2-m_j^2)) s\right.\nn
&+&\left.l_{\phi_2\phi_1}\, m_{\phi}^2\, (2\,\taua + 3\, s))\right)+\frac{1}{4\, m_{\phi}^2}\left((\taua + 2\, m_i^2)  m_{\phi}^2 + \ka\right.\nn&&\left.\ia\right)
-\frac{s^2\,L(\chi,\phi_1,\phi_2)}{4\,b_{\phi_1\phi_2}\, m_{\phi}^2\,s}\left(\ka\, m_i^2\, (\ka + m_{\phi}^2 \right.\nn&+&\left. \ia)+ 
m_\phi^2\, (2\, d (-d_{\phi_1\phi_2}
+m_1^2) + 2\,l_{\phi_1\phi_2}\, (l_i + m_i^2) + m_i^2 \right.\nn&&\left.\ia\right)+\frac{l_{ia}}{4\,s} (2\, \ka + s)\nn
\mathcal{B}_6(\phi,\chi,\phi_1,\phi_2)&=&\ba\, m_{\phi}^2 +\left(2\,\ka\, (l_{\phi_1\phi_2} -l_i)\, s +m_{\phi}^2\, (\rua \right.\nn&+&\left. 2\, (d_{\phi_1\phi_2} -m_1^2 -l_i) s)\right)\,\mathcal{A}(\chi,\phi_1,\phi_2)\nn
\mathcal{B}_7(\chi,\phi_1,\phi_2)&=&-\frac{1}{48\,m_\chi^4\,s^2}\left(b_{\phi_1\phi_2}^2\, m_\chi^2+24\,d\,(m_\chi^2 - s)^2\,s^2+12\,l_i\,l_{ia}\,(2\, m_\chi^2 - s)\,s^2\right)\nn
&&-\frac{l_{\phi_1\phi_2}^2}{48\,m_\chi^6\,s^2}\,\left(l_i\,l_{ia}\, (m_\chi^2 - s)^2 + 2\, d\, (2\, m_\chi^2 - s)\, s^2\right)\nn
\mathcal{B}_8(\chi,\phi_1,\phi_2)&=&\frac{1}{8\,m_\chi^6}\left(l_{\phi_1\phi_2}^2\, (-2\, m_\chi^2 + s) - 4\, m_\chi^2\, (3\, m_\chi^4 - 2\, m_\chi^2\, s + s^2)\right)\nn
\mathcal{B}_9(\phi,\chi,\phi_1,\phi_2)&=&-\frac{m_j}{8\,m_\phi^4\,s}\,\left(l_{\phi_1\phi_2}\, l_i\, (m_\phi^2 - s) +m_\phi^2\, (R_{\phi_1\chi} + 2\, s\, (-3\, m_\phi^2 + s))\right)\nn
&+&\frac{m_j\,L(\chi,\phi_1,\phi_2)}{16\,b_{\phi_1\phi_2}\,m_\phi^4\,s}\,\left(2\, m_\phi^4\, s\, (-3\, R_{\phi_1\chi} + 4 \,m_i^2\, s)+l_{\phi_1\phi_2}\, s\, (-l_i\,\rua\right.\nn&+&\left. 2\, l_{\phi_1\phi_2}\, m_i^2\, s)+
m_\phi^2\, (\rua^2 + 2\, (-\tau_{\phi_1\phi_2} + 2\, l_i^2 + \rua) \,s^2 - 8\, m_i^2\, s^3)\right)\nonumber
\end{eqnarray}
\begin{eqnarray}
\mathcal{B}_{10}(\phi,\chi,\phi_1,\phi_2)&=&\frac{m_i}{16\,m_\phi^4}\,\left(\tau_{\phi_1\phi_2}-\omega_{\phi_1\phi_2} + 12\, m_\phi^4 - 4\,m_\phi^2\, s\right)+\frac{m_i\,L(\chi,\phi_1,\phi_2)}{32\,b_{\phi_1\phi_2}\,m_\phi^4\,s}\nn&&\left(\tau_{\phi_1\phi_2}-\omega_{\phi_1\phi_2}\, (12\, m_{\phi}^4 - \rua)-4\,\tau_{\phi_1\phi_2}\, m_\phi^2\, s + l_{\phi_1\phi_2}^2\, (l_i^2 - l_i \,l_{ia}\right.\nn&+&\left. 4\, d\, s)+4\, m_\phi^2\, (3\,m_\phi^2 - s)\, (-\rua + 4\, d\, s)\right)\nn
\mathcal{B}_{11}(\phi,\chi,\phi_1,\phi_2)&=&\frac{m_\chi}{8\,m_\phi^4\,s}\,\left(m_\phi^2\, \rua - l_{\phi_1\phi_2}\, (l_i\, m_\phi^2 + s\, (-2\,m_\phi^2 + s))\right)+\frac{m_\chi\,L(\chi,\phi_1,\phi_2)}{16\,b_{\phi_1\phi_2}\,m_\phi^4\,s}\nn
&&\left(l_{\phi_1\phi_2}^2\, s\, (l_i\, l_{ia} - (2\, d + l_i)\, s)+l_{\phi_1\phi_2}\, \rua\, (2\, l_i\,m_\phi^2 + s\, (-2\, m_\phi^2 + s))\right.\nn&-&\left.m_\phi^2\, (\rua^2 + 4\, s^2\, (l_i l_{ia} + 2\, d\, m_\phi^2 - 2\, d\, s))\right)\nn
\mathcal{B}_{12}(\phi,\chi,\phi_1,\phi_2)&=&-\frac{m_i\,m_j\,m_\chi\,s\,L(\chi,\phi_1,\phi_2)}{8\,b_{\phi_1\phi_2}\,m_\phi^4}\,\left(l_{\phi_1\phi_2}^2 + 12\, m_\phi^4 - 4\,m_\phi^2\, s\right)\nn
\mathcal{B}_{13}(\phi,\chi,\phi_1,\phi_2)&=&\frac{m_\chi\,L(\chi,\phi_1,\phi_2)}{16\,b_{\phi_1\phi_2}\,m_\phi^4}\,\left(l_{\phi_2\phi_1}\,\rho(\chi,\phi_1,\phi_2)\, (-2\, m_\phi^2 + s) + l_{\phi_2\phi_1}^2\, (l_i\, l_{ia} - 2\, d\, s)\right. \nn && \left.- 
4\, m_\phi^2\, s\, (l_i\, l_{ia} + 2\, d\, m_\phi^2 - 2\, d\, s)\right)+\frac{m_\chi\,l_{\phi_2\phi_1}}{4\,b_{\phi_1\phi_2}\,m_\phi^2}\nonumber
\end{eqnarray}
\begin{eqnarray}
\mathcal{B}_{14}(\phi,\chi,\phi_1,\phi_2)&=&-\frac{m_i\,m_j\,m_\chi\,s\,L(\chi,\phi_1,\phi_2)}{8\,b_{\phi_1\phi_2}\,m_\phi^4}\,\left(l_{\phi_2\phi_1}^2 + 12\, m_\phi^4 - 4\, m_\phi^2\, s\right)\nn
\mathcal{B}_{15}(\phi,\chi,\phi_1,\phi_2)&=&\frac{m_j\,L(\chi,\phi_1,\phi_2)}{32\,b_{\phi_1\phi_2}\,m_\phi^4}\,\left(l_{\phi_2\phi_1}\, (-l_{\phi_1\phi_2}\,\rho(\chi,\phi_1,\phi_2) + 4\, d_{\phi_1\phi_2}\, l_i\, s)\right.\nn&&\left. +l_{\phi_2\phi_1} ^2\, (\rho(\chi,\phi_1,\phi_2) - 4\, m_i^2\, s) + 
4\, m_\phi^2\, (3\,m_\phi^2 - s)\, (\rua - 4\, m_i^2\, s)\right)\nn&-&\frac{m_j}{2\,m_\phi^4}\,\left(l_{\phi_2\phi_1}^2 - l_{\phi_2\phi_1}\, l_{\phi_1\phi_2} + 12\, m_\phi^4 - 4\, m_\phi^2 \,s\right)\nn
\mathcal{B}_{16}(\phi,\chi,\phi_1,\phi_2)&=&\frac{m_i\,L(\chi,\phi_1,\phi_2)}{32\,b_{\phi_1\phi_2}\,m_\phi^4}\left(2\,l_{\phi_2\phi_1}^2\, l_i\, l_{ia} + 8\, d_{\phi_1\phi_2}\,l_{\phi_2\phi_1}\, l_{ia}\,m_\phi^2 +(l_{\phi_2\phi_1}^2\right.\nn&-&\left.l_{\phi_2\phi_1}\, l_{\phi_1\phi_2})\,\rho(\chi,\phi_1,\phi_2)-  
4\,m_2^2\,(l_{\phi_2\phi_1} - m_1^2)\,R_{\phi_1\chi}- 
2\, (l_{\phi_2\phi_1}^3 \right.\nn&+&\left.4\, l_i\, l_{ia}\,m_\phi^2 + 8\, d\, m_\phi^4 + 12\,l_{\phi_2\phi_1} \, m_\phi^4-  4\,l_{\phi_1\phi_2}\,m_\phi^4 - 
l_{\phi_2\phi_1}\,\rua)\, s \right.\nn&+&\left. 8\, (2\, d + 2\, s)\, m_\phi^2\, s^2) \right)+\frac{m_i}{32\,m_\phi^4}\,\left(2\,l_{\phi_2\phi_1}\, (-l_{\phi_2\phi_1} + l_{\phi_1\phi_2}) + 8\,m_\phi^2\, (m_\phi^2 + s)\right)\nn
\mathcal{B}_{17}(\phi,\chi,\phi_1,\phi_2)&=& s\, (m_2^4 + (m_1^2 - s)^2 - 2\, m_2^2\, (m_1^2 + s))\frac{\mathcal{A}(\chi,\phi_1,\phi_2)}{4\,b_{\phi_1\phi_2}\, m_1^2\, (m_2^2 - s)\, s}\nn
\mathcal{B}_{18}(\phi,\chi,\phi_1,\phi_2)&=& (6\,m_\phi^2 - s)\, s -(m_1^2-m_2^2) (2\,m_\phi^2 + s)\nn
\mathcal{B}_{19}(\phi,\chi,\phi_1,\phi_2)&=& 4\, d_{\phi_1\phi_2}\, s^2 - 3\,m_\phi^2\, (m_2^2-m_1^2 + 2\, s^2)\nonumber
\end{eqnarray}
\begin{eqnarray}
\mathcal{B}_{20}(\phi,\chi,\phi_1,\phi_2)&=&\frac{m_j}{8\,m_1^4}\,\left(d_{\phi_1\phi_2}\, l_i + 2\, (-\da + l_i)\, m_1^2 + 4\, m_1^4\right)+\frac{s\,m_j\,L(\chi,\phi_1,\phi_2)}{4\,\ba\,m_1^2}\nn&&\left(2\, \da^2\, m_i^2 + l_i\, (-\da + l_i)\, m_1^2\right)-\frac{m_j\,\rua}{8\,\ba\,m_1^4\,s}\,\left(\ba\, (\da + 2\,m_1^2)\right.\nn &+&\left. (\da\, (l_i - m_1^2) + 2\, m_1^2\, (l_i + m_1^2))\,s\,L(\chi,\phi_1,\phi_2)\right)\nonumber
\end{eqnarray}
\begin{eqnarray}
\mathcal{B}_{21}(\phi,\chi,\phi_1,\phi_2)&=&\frac{m_i}{8\,\ba\,m_1^4}\,\left(\da\, l_i + 2\, (-\da + l_i)\,m_1^2 + 4\, m_1^4\right)+\frac{m_i\,s\,L(\chi,\phi_1,\phi_2)}{8\,\ba\,m_1^4}\nn&&\left(-4\, d\, \da^2 - (\da - l_i)\, (l_i - l_{ia})\,m_1^2 + (\da\, l_i + 4\,m_1^4)\,s\right)\nn&-&\frac{m_i\,\rua}{8\,\ba\,s\,m_1^4}\,\left(\ba\, (\da + 2\,m_1^2) + (\da\, (l_i - m_1^2) + 2\,m_1^2 (l_i + m_1^2))\right.\nn&&\left.s\,L(\chi,\phi_1,\phi_2)\right)\nn
\mathcal{B}_{22}(\phi,\chi,\phi_1,\phi_2)&=&\frac{m_\chi\,s}{8\,\ba\,m_1^4}\,\left(-2\, \ba\, \da + (4\, d\, (\da^2 + 2\, m_1^4) + \da (\rua - l_i\,s))\right.\nn&&\left.L(\chi,\phi_1,\phi_2)\right)\nn
\mathcal{B}_{23}(\phi,\chi,\phi_1,\phi_2)&=&-\frac{m_i\,m_j\,m_\chi\,s\,L(\chi,\phi_1,\phi_2)}{2\,\ba\,m_1^4}\,(\da^2 + 2\, m_1^4)\nonumber
\end{eqnarray}
\begin{eqnarray}
\mathcal{B}_{24}(\phi,\chi,\phi_1,\phi_2)&=&\frac{1}{192\,\ba\,m_1^4\,m_{\phi}^2\,s^2}\,\left(2\, \ba\, (2\, \ba^2\,m_{\phi}^2\, (2\,m_1^2 - s) + 24\,\da \,(d - m_i^2)\right.\nn&&\left.\,m_{\phi}^2\, s^3- 
12\, m_1^4\,s^2\, (l_i^2 - 3\, l_i\, l_{ia} + 4\, d\,s + 
2\, m_{\phi}^2\, (-2\, (2\, d + l_i + m_i^2) + 5\, s))\right.\nn&+&\left. 
6\, m_1^2\, s^2 ((-2\, \da + s)\, (l_i^2 - 3\, l_i\, l_{ia} + 4\, d\,s) + 
m_{\phi}^2\, (-l_i^2 + l_i\, (7\, l_{ia} - 2\,s) \right.\nn&+&\left. 2\,s\, (4\, \da + 2\, m_i^2 + s)))) + 
3\, s^2\, (-m_{\phi}^2\, (l_i^3 + 8\,\da\,l_i\, m_i^2 - 8\, \da \,l_{ia}\, m_i^2 \right.\nn&+&\left. 
l_i^2\, (l_{ia} - 2\,s))\,s^2 + 
16\,m_1^6\, s \,(l_i\, l_{ia} + 6\, d\,m_{\phi}^2 - 2\, d\, s) + 
4\, m_1^4\, s\, (-l_i^3 + 3\, l_i^2\, l_{ia}\right.\nn &-&\left. 2\, l_i\,(2\, d + l_{ia})\, s + 4\, d\, s^2 + 
2\,m_{\phi}^2\, (2\, (l_i - 4\, l_{ia})\, m_i^2 - l_i\,s + 
2\, d\, (-6\, \da + 6\, l_i \right.\nn&+&\left. s))) + 
2\, m_1^2\, ((-2\, \da +s)\,s (l_i\, (l_i^2 + 4\, \da\, l_{ia} - 3\, l_i\, l_{ia}) + 
4\, d\, (-2\,\da \right.\nn&+&\left. l_i)\,s) + 
m_{\phi}\, s\, (4\, \da\, (l_i^2 - l_i\, l_{ia} + 4\, l_{ia}\, m_i^2) - 
8\, d\, \da \,(2\, l_i + s) + 
l_i\, (-l_i^2 \right.\nn&+&\left. 7\, l_i\, l_{ia} - 2\, l_i\,s + 4\, m_i^2\,s + 2\, s^2))))\,L(\chi,\phi_1,\phi_2)\right)+\frac{\rua}{64\,\ba\,m_1^2\,m_{\phi}^2\,s}\nn&&\left(4\, \ba\,m_{\phi}^2\, (-4\,m_1^2 (-\da + l_i + 2\,m_1^2) + 2\, (l_i + m_1^2)\,s - s^2) - 
s\, \right.\nn&&\left.(2\, m_1^2\, (-2\, l_i\, (l_i - 3\, l_{ia})\, (\da + m_1^2) + (3\, l_i^2 + 4\, \da\, l_{ia} + 
7\, l_i\, l_{ia} \right.\nn&+&\left. 8\, (2\,d +2\, l_i + m_i^2)\, m_1^2)\,m_{\phi}^2) + 
4\, (2\, d\, m_1^2 + (l_i + m_1^2)\, m_{\phi}^2)\, s^2 - 
s\, (2\, m_1^2\, (-l_i\right.\nn&&\left. (l_i - 3\, l_{ia}) + 8\, d\, (\da + m_1^2)) + 
m_{\phi}^2\, (-8\, d\, \da + 8\,\da\, m_i^2 + 8\, (l_i - m_i^2)\right.\nn&&\left.m_1^2 + 40\,m_1^4 + 
l_i\, (4\, l_i + 2\, s))))\,L(\chi,\phi_1,\phi_2)\right)+\frac{\rua^2}{32\,\ba\,m_1^2\,s}\,\left(\ba\right.\nn&&\left.(4\,m_1^2 - 2\,s) + 
s\, (-4\,\da\,m_1^2 + 6\, l_i\,m_1^2 + 8\,m_1^4 - 3\, l_i\,s - 2\, m_1^2\, s + s^2)\right.\nn&&\left.L(\chi,\phi_1,\phi_2)\right)+\frac{6\,\rua^3\,L(\chi,\phi_1,\phi_2)}{192\,\ba\,m_1^4\,s^2}\,\left(s-2\,m_1^2\right)\nn
\mathcal{B}_{25}(\phi,\chi,\phi_1,\phi_2)&=&\frac{m_i\,m_j}{8\,\ba\,m_1^2\,m_{\phi}^2}\,\left(2\,\da\, (2\, m_{\phi}^2 - s) + 2\,m_1^2\, (5\,m_{\phi}^2 -s) + 
s\, (3\,m_{\phi}^2 + s)\right)\nn&& \left(2\, \ba - (\rua + (2\,\da - l_i - 2\,m_1^2)\, s)\,L(\chi,\phi_1,\phi_2)\right)\nonumber
\end{eqnarray}
\begin{eqnarray}
\mathcal{B}_{26}(\phi,\chi,\phi_1,\phi_2)&=&\frac{m_\chi\,m_j}{8\,\ba\,m_1^4}\,\left(2\,\ba (-4\,m_1^4 + \da\, (2\,m_1^2 + s)) - (l_i\,m_1^2\,s^2 - 
2\,m_1^4\, (2\,\rua \right.\nn&+&\left. 5\, l_i\,s) + \da\, (\rua\,s + 2\, m_1^2\, (\rua + 2\, l_i\,s)))\,L(\chi,\phi_1,\phi_2)\right)\nonumber
\end{eqnarray}
\begin{eqnarray}
\mathcal{B}_{27}(\phi,\chi,\phi_1,\phi_2)&=&\frac{m_\chi\,m_i}{8\,\ba\,m_1^4}\,\left(-2\,\ba (-4\,m_1^4 +\da\, (2\,m_1^2 + s)) + (l_{ia}\,m_1^2\,s^2 - 
2\,m_1^4\, (2\,\rua\right.\nn&-&\left. 2\, l_i\,s + 7\, l_{ia}\,s) + 
\da (s\, (\rua - l_i\,s + l_{ia}\,s) + 2\,m_1^2\, (\rua - l_i\,s + 3\, l_{ia}\,s)))\right.\nn&&\left.L(\chi,\phi_1,\phi_2)\right)\nonumber
\end{eqnarray}
\begin{eqnarray}
\mathcal{C}_1(\chi_1,\chi_2,\phi_1,\phi_2)&=&\frac{1}{8\,b_{\phi_1\phi_2}\,m_1^2\,(R_{\phi_1\chi_1}-R_{\phi_1\chi_2})}\left((\omega_{\phi_1\phi_2}-\tau_{\phi_1\phi_2})\,(m_i^2-2\,m_1^2)+4\, (2\, d + m_i^2)\right.\nn&&\left. m_1^2\, s\right)\,\mathcal{X}_1(\chi_1,\chi_2,\phi_1,\phi_2)+
\frac{s\,\mathcal{Y}_1(\chi_1,\chi_2,\phi_1,\phi_2)}{2\,b_{\phi_1\phi_2}\,(R_{\phi_1\chi_1}-R_{\phi_1\chi_2})}\left((\tau_{\phi_1\phi_2} -\omega_{\phi_1\phi_2})\, m_i^2\right.\nn
&&\left. + 2\, d\, (m_i^2 - m_1^2)\, s\right)+\frac{2\, d - m_i^2 + 2 m_1^2}{8\,b_{\phi_1\phi_2}\,m_1^2\,(R_{\phi_1\chi_1}-R_{\phi_1\chi_2})}\left(2\, b_{\phi_1\phi_2} (R_{\phi_1\chi_2}-R_{\phi_1\chi_1}) \right.\nn
&&\left.+\mathcal{Z}_1(\chi_1,\chi_2,\phi_1,\phi_2)\right)\nn
\mathcal{C}_2(\chi_1,\chi_2,\phi_1,\phi_2)&=&\frac{m_{\chi_1}\,m_{\chi_2}\,(\tau_{\phi_1\phi_2}-\omega_{\phi_1\phi_2})}{8\,b_{\phi_1\phi_2}\,m_1^2\,(R_{\phi_1\chi_1}-R_{\phi_1\chi_2})}\,\mathcal{X}_1(\chi_1,\chi_2,\phi_1,\phi_2)\nn&+&
\frac{s^2\,d\,m_{\chi_1}\,m_{\chi_2}\,\mathcal{Y}_1(\chi_1,\chi_2,\phi_1,\phi_2)}{b_{\phi_1\phi_2}\,(R_{\phi_1\chi_1}-R_{\phi_1\chi_2})}\,+\frac{m_{\chi_1}\,m_{\chi_2}}{8\,b_{\phi_1\phi_2}\,m_1^2\,(R_{\phi_1\chi_1}-R_{\phi_1\chi_2})}\nn
&&\left(2\, b_{\phi_1\phi_2} (R_{\phi_1\chi_2}-R_{\phi_1\chi_1})+\mathcal{Z}_1(\chi_1,\chi_2,\phi_1,\phi_2)\right)\nn
\mathcal{C}_3(\chi_1,\chi_2,\phi_1,\phi_2)&=&\frac{\mathcal{Y}_1(\chi_1,\chi_2,\phi_1,\phi_2)}{b_{\phi_1\phi_2}\,(R_{\phi_1\chi_1}-R_{\phi_1\chi_2})}\nn
\mathcal{C}_4(\chi_1,\chi_2,\phi_1,\phi_2)&=&\frac{m_j}{8\,b_{\phi_1\phi_2}\,m_1^2\,(R_{\phi_1\chi_1}-R_{\phi_1\chi_2})}\left(2\, b_{\phi_1\phi_2} (R_{\phi_1\chi_2}-R_{\phi_1\chi_1}) \right.\nn
&&\left.+ \mathcal{Z}_1(\chi_1,\chi_2,\phi_1,\phi_2)\right)\nn
\mathcal{C}_5(\chi_1,\chi_2,\phi_1,\phi_2)&=&\frac{3\,m_i\,s}{4\,b_{\phi_1\phi_2}\,(R_{\phi_1\chi_1}-R_{\phi_1\chi_2})}\left[\left((\omega_{\phi_1\phi_2}-\tau_{\phi_1\phi_2}) -R_{\phi_1\chi_1} + 4\, d\, s\right)\right.\nn
&&\left.L(\chi_1,\phi_1,\phi_2)-\left((\omega_{\phi_1\phi_2}-\tau_{\phi_1\phi_2}) -R_{\phi_1\chi_2} + 4\, d\, s\right)\right.\nn
&&\left.L(\chi_2,\phi_1,\phi_2)\right]\nn
\mathcal{C}_6(\chi_1,\chi_2,\phi_1,\phi_2)&=&\frac{3\,m_j\,s}{4\,b_{\phi_1\phi_2}\,(R_{\phi_1\chi_1}-R_{\phi_1\chi_2})}\left[\left(3\,R_{\phi_1\chi_1} - 4\, m_i^2\, s\right)\,L(\chi_1,\phi_1,\phi_2)\right.\nn
&&\left.-\left(3\,R_{\phi_1\chi_2} - 4\, m_i^2\, s\right)\,L(\chi_2,\phi_1,\phi_2)\right]\nn
\mathcal{C}_7(\chi_1,\chi_2,\phi_1,\phi_2)&=&\frac{3\,m_i\,m_j\,s}{2\,b_{\phi_1\phi_2}\,(R_{\phi_1\chi_1}-R_{\phi_1\chi_2})}\left[\left(-\,R_{\phi_1\chi_1}+ 2\, (m_1^2 - m_i^2) \,s\right)\,L(\chi_1,\phi_1,\phi_2)\right.\nn
&&\left.-\left(-\,R_{\phi_1\chi_2}+ 2\, (m_1^2 - m_i^2) \,s\right)\,L(\chi_2,\phi_1,\phi_2)\right]\nonumber
\end{eqnarray}
\begin{eqnarray}
\mathcal{C}_8(\chi_1,\chi_2,\phi_1,\phi_2)&=&\frac{\mathcal{X}_1(\chi_1,\chi_2,\phi_1,\phi_2)}{4\,m_2^2,b_{\phi_1\phi_2}\,(R_{\phi_1\chi_1}-R_{\phi_1\chi_2})}\left((m_1^2-m_i^2) (-\tau_{\phi_1\phi_2} + l_{\phi_1\phi_2}\,s) +m_2^2\, ( \tau_{\phi_1\phi_2} \right.\nn
&&\left.- l_{\phi_2\phi_1}\, s+ 2\, m_i^2\, s) + 
d_{\phi_1\phi_2} (\tau_{\phi_1\phi_2} -\omega_{\phi_2\phi_1} + 4 m_i^2\, s)\right)\nn
&&+\frac{\mathcal{Y}_1(\chi_1,\chi_2,\phi_1,\phi_2)}{8\,m_2^2,b_{\phi_1\phi_2}\,(R_{\phi_1\chi_1}-R_{\phi_1\chi_2})}\,\left(8\, d_{\phi_1\phi_2}\, m_i^2\, s\, ( \tau_{\phi_1\phi_2}-\omega_{\phi_2\phi_1})\right.\nn
&&\left. + 4\, m_i^2\, m_2^2\, s\, (\tau_{\phi_1\phi_2}-\omega_{\phi_1\phi_2}) - 
(m_1^2-m_i^2)\, (8\, d\, m_2^2\, s^2 + \tau_{\phi_2\phi_1}\,\omega_{\phi_2\phi_1}\right.\nn
&&\left. - 2\, l_{\phi_2\phi_1}\, s\,\tau_{\phi_1\phi_2} + \tau_{\phi_1\phi_2}^2)\right)-\frac{s-2\,m_1^2+m_i^2}{8\,m_2^2,b_{\phi_1\phi_2}\,(R_{\phi_1\chi_1}-R_{\phi_1\chi_2})}\left(2\, b_{\phi_1\phi_2}\right.\nn
&&\left. (R_{\phi_1\chi_2}-R_{\phi_1\chi_1}) + \mathcal{Z}_1(\chi_1,\chi_2,\phi_1,\phi_2)\right)\nonumber
\end{eqnarray}
\begin{eqnarray}
\mathcal{C}_9(\chi_1,\chi_2,\phi_1,\phi_2)&=&\frac{\mathcal{X}_1(\chi_1,\chi_2,\phi_1,\phi_2)}{4\,m_2^2\,b_{\phi_1\phi_2}\,(R_{\phi_1\chi_1}-R_{\phi_1\chi_2})}\left(\tau_{\phi_1\phi_2}-l_{\phi_2\phi_1}\,s\right)\nn&+&\frac{\mathcal{Y}_1(\chi_1,\chi_2,\phi_1,\phi_2)}{8\,m_2^2\,b_{\phi_1\phi_2}\,(R_{\phi_1\chi_1}-R_{\phi_1\chi_2})}\,\left(8\, d\, m_2^2\, s^2 +\omega_{\phi_2\phi_1}\,\tau_{\phi_2\phi_1} \right.\nn&-&\left. 2\, l_{\phi_2\phi_1}\, s\,\tau_{\phi_1\phi_2} +\tau_{\phi_1\phi_2}^2\right)+\frac{1}{8\,m_2^2\,b_{\phi_1\phi_2}\,(R_{\phi_1\chi_1}-R_{\phi_1\chi_2})}\nn&&\left(2\, b_{\phi_1\phi_2} (R_{\phi_1\chi_2}-R_{\phi_1\chi_1})+ \mathcal{Z}_1(\chi_1,\chi_2,\phi_1,\phi_2)\right)\nonumber
\end{eqnarray}
\begin{eqnarray}
\mathcal{C}_{10}(\chi_1,\chi_2,\phi_1,\phi_2)&=&-\frac{m_i\,s\,L(\chi_1,\phi_1,\phi_2)}{4\,m_2^2\,b_{\phi_1\phi_2}\,(R_{\phi_1\chi_1}-R_{\phi_1\chi_2})}\left (2\, d_{\phi_1\phi_2}\, (\omega_{\phi_2\phi_1} + R_{\phi_1\chi_1}- \tau_{\phi_1\phi_2})\right.\nn
&+&\left. m_2^2\, (\omega_{\phi_1\phi_2} +R_{\phi_1\chi_1} - \tau_{\phi_1\phi_2})\right)+\frac{m_i\,s\,L(\chi_2,\phi_1,\phi_2)}{4\,m_2^2\,b_{\phi_1\phi_2}\,(R_{\phi_1\chi_1}-R_{\phi_1\chi_2})}\nn&&\left (2\, d_{\phi_1\phi_2}\, (\omega_{\phi_2\phi_1} + R_{\phi_1\chi_2}- \tau_{\phi_1\phi_2})+ m_2^2\, (\omega_{\phi_1\phi_2}+R_{\phi_1\chi_2} - \tau_{\phi_1\phi_2})\right)\nn
\mathcal{C}_{11}(\chi_1,\chi_2,\phi_1,\phi_2)&=&-\frac{m_j\,s\,L(\chi_1,\phi_1,\phi_2)}{4\,b_{\phi_1\phi_2}\,(R_{\phi_1\chi_1}-R_{\phi_1\chi_2})}\left(3\, R_{\phi_1\chi_1} - 4\, m_i^2\, s\right)\nn
&&+\frac{m_j\,s\,L(\chi_1,\phi_1,\phi_2)}{4\,b_{\phi_1\phi_2}\,(R_{\phi_1\chi_1}-R_{\phi_1\chi_2})}\left(3\, R_{\phi_1\chi_2} - 4\, m_i^2\, s\right)\nn
\mathcal{C}_{12}(\chi_1,\chi_2,\phi,\phi)&=&-\frac{1}{48\,m_\phi^4\,S^2}\,\left(b_{\phi\phi}^2 + 3\, (R_{\phi\chi_1}^2 + R_{\phi\chi_1}\,R_{\phi\chi_2}+ 
R_{\phi\chi_2}^2 - (2\, l_i - l_{ia} + 4\, m_\phi^2)\right.\nn&&\left. (R_{\phi\chi_1} +R_{\phi\chi_2})\, S - (l_i\, l_{ia} - 
8\, d_{\phi\phi}\, m_i^2 + 4\, (2\, d + 4\, d_{\phi\phi} - 2\, l_i + l_{ia})\right.\nn&&\left.m_\phi^2 - 4\, m_\phi^4)\, S^2)\right)+\frac{S^2\,\mathcal{X}_1(\chi_1,\chi_2,\phi,\phi)}{4\,b_{\phi\phi}\,m_\phi^2\,(R_{\phi\chi_1}-R_{\phi\chi_2})}\,\left(-(4\, d_{\phi\phi} - l_i)\, (l_i - l_{ia})\right.\nn&&\left. m_i^2 + 2\, (l_i + 2\, m_i^2)\, m_j^2\, m_\phi^2 + 4\, d\, m_\phi^4\right)-\frac{S\,\mathcal{Y}_1(\chi_1,\chi_2,\phi,\phi)}{4\,b_{\phi\phi}\,m_\phi^4\,(R_{\phi\chi_1}-R_{\phi\chi_2})}\,\left(d_{\phi\phi}\right.\nn&&\left. (l_i - l_{ia})\, m_i^2 + 4\, d_{\phi\phi}\, (-2\, m_i^2 + m_j^2)\,m_\phi^2 + 
m_\phi^2\, ((l_i - l_{ia})\, (l_i + m_i^2)\right.\nn&+&\left.m_\phi^2\, (-4\, d + 2\, m_j^2 + S))\right)
+\frac{\mathcal{Z}_1(\chi_1,\chi_2,\phi,\phi)}{16\,b_{\phi\phi}\,m_\phi^4\,(R_{\phi\chi_1}-R_{\phi\chi_2})}\,\left(l_i^2 - l_i\, l_{ia} + 4\, d_{\phi\phi}\, m_i^2 \right.\nn&-&\left. 2\, (2\, d + 4\, d_{\phi\phi} - 3\, l_i + l_{ia})\, m_\phi^2 + 
2\, m_\phi^4\right)+\frac{\mathcal{R}_1(\chi_1,\chi_2,\phi,\phi)}{32\,b_{\phi\phi}\,m_\phi^4\,S\,(R_{\phi\chi_1}-R_{\phi\chi_2})}\nn&&\left(-3\, l_i + l_{ia} - 4\, m_\phi^2\right)+\frac{\mathcal{Q}_1(\chi_1,\chi_2,\phi,\phi)}{32\,b_{\phi\phi}\,m_\phi^4\,S^2\,(R_{\phi\chi_1}-R_{\phi\chi_2})}\nonumber
\end{eqnarray}
\begin{eqnarray}
\mathcal{C}_{13}(\chi_1,\chi_2,\phi,\phi)&=&\frac{m_{\kk}\, m_{\kl}\, (d_{\phi\phi} -m_\phi^2)}{2\,m_\phi^4}-\frac{m_{\kk}\,m_{\kl}\,S\,\mathcal{X}_1(\chi_1,\chi_2,\phi,\phi)}{4\,b_{\phi\phi}\,m_\phi^2\,(R_{\phi\chi_1}-R_{\phi\chi_2})}\,\left(l_i\, (-l_i + l_{ia})\right.\nn&+&\left. 4\, d\, m_\phi^2\right)+\frac{m_{\kk}\,m_{\kl}\,S\,\mathcal{Y}_1(\chi_1,\chi_2,\phi,\phi)}{4\,b_{\phi\phi}\,m_\phi^2\,(R_{\phi\chi_1}-R_{\phi\chi_2})}\,\left(d_{\phi\phi}\, (l_i - l_{ia}) - 2\, l_i\,m_\phi^2\right)\nn&+&\frac{m_{\kk}\,m_{\kl}\,(d_{\phi\phi}-m_\phi^2)}{4\,b_{\phi\phi}\,m_\phi^2\,(R_{\phi\chi_1}-R_{\phi\chi_2})}\nn
\mathcal{C}_{14}(\chi_1,\chi_2,\phi,\phi)&=&\frac{9\,m_i\,m_j\,S}{2\,b_{\phi\phi}\,(R_{\phi\chi_1}-R_{\phi\chi_2})}\,\left(2\, m_i^2\, S\, \mathcal{X}_1(\chi_1,\chi_2,\phi,\phi) + 2\, m_\phi^2\, S\, \mathcal{X}_1(\chi_1,\chi_2,\phi,\phi)\right.\nn&-&\left. \mathcal{Y}_1(\chi_1,\chi_2,\phi,\phi)\right)\nn
\mathcal{C}_{15}(\chi_1,\chi_2,\phi,\phi)&=&\frac{9\,m_i\,m_j\,m_{\kk}\,m_{\kl}\,S}{b_{\phi\phi}\,(R_{\phi\chi_1}-R_{\phi\chi_2})}\nonumber
\end{eqnarray}
\begin{eqnarray}
\mathcal{C}_{16}(\chi_1,\chi_2,\phi,\phi)&=&\frac{3\,m_j}{4\,m_\phi^2}+\frac{3\,m_j}{8\,m_\phi^2\,b_{\phi\phi}\,(R_{\phi\chi_1}-R_{\phi\chi_2})}\,\left(-8\, m_i^2\,m_\phi^2\, S^2\, \mathcal{X}_1(\chi_1,\chi_2,\phi,\phi)\right.\nn &+&\left. 6\, m_\phi^2\, S\, \mathcal{Y}_1(\chi_1,\chi_2,\phi,\phi) - \mathcal{Z}_1(\chi_1,\chi_2,\phi,\phi)\right)\nn
\mathcal{C}_{17}(\chi_1,\chi_2,\phi,\phi)&=&\frac{3\,m_i}{4\,m_\phi^2}+\frac{3\,m_i\,S^2\,\mathcal{X}_1(\chi_1,\chi_2,\phi,\phi)}{4\,b_{\phi\phi}\,m_\phi^2\,(R_{\phi\chi_1}-R_{\phi\chi_2})}\,\left((2\, d_{\phi\phi} - l_i)\, (l_i - l_{ia}) \right.\nn&-&\left. m_\phi^2\, (-4 \,d + 2\, S)\right)
-\frac{3\,m_i\,S\,\mathcal{Y}_1(\chi_1,\chi_2,\phi,\phi)}{8\,b_{\phi\phi}\,m_\phi^2\,(R_{\phi\chi_1}-R_{\phi\chi_2})}\,\left((4\, d_{\phi\phi}- 3\, l_i + l_{ia} - 2\, m_\phi^2\right)\nn&-&\frac{3\,m_i\,\mathcal{Z}_1(\chi_1,\chi_2,\phi,\phi)}{8\,b_{\phi\phi}\,m_\phi^2\,(R_{\phi\chi_1}-R_{\phi\chi_2})}\nonumber
\end{eqnarray}
Where $m_i$,$m_j$ are masses of $\chi^0_i$ and $\chi_j^0$ respectively. 
\bibliographystyle{ieeetr}
\bibliography{cross.bib}
\end{document}